\documentclass[12pt]{article}
\usepackage{amsmath}
\usepackage{graphicx}
\usepackage{enumerate}
\usepackage{natbib}
\usepackage{url} 
\usepackage{tabularx}
\usepackage{rotating}
\usepackage{threeparttable}
\usepackage{multirow}
\usepackage{booktabs}
\usepackage{algorithm}
\usepackage{algorithmic}
\usepackage{float}
\usepackage{color}
\usepackage{pgfplots}
\usepackage{tikz}
\usetikzlibrary{positioning, shapes.geometric}
\usetikzlibrary{shapes,snakes}
\newcommand{\blind}{1}
\usepackage{amsfonts}
\usepackage[bookmarks=true]{hyperref}
\usepackage{ragged2e,subfigure}
\renewcommand{\raggedright}{\leftskip=0pt \rightskip=0pt plus 0cm}
\newtheorem{theorem}{Theorem}

\newtheorem{lemma}{Lemma}
\newtheorem{proposition}{Proposition}
\newtheorem{corollary}{Corollary}
\def\beq{\begin{equation}}
\def\eeq{\end{equation}}
\def\beqr{\begin{eqnarray}}
\def\eeqr{\end{eqnarray}}
\def\beqrs{\begin{eqnarray*}}
\def\eeqrs{\end{eqnarray*}}
\def\bet{\begin{theorem}}
\def\eet{\end{theorem}}
\def\bel{\begin{lemma}}
\def\eel{\end{lemma}}
\def\bep{\begin{proposition}}
\def\eep{\end{proposition}}
\def\bec{\begin{corollary}}
\def\eec{\end{corollary}}

\begin{document}

\def\spacingset#1{\renewcommand{\baselinestretch}%
{#1}\small\normalsize} 
\spacingset{1.34}

\date{}
\if1\blind
{
  \title{Low-rank Latent Matrix-factor Prediction Modeling for Generalized High-dimensional Matrix-variate Regression}
\author{Yuzhe Zhang\thanks{School of Management, University of Science and Technology of China, China. Co-first author.}, 
Xu Zhang\thanks{School of Mathematical Sciences, South China Normal University, China.  Co-first author.}, 
Hong Zhang\thanks{School of Management, University of Science and Technology of China, China.}, 
Aiyi Liu\thanks{National Institute of Child Health and Human Development, National Institutes of Health, 
6710B Rockledge Drive
Bethesda MD 20817, USA.} 
and Catherine C. Liu\thanks{Department of Applied Mathematics, The Hong Kong Polytechnic University, Hong Kong SAR. Corresponding author: macliu@polyu.edu.hk.}}
  \maketitle
} \fi

\if0\blind
{
  \bigskip
  \bigskip
  \bigskip
  \begin{center}
    {\LARGE\bf Low-rank Latent Matrix-factor Prediction Modeling for Generalized High-dimensional Matrix-variate Regression}
\end{center}
  \medskip
} \fi

\vspace{-1cm}
\bigskip
\begin{abstract}
Motivated by diagnosing the COVID-19 disease using 2D image biomarkers from computed tomography (CT) scans, we propose a novel latent matrix-factor regression model to predict responses that may come from an exponential distribution family, where covariates include high-dimensional matrix-variate biomarkers. 
    A latent generalized matrix regression (LaGMaR) is formulated, where the latent predictor is a low-dimensional matrix factor score extracted from the low-rank signal of the matrix variate through a cutting-edge matrix factor model. 
    Unlike the general spirit of penalizing vectorization plus the necessity of tuning parameters in the literature, instead, our prediction modeling in LaGMaR conducts dimension reduction that respects the geometry characteristic of intrinsic two-dimensional structure of the matrix covariate and thus avoids iteration. This greatly relieves the computation burden, and meanwhile maintains structural information so that the latent matrix factor feature can perfectly replace the intractable matrix-variate owing to high-dimensionality.
    The estimation procedure of LaGMaR is subtly derived by transforming the bilinear form matrix factor model onto a high-dimensional vector factor model, so that the method of principle components can be applied.
    We establish bilinear-form consistency of the estimated matrix coefficient of the latent predictor and consistency of prediction. The proposed approach can be implemented conveniently.
    Through simulation experiments, prediction capability of LaGMaR is shown to outperform existing penalized methods under diverse scenarios of generalized matrix regressions. 
    Through the application to a real COVID-19 dataset, the proposed approach is shown to predict efficiently the COVID-19. 
 \end{abstract}

\vfill

\spacingset{1.59} 

\section{Introduction}
\label{sec1}

Matrix-variates are commonly encountered in contemporary biometry, genetic image clinics, and a wide range of medical studies. For example, two-dimensional image biomarkers in medical digital image processing are usually restored in high-dimensional matrices. Such image biomarkers have been playing an important role in aiding diagnostics, medical care plan, prognosis, and monitoring of treatment outcomes in clinical trials \citep{Suetens2017fundamentals}. Research into regressions involving high-dimensional matrix-variates as predictors or responses has gained considerable interest in the past decade or so (\citet{li2010dimension, hung2013matrix, zhou2014regularized, DingCook2018, jiang2020bayesian, yu2022mapping}; among others). 
In this paper, we target to provide and fit a novel working regression model that can efficiently predict various responses, where the high-dimensional matrix covariate is effectively dealt with from the perspective of the matrix factor model tool.

\begin{figure}
    \centering                                                    
    \subfigure[]{
        \begin{minipage}[t]{0.48\textwidth}
        \centering                                     \includegraphics[width=5.5cm,height=4cm]{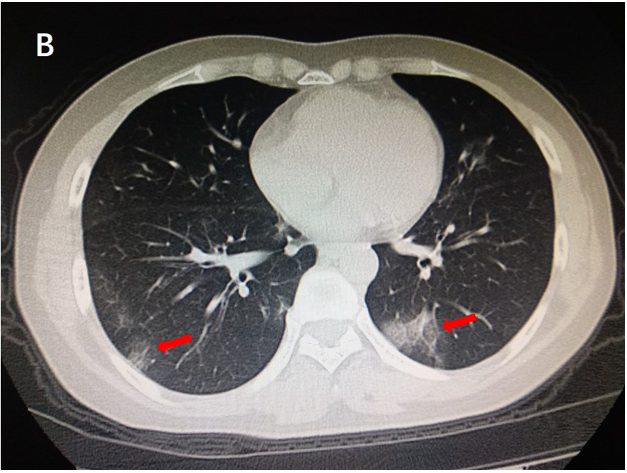}
        \end{minipage}}
        \subfigure[]{
        \begin{minipage}[t]{0.48\textwidth}
        \centering                                                      \includegraphics[width=5.5cm,height=4cm]{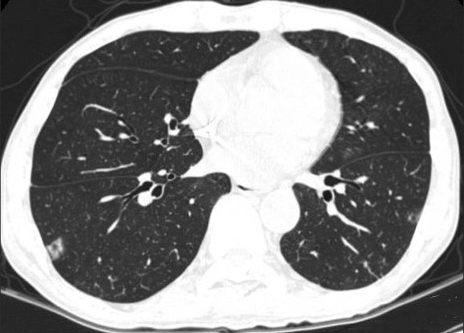}
        \end{minipage}}
        \subfigure[]{
        \begin{minipage}[t]{0.48\textwidth}
        \centering                                                      \includegraphics[width=5.5cm,height=4cm]{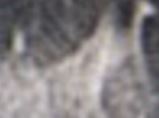}
        \end{minipage}}
        \subfigure[]{
        \begin{minipage}[t]{0.48\textwidth}
        \centering                                                      \includegraphics[width=5.5cm,height=4cm]{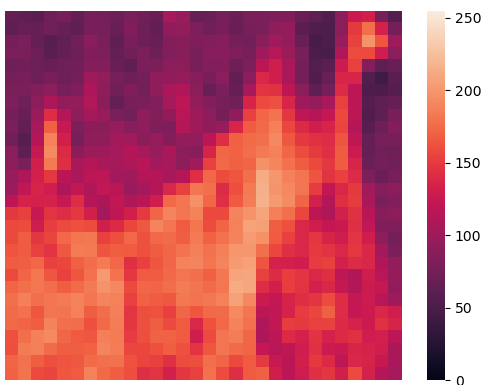}
        \end{minipage}}
        \caption{(a): COVID-19 pneumonia with typical imaging features according to the Radiological Society of North America 
(RSNA) chest CT classification system \citep{kwee2020chest}; (b): CT Images of a non-COVID patient. (c): The locally enlarged region pointed by the right red arrow in Subfigure (a). (d): The heatmap of the pixel value of Subfigure (c). The brighter the original region is, the larger pixel value it has. The darker the original region is, the smaller pixel value it has.}
        \label{COVID}                                   
\end{figure}

Our work is motivated by the imperative issue of discriminating infection of novel coronavirus (COVID-19) subjects from slices of their axial volumetric scanning of chest computed tomography (CT) \citep{he2020sample}. 
Identifying and isolating people infected with COVID-19 are important steps in prevention and control of this global pandemic.
Specialists usually identify COVID-19 patients by observing light grey tissues in a CT scan image, refer to Figure \ref{COVID} for an example.
Such mainstream means of diagnostics of COVID-19 is very resource demanding due to a large scale CT scans during the outbreak of the pandemic. 
However, the worldwide medical and hospitality resources are limited. 
This calls an urgent need for fast automatic apps to conveniently and accurately screen COVID-19 patients based on CT scans. 

\centerline{[Figure \ref{COVID} about here.]}

CT scan images, as two-dimensional (2D) medical digital images, are intrinsically matrices. 
Pixels of a 2D image are stored in a rectangular grid in digital image processing, where the height and the width of the image correspond to the grid-row size and the grid-column size, respectively, and the pixel value is assigned to each cross location of the grid. 
Therefore, the pixel value, the height and the width of the image correspond to the entry, the row size and the column size of a matrix, respectively.
Note that, the induced matrix is high dimensional since the heights and the widths of all raw image samples range from 153 to 1853 and 124 to 1485, respectively. 
In terms of explanation, the entry of the matrix, or equivalently the pixel value of the 2D image, may represent brightness, signal intensity, color characteristics such as hue and saturation, or any other derived quantity. 
Furthermore, regions with larger pixel values in an image are visually brighter, while those with smaller pixel values are visually darker.
Take the grayscale image for instance, the gray values of pixels generally range from 0 to 255, with 255 for white and 0 for black.

There is a wealth of literature investigating regression analysis involving matrix-variate covariates. 
To name a few, early works in statistical community traced to dimension-folding methods, which take the spirit of preserving the matrix structure \citep{li2010dimension}. 
However, such methods need data preprocessing and it is hard to tune the dimension of the associated central dimension-folding subspace. 
Then, in the next decade, researchers focused on generalized matrix regressions using regularization schemes. 
\citet{zhou2014regularized} estimated matrix coefficients directly by adding penalties on singular values within the matrix regression setting.
A class of tensor regression methods can be solutions for the matrix regression as well. For example, the methods of \citet{zhou2013tensor} and \citet{li2018tucker} were based on CP and Tucker decompositions, respectively, which need penalization on vectorized tensor coefficients.
\citet{hung2013matrix} studied the specific logistic matrix regression by penalizing the CP tensor decomposition but it still needs data preprocessing.
All of the aforementioned methods need tuning parameters and some even need data preprocessing, making analysis results sensitive or computing intensive. Recently, there comes the third research line under the Bayesian paradigm, which first extracts vector-type features from a tensor-variate covariate
and assign priors on the coefficients of the features, and then conducts regression  \citep{Miranda2018AOAS} or classification \citep{jiang2020bayesian}. They may suffer from non-automatic sampling and slow MCMC convergence owing to tuning many hyperparameters. 

Inspired by the pros and cons of the existing literature, we target to achieve high prediction capability by respecting structural geometry and implement efficient computing by circumventing parameter tuning. The spirit is implemented by extracting \textit{low-dimensional matrix features} from the low-rank approximation signal of a high-dimensional matrix-variate as doable covariates in a regression, yielding a latent generalized matrix regression called LaGMaR thereafter. Our modeling and methodology are actually driven by the cutting-edge unsupervised learning for the bilinear-form matrix factor models  (\citet{wang2019factor} for two-dimensional time series and  \citet{chen2021statistical} for independent matrix-variate sequences), which recently have drawn arising research interest as extensions of the high-dimensional vector factor model 
(\citet{bai2003inferential}; among others). 
The common characteristic is to use the bilinear-form of a low-dimensional matrix factor to separate the signal from the noise for a random matrix object. Then, the classical method of principle components can estimate each component in the low-rank approximation and the number of factors. Although low-rank approximation of matrix objects has been broadly investigated in applied mathematics and machine learning community (\cite{ye2005generalized,2014Low}; among others), its applications in supervised learning and unsupervised learning are sporadic.

The rest of this paper is organized as follows. Section \ref{sec2} formulates the latent matrix generalized linear model and develops an estimation procedure and its algorithm. Section \ref{sec3} displays the finite sample performance of LaGMaR compared with several existing methods. Section \ref{sec4} illustrates a real-world application of LaGMaR on the COVID-19 CT dataset. Some conclusion remarks and discussions are provided in Section \ref{sec5}.

\section{Latent-factor Generalized Matrix Regression}
\label{sec2}

Let $(Y, \mathbf{X}, \mathbf{v})$ be generic triplet variables, where $Y$ is a response variable that may come from an exponential family of distributions, $\mathbf{X} \in \mathbb{R}^{p_1 \times p_2}$ and $\mathbf{v} \in \mathbb{R}^{m}$ are covariates of matrix-variate and vector-variate respectively.
Here $p_1$ and $p_2$ may be sufficiently large and $m$ is finite. Under a classical generalized linear regression setting, the response $Y$ can be related to the covariates $(\mathbf{X}, \mathbf{v})$ through a prespecified link function $g(\cdot)$,    
\begin{equation}\label{glm}
	g\left\{E(Y|\mathbf{X}, \mathbf{v})\right\} = \gamma + \langle\mathbf{B},\mathbf{X}\rangle + \beta^{\top}\mathbf{v},
\end{equation}
where $\gamma$ is an intercept term, $\mathbf{B}=(b_{ij})\in\mathbb{R}^{p_1\times p_2}$ is a matrix coefficient, $\beta\in\mathbb{R}^m$ is a vector  coefficient, and $\left<\cdot, \cdot \right>$ represents an inner product such that
$\left<\mathbf{B}, \mathbf{X} \right> =  \text{vec}(\mathbf{B})^{\top}\text{vec}(\mathbf{X})$ with $\text{vec}(\mathbf{B})$ being a
$(p_1p_2)$-dimensional column vector stacking  all column vectors of the matrix $\mathbf{B}$.
The high-dimensional matrix $\mathbf{X}$ generates the high-dimensionality of $\text{vec}(\mathbf{B})$ under the conventional regression model (\ref{glm}). Therefore, traditional vectorization may lose information and incurs a biased or even inefficient fitting.

\subsection{Model formulation}
In this subsection, we propose a latent matrix factor regression, called the LaGMaR, as the working model for the generalized matrix regression (\ref{glm}).
Like that the $k$th entry of the finite-dimensional covariate vector $\mathbf{v}$ has its own effect $\beta_{k}$ on the response, 
the $(i,j)$th entry of the matrix-variate covariate $\mathbf{X}$ possesses its interaction effect $b_{ij}$ from both row $i$ and column $j$ on the response. Therefore, intuitively, to preserve the two-dimensional structure of the matrix-variate will maintain more interrelate information, leading to higher prediction capability.
In literature, the common strategy under the matrix regression is to penalize the high-dimensionality of the matrix covariate directly. Instead, inspired by the spirit of integration of separable covariance structure and factor models in \citet{fosdick2014separable}, we integrate the generalized regression framework with an extracted low-dimensional matrix factor predictor, formulated by  
\begin{equation}
\label{famglm}
\begin{aligned}
\left\{\begin{array}{l}
    g\left\{E(Y|\mathbf{Z}, \mathbf{v})\right\} = \gamma + \left<\mathbf{A}, \mathbf{Z}\right> + \beta^{\top} \mathbf{v} \quad \text{(generalized regression)},\\ 
    \mathbf{X} = \mathbf{R} \mathbf{Z} \mathbf{C}^{\top} + \mathbf{E} \quad \text{(matrix factor model)},
\end{array}\right.
\end{aligned}
\end{equation}
where $\mathbf{Z} \in \mathbb{R}^{k_1 \times k_2}$ ($k_1 \ll p_1$ and $k_2 \ll p_2$) is an unobserved common fundamental matrix factor,
$\mathbf{A} \in \mathbb{R}^{k_1 \times k_2}$ is the matrix coefficient of the latent factor covariate $\mathbf{Z}$, 
$\mathbf{R} \in \mathbb{R}^{p_1 \times k_1}$ and $\mathbf{C} \in \mathbb{R}^{p_2 \times k_2}$ are row and column factor loading matrices with $k_1$ and $k_2$ being the numbers of row and column factors, respectively, 
and $\mathbf{E} \in \mathbb{R}^{p_1 \times p_2}$ is a random error matrix.
The latent matrix factor
$\mathbf{Z}$ becomes an excellent low-dimensional structural regressor that catches the intrinsic interrelationship among rows and columns of $\mathbf{X}$,
since $\mathbf{R} $  and $\mathbf{C} $ reflect the dependencies between rows and columns.
Without loss of generality, we still follow the same notation of $\gamma$ and $\beta$ in (\ref{glm}) since the meaning is invariant. 

With i.i.d. sample data set $\left\{(Y_i, \mathbf{X}_i, \mathbf{v}_i)\right\}_{i=1}^{n}$, the interpretation of model (\ref{famglm}) can be demonstrated. On one hand, one can view the factor model part in model (\ref{famglm}) as 
    \[x_{k j,i} = \sum_{h,l}r_{k h} z_{h l, i} c_{j l} + e_{k j, i},\]
where $x_{kj,i}$ is the $(k,j)$th element of the $i$th matrix variate observation $\mathbf{X}_i$, $r_{kh}$ is the $(k,h)$th element of the row loading matrix $\mathbf{R}$, $c_{jl}$ is the $(j,l)$th element of the column loading matrix $\mathbf{C}$, $z_{hl,i}$ is the $(h,l)$th element of the $i$th latent matrix factor, and $e_{kj,i}$ is the $(k,j)$th element of the $i$th error matrix $\mathbf{E}_i$. Each summand $r_{kh} z_{hl,i} c_{jl}$ can be interpreted as the latitudinal contribution from the $k$th row to the $h$th row, and the longitudinal contribution of column $j$ to column $l$ in the interaction effect from row $h$ to column $l$. The total volume $x_{kj,i}$ is the aggregation of the interaction volumes from row $k$ to column $j$ through all the latent factor scores.

On the other hand, the low-dimensional regressor matrix $\mathbf{Z}$, which replaces the original matrix-variate predictor $\mathbf{X}$, plays a critical role in the latent matrix factor modeling in that it extracts and maintains the inherent information contained in the matrix-variate covariate $\mathbf{X}$. Meanwhile, the number of parameters to be estimated reduce significantly from $1 + p_1p_2 + m$ to $1 + k_1 k_2 + m$. Once we can estimate the latent matrix factor $\mathbf{Z}$ accurately, the response in both the working model and original model can be fitted effectively through the general methods.

\subsection{Estimation and Algorithms}

In this subsection, we provide a two-step statistical estimation procedure to fit model (\ref{famglm}).
In the first step, we extract the low-dimensional matrix feature based on the low-rank approximation of the high-dimensional matrix-variate covariate. Note that the matrix factor model in (\ref{famglm}) is not identifiable since it is unchanged if one replaces the triplet $(\mathbf{R}, \mathbf{C}, \mathbf{Z})$ on the right-hand side by $(\mathbf{R}\mathbf{H}_1, \mathbf{C}\mathbf{H}_2, \mathbf{H}_1^{\top}\mathbf{Z}\mathbf{H}_2)$ for any orthogonal matrices $\mathbf{H}_1$ and $\mathbf{H}_2$. 
One may assume that the columns of the factor loading matrices $\mathbf{R}$ and $\mathbf{C}$ are orthogonal, that is 
\begin{equation}\label{iden}
	\mathbf{R}^{\top}\mathbf{R} = p_1 \mathbf{I}_{k_1}, \mathbf{C}^{\top}\mathbf{C} = p_2 \mathbf{I}_{k_2}.
\end{equation}
Under the constraint (\ref{iden}), the linear space spanned by the columns of $\mathbf{R}$ or $\mathbf{C}$, denoted by $\mathcal{M}(\mathbf{R})$ or $\mathcal{M}(\mathbf{C})$, is uniquely defined, that is,  $\mathcal{M}(\mathbf{R})=\mathcal{M}(\mathbf{R}\mathbf{H}_1)$ and $\mathcal{M}(\mathbf{C})=\mathcal{M}(\mathbf{C}\mathbf{H}_2)$. 
Once such $\mathbf{R}$ and $\mathbf{C}$, or equivalently, $\mathcal{M}(\mathbf{R})$ and $\mathcal{M}(\mathbf{C})$, are estimated, the matrix factor score $\hat{\mathbf{Z}}$ can be uniquely determined (refer to (\ref{estZ})).
{When consider the case of $p_1$ and $p_2$ tending to infinity, condition (\ref{iden}) should be $p_1^{-1}\mathbf{R}^{\top}\mathbf{R}$ and $p_2^{-1}\mathbf{C}^{\top}\mathbf{C}$ tending to $\mathbf{I}_{k_1}$ and $\mathbf{I}_{k_2}$ respectively, 
and $\mathbf{H}_1$ and $\mathbf{H}_2$ should be asymptotic orthogonal matrices consequently.}

Denote  $\mathbf{F}_i=\mathbf{Z}_i\mathbf{C}^{\top}\in\mathbb{R}^{k_1\times p_2}$. One has 
\[\mathbf{X}_i=\mathbf{R} \mathbf{F}_i+\mathbf{E}_i, i=1 \dots n,\]
which has the form of classical $p_1$-dimensional \textit{vector factor model} for each column of $\mathbf{X}_i$ with sample size $(np_2)$
\citep{bai2003inferential}. 
Then one may estimate the row factor loading $\mathbf{R}$ based on the spectral decomposition on $\hat{\mathbf{M}}_R$, the column-wise sample variance-covariance matrix of the column of $\mathbf{X}$, yielding the principle component estimator  $\hat{\mathbf{R}}$, formulated as 
\begin{equation}
    \label{ms1}
      \hat{\mathbf{R}}=\sqrt{p_1}~ eig(\hat{\mathbf{M}}_R,k_1)
    ~\text{with}~
    \hat{\mathbf{M}}_R = \frac{1}{np_1 p_2} \sum_{i=1}^{n} \mathbf{X}_i \mathbf{X}_i^{\top},
\end{equation}
where $eig(U, r)$ represents the matrix with the top $r$ eigenvectors of $U$ as columns.
Analog to the above method of principle component, one  has 
\[\mathbf{X}_i^{\top}= \mathbf{C}\tilde{\mathbf{F}}_i^{\top}+\mathbf{E}_i^{\top}, i=1 \dots n,\]
where $\tilde{\mathbf{F}}_i=\mathbf{R}\mathbf{Z}_i\in\mathbb{R}^{p_1\times k_2}$, and the right-hand side of which has the form of classical $p_2$-dimensional vector factor model for each row of $\mathbf{X}_i$ with sample size $(np_1)$. The corresponding principle component estimator $\hat{\mathbf{C}}$ with companion
row-wise sample variance-covariance matrix are expressed as
\begin{equation}
    \label{ms2}
    \hat{\mathbf{C}}= \sqrt{p_2}~eig(\hat{\mathbf{M}}_C,k_2)
    ~\text{with}~
    \hat{\mathbf{M}}_C = \frac{1}{np_1p_2} \sum_{i=1}^{n} \mathbf{X}_i^{\top}\mathbf{X}_i.
\end{equation}
Then one can obtain an explicit estimator of the matrix factor $\mathbf{Z}_i$
\begin{equation}
    \label{estZ}
    \hat{\mathbf{Z}}_i = \frac{1}{p_1 p_2}\hat{\mathbf{R}} ^{\top}\mathbf{X}_i \hat{\mathbf{C}}.
\end{equation}
Now we come to the estimation of the factors $k_1$ and $k_2$ in (\ref{ms1}) and (\ref{ms2}).
We follow the popular method of ratio-based estimator, 
which was first raised by \citet{lam2012factor} with the spirit of \citet{wang2010factor}, 
\begin{equation}
    \label{rk1}
	\hat{k}_1 = \arg \max_{1 \leq j \leq \left \lceil p_1/2 \right \rceil} \frac{\hat{\lambda}_{1,j}}{\hat{\lambda}_{1,j+1}},\quad	\hat{k}_2 = \arg \max_{1 \leq j \leq \left \lceil p_2/2 \right \rceil} \frac{\hat{\lambda}_{2,j}}{\hat{\lambda}_{2,j+1}},
\end{equation}
where $\hat{\lambda}_{1,1} \geq \hat{\lambda}_{1,2} \geq \cdots \geq \hat{\lambda}_{1,p_1} \geq 0$ are eigenvalues of $\hat{\mathbf{M}}_R$ in descending order, and $\hat{\lambda}_{2,1} \geq \hat{\lambda}_{2,2} \geq \cdots \geq \hat{\lambda}_{2,p_2} \geq 0$ 
are eigenvalues of $\hat{\mathbf{M}}_C$ in descending order, respectively.

In the second step, we fit the low-dimensional generalized regression model based on the triplet data $\{Y_i, \hat{\mathbf{Z}}_i, \mathbf{v}_i\}_{i=1}^{n}$ by estimating  $\hat{\gamma}$, $\hat{\mathbf{A}}$, $\hat{\beta}$, that is,  the intercept, matrix coefficient, and the vector coefficient together.
The fitting and prediction procedures based on model (\ref{famglm}) are summarized in Algorithms \ref{lagmar_fit} and \ref{lagmar_pred}, respectively.

\begin{algorithm}[htbp]
\caption{LaGMaR Fitting}
\label{lagmar_fit}
\begin{algorithmic}[1]
\REQUIRE Response variables $\left\{Y_i\right\}_{i=1}^{n}$, matrix-valued covariates $\left\{\mathbf{X}_i\right\}_{i=1}^{n}$, vector-valued covariates $\left\{\mathbf{v}_i\right\}_{i=1}^{n}$.  
\STATE Calculate matrices $\hat{\mathbf{M}}_R$ and $\hat{\mathbf{M}}_C$ according to (\ref{ms1}) and (\ref{ms2}). 
\STATE Choose the number of row factors, $k_1$, and the number of column factors,  $k_2$, according to (\ref{rk1}). 
\STATE Calculate loading matrices according to (\ref{ms1}) and (\ref{ms2}).
\STATE Calculate latent factor matrices according to (\ref{estZ}).
\STATE Fit GLM using the data set $\big\{(Y_i, \hat{\mathbf{Z}}_i, \mathbf{v}_i)\big\}_{i=1}^{n}$.
\end{algorithmic}
\end{algorithm}

\begin{algorithm}[htbp]
\caption{LaGMaR Prediction}
\label{lagmar_pred}
\begin{algorithmic}[1]
\REQUIRE A newly observed  data point $(\mathbf{X}_{\text{new}}, \mathbf{v}_{\text{new}})$, coefficient estimate $\left\{\hat{\gamma}, \hat{\mathbf{A}}, \hat{\beta}\right\}$ using Algorithm \ref{lagmar_fit}, and estimated loading matrices $\hat{\mathbf{R}}$ and $\hat{\mathbf{C}}$ using Algorithm \ref{lagmar_fit}. 
\STATE Calculate the predicted latent factor matrix $\hat{\mathbf{Z}}_{\text{new}} = \hat{\mathbf{R}}^{\top}\mathbf{X}_{\text{new}} \hat{\mathbf{C}}/(p_1p_2)$
\STATE Calculate the predicted regression mean $\hat{\mu}=\hat{\gamma} + \big<\mathbf{\hat{A}, \hat{\mathbf{Z}}_{\text{new}}}\big> + \hat{\beta}^{\top} \mathbf{v}_{\text{new}}$.
\STATE Calculate the predicted response $\hat{Y}_{\text{new}}=g^{-1}(\hat{\mu})$.
\end{algorithmic}
\end{algorithm}

\textbf{Remark}: Here we relate LaGMaR to some existing works. To estimate $\mathbf{Z}$, we have to first estimate the row-wise and column-wise loading matrice parameters $\mathbf{R}$ and $\mathbf{C}$ in the bilinear form low-rank signal.
There are a few works on unsupervised learning on the exact bilinear form matrix factor models, where the matrix-variate observations may be time series or independent data. 
Their line of estimation is to conduct spectral decomposition on either sample auto-covariance matrices \citep{wang2019factor} or sample variance-covariance matrices \citep{chen2021statistical,yu2021projected}.
In contrast, we subtly transform the bilinear matrix factor model onto a classical high-dimensional vector factor model, so that we can directly apply the results of the seminal work of \citet{bai2003inferential}.

\subsection{Bilinear-form Consistency and Asymptotic Prediction Invariance}
In this subsection, we derive that good prediction capability of LaGMaR can still be retained without consistent estimator of the coefficient matrix $\mathbf{A}$ in the generalized matrix regression (\ref{famglm}) because of the characteristic of the tool of the matrix factor model. 
{Note that when derive the asymptotic property, we assume $n$, $p_1$ and $p_2$ tend to infinity.
Hence, the constraint (\ref{iden}) for estimation in Subsection 2.2 turns to $p_1^{-1}\mathbf{R}^{\top}\mathbf{R}$ and $p_2^{-1}\mathbf{C}^{\top}\mathbf{C}$ tending to $\mathbf{I}_{k_1}$ and $\mathbf{I}_{k_2}$ respectively. 
Due to the lack of uniqueness of $\mathbf{R}$ and $\mathbf{C}$, we may rotate the estimated factor loading matrices $\hat{\mathbf{R}}$ and $\hat{\mathbf{C}}$ to facilitate the estimation procedure so that the factor loading matrices can be uniquely defined up to \textit{appropriate asymptotic orthogonal rotations}, in the sense that $\mathcal{M}(\hat{\mathbf{R}})=\mathcal{M}(\hat{\mathbf{R}}\mathbf{H}_1)$ and $\mathcal{M}(\hat{\mathbf{C}})=\mathcal{M}(\hat{\mathbf{C}}\mathbf{H}_2)$.}
Once we have obtained the pair estimators $(\hat{\mathbf{R}}, \hat{\mathbf{C}})$ under the constraint (\ref{iden}), 
we can derive the estimated matrix factor $\hat{\mathbf{Z}}_i$ determined by equation (\ref{estZ}), 
which can be further written as 
$$\hat{\mathbf{Z}}_i=\frac{1}{p_1p_2}\mathbf{H}_1 \mathbf{H}_1^{\top} (\hat{\mathbf{R}} ^{\top}\mathbf{X}_i \hat{\mathbf{C}}) \mathbf{H}_2 \mathbf{H}_2^{\top}  \quad (\text{or equivalently} \quad \mathbf{H}_1^{\top}\hat{\mathbf{Z}}_i\mathbf{H}_2=\frac{1}{p_1p_2}(\hat{\mathbf{R}}\mathbf{H}_1)^{\top}\mathbf{X}_i(\hat{\mathbf{C}}\mathbf{H}_2)).$$
That is, as ${\mathbf{Z}}_i$ will change along the orthogonal transformations of $\mathbf{R}$ and $\mathbf{C}$ owing to not being separately identifiable, accordingly $\hat{\mathbf{Z}}_i$ will change along the orthogonal transformations of $\hat{\mathbf{R}}$ and $\hat{\mathbf{C}}$ as well.  
This yields as stated in Lemma \ref{lemma} that $\hat{\mathbf{Z}}_i$ is consistent up to a bilinear-form transformation.

\begin{lemma}[Bilinear-form consistency]
\label{lemma}
Under some regular assumptions for matrix factor models, there exist asymptotic orthogonal matrices $\mathbf{H}_1\in\mathbb{R}^{k_1\times k_1}$ and $\mathbf{H}_2\in\mathbb{R}^{k_2\times k_2}$ such that 
\begin{equation}
    \label{lemma21}
    \hat{\mathbf{Z}}_i-\mathbf{H}_1^{\top}\mathbf{Z}_i \mathbf{H}_2=o_p(1)\mathbf{1}_{k_1\times k_2},
\end{equation}
where $\otimes$ is the Kronecker product, and $\mathbf{1}_{k_1 k_2}$ is a column vector of length $k_1 k_2$ with all entries being 1.
\end{lemma}

The regular assumptions can be found in the recent literature of the matrix factor model \citep{chen2021statistical, yu2021projected}.
This bilinear-form consistency result regarding $\hat{\mathbf{Z}}_i$ has been established in Theorem 3 of \citet{chen2021statistical}. 
Note that because $\hat{\mathbf{Z}}_i$ is consistent up to a bilinear transformation, the estimator of the matrix coefficient $\mathbf{A}$ will be consistent up to a bilinear-form orthogonal rotation as well (see Corollary 1 below). 
The existence of $\mathbf{H}_1$ and $\mathbf{H}_2$ guarantees the column space consistency of $\hat{\mathbf{R}}$ and $\hat{\mathbf{C}}$, and the bilinear-form consistency of $\hat{\mathbf{Z}}_i$ and $\hat{\mathbf{A}}$, while the exact values of  $\mathbf{H}_1$ and $\mathbf{H}_2$ can not be obtained. 
Hence $\hat{\mathbf{A}}$ is not an accurate estimation of $\mathbf{A}$, but we will show that the prediction $\hat{Y}_i$ is asymptotic invariant as if the latent factors were known (see Theorem \ref{theorem} below).

Without loss of generality, we consider the matrix regression model which only involves the matrix covariate:
\[Y_i = \left<\mathbf{A},\mathbf{Z}_i\right> + \epsilon_i,~ i=1,\cdots,n.\]
Denote $\mathbf{Y}=(Y_1,\cdots,Y_n)^{\top} \in \mathbb{R}^{n}$, $\mathcal{Z}=\{\text{vec}(\mathbf{Z}_1),\cdots,\text{vec}(\mathbf{Z}_n)\}^{\top} \in \mathbb{R}^{n \times (k_1k_2)}$, and $\hat{\mathcal{Z}}=\{\text{vec}(\mathbf{\hat{Z}}_1),$ $\cdots,\text{vec}(\mathbf{\hat{Z}}_n)\}^{\top}$.
Then the least squares estimator of $\mathbf{A}$ is in generic denoted as $$\text{vec}(\tilde{\mathbf{A}})=(\mathcal{Z}^{\top}\mathcal{Z})^{-1}\mathcal{Z}^{\top}\mathbf{Y}$$ 
if $\mathcal{Z}$ were known. 
However, under the LaGMaR estimation strategy, it is the matrix factor score $\hat{\mathcal{Z}}$ that plays the role of predictors. 
Hence, $\text{vec}(\tilde{\mathbf{A}})$ is replaced by the available $\text{vec}(\hat{\mathbf{A}})$
\[\text{vec}(\hat{\mathbf{A}})=(\hat{\mathcal{Z}}^{\top}\hat{\mathcal{Z}})^{-1}\hat{\mathcal{Z}}^{\top}\mathbf{Y}.\]
Next, we show that $\hat{\mathbf{A}}$ is a consistent estimator of $\mathbf{A}$ under a bilinear-form orthogonal transformation.

\begin{corollary}[Bilinear-form consistency]
\label{corollary} 
 Under some regular assumptions for matrix factor models and linear regression models, there exist asymptotic orthogonal matrices $\mathbf{H}_1\in\mathbb{R}^{k_1\times k_1}$ and $\mathbf{H}_2\in\mathbb{R}^{k_2\times k_2}$ such that  $$\mathbf{H}_1\hat{\mathbf{A}}\mathbf{H}_2^{\top}=\mathbf{A}+o_p(1)\mathbf{1}_{k_1\times k_2},$$
where $\mathbf{1}_{k_1 \times k_2}$ is a matrix of dimension $k_1 \times k_2$ with all entries being 1. 
\end{corollary}
\textit{Proof:} 
By equation (\ref{lemma21}) in Lemma \ref{lemma} and the representations of $\text{vec}(\tilde{\mathbf{A}})$ and $\text{vec}(\hat{\mathbf{A}})$, we have 
\[\text{vec}(\hat{\mathbf{A}})=(\mathbf{H}_2\otimes \mathbf{H}_1)^{\top}\text{vec}(\tilde{\mathbf{A}})+o_p(1)\mathbf{1}_{k_1k_2} \quad \text{i.e.,} \quad 
(\mathbf{H}_2\otimes \mathbf{H}_1)\text{vec}(\hat{\mathbf{A}})=\text{vec}(\tilde{\mathbf{A}})+o_p(1)\mathbf{1}_{k_1k_2},\]
where $\mathbf{H}_1$ and $\mathbf{H}_2$ are the asymptotic orthogonal matrices in Lemma \ref{lemma}.
It is readily seen
$$\text{vec}(\tilde{\mathbf{A}})=\text{vec}({\mathbf{A}})+o_p(1)\mathbf{1}_{k_1k_2},$$ 
under some general conditions for linear model.
Hence, 
$$(\mathbf{H}_2\otimes \mathbf{H}_1)\text{vec}(\hat{\mathbf{A}})=\text{vec}({\mathbf{A}})+o_p(1)\mathbf{1}_{k_1k_2}.~\Box$$ 

Finally, we have the result of the asymptotic prediction invariance.
\begin{theorem}[Consistency of Prediction]\label{theorem}
Under some regular assumptions for matrix factor models and linear regression models, we have
$$\hat{Y}_i=\left<{\mathbf{A}},\mathbf{Z}_i\right>+o_p(1).$$
\end{theorem}
\textit{Proof:} The theorem follows from Lemma \ref{lemma} and Corollary \ref{corollary}:
$$\hat{Y}_i=\big<\hat{\mathbf{A}},\hat{\mathbf{Z}}_i\big>=\big<\tilde{\mathbf{A}},\mathbf{Z}_i\big>+o_p(1)=\left<{\mathbf{A}},\mathbf{Z}_i\right>+o_p(1).~\Box$$ 
For the unknown matrix regression coefficient $\mathbf{A}$ and the unobservable latent matrix $\mathbf{Z}$, once both are consistently estimated up to a bilinear-form transformation by the LaGMaR estimation approach, the predicted response will be asymptotically equivalent to the true signal of the mean matrix regression. 

\section{Simulation Studies}
\label{sec3}
\subsection{Considered Methods, Evaluation Metrics, and Implementation}
In this section, we compare predictive capability of the proposed LaGMaR with several existing penalized statistical methods including two tensor regression methods and two matrix regression methods. The two tensor regression methods are the CP Tensor Regression by \citet{zhou2013tensor} (CPTR) (i.e., CPTR($r$), $r=1, 2$, where $r$ is the rank of CP decomposition of the matrix coefficient) and the Tucker Tensor Regression by \citet{li2018tucker} (TTR) (i.e., TTR($r_1, r_2$), $(r_1, r_2)=(1, 2), (2, 2)$, where $r_1$ and $r_2$ represent the rank of Tucker decomposition of the matrix coefficient). The two matrix regression methods are rSVD by \citet{zhou2014regularized} and MVLogistic by \citet{hung2013matrix}, respectively.

The considered methods are compared and assessed by various metrics in four simulation studies. In the first three simulation studies, the outcomes are binary, count, and continuous, respectively. In the fourth study, the outcomes mimick real COVID-19 CT data. With binary outcomes, the metrics include classification accuracy (CA), Kappa coefficient, ROC curve and area under the curve (AUC), sensitivity, and F1 score. With count outcomes (Poission distributed), the metrics include root mean squared error (RMSE), normalized mean squared error (NMSE), and mean absolute error (MAE). With continuous outcomes, the metrics include RMSE and MAE. With COVID-19 CT outcomes, the metrics include CA, Kappa, Sensitivity, AUC, and F1 score.

We implement CPTR, rSVD, and TTR using the Matlab toolbox TensorReg  \texttt (\url{https://hua-zhou.github.io/TensorReg}). Our LaGMaR is implemented in R language and the R code for simulation studies can be downloaded from GitHub \texttt (\url{https://github.com/zyz0000/LaGMaR}).

\subsection{Finite Sample Performance of LaGMaR}
\label{sub33}

The matrix-valued covariate $\mathbf{X}$ is generated according to $\mathbf{X}=\mathbf{R}\mathbf{Z}\mathbf{C}^{\top} + \mathbf{E}$. We choose $(p_1, p_2)$ among (20,20), (20,50) and (50,50), and let sample size be $n=\rho p_1p_2, \rho \in \left\{0.5, 1, 1.5, 2 \right\}$. For the latent matrix factor $\mathbf{Z}$, let the dimension of global latent matrix factor $\mathbf{Z}$ be $(k_1, k_2)=(3,3)$ and $\text{vec}(\mathbf{Z}) \sim MVN(\mathbf{0}, \mathbf{\Sigma})$, where $\mathbf{\Sigma}_{ij}=0.5^{|i-j|},1 \leq i,j \leq k_1  k_2$. The entries of true loading matrices $\mathbf{R}$ and $\mathbf{C}$ are independently sampled from the uniform distribution $U(-\sqrt{p_1}, \sqrt{p_1})$ and $U(-\sqrt{p_2}, \sqrt{p_2})$, respectively. 
Each entry of the noise matrix $\mathbf{E}$ is generated from the standard normal distribution $N(0,1)$. In addition, we take the usual vector of covariates $\mathbf{v}$ into consideration for model (\ref{famglm}), where $\mathbf{v}=(v_1, v_2, v_3)^{\top} \sim MVN(\mathbf{0}, \mathbf{I}_3)$.

Denote the mean regression $\mu=\gamma + \left<\mathbf{A}, \mathbf{Z}\right> + \beta^{\top} \mathbf{v}$, where $\gamma=1$, $\text{vec}(\mathbf{A})=(2, -2, \mathbf{1}_{3}^{\top}, -\mathbf{1}_4^{\top})^{\top}$ and $\beta=\mathbf{1}_3$. We then generate the response through three submodels of GLM: for binomial model, $Y \sim \text{Bernoulli}(p)$, with the link function $g(x)=\log\left\{x / (1-x)\right\}$ and $p=\text{logit}^{-1}(\mu)$; for normal model, $Y \sim N(\mu, 1)$, with the identical link function $g(x)=x$; for Poisson model, $Y \sim \text{Poisson}(\max(\exp(\mu), 1))$, with link function $g(x)=\log(x)$,  where the $\max(\cdot)$ function is used to prevent the possible numerical instability in the case when $\mu$ is very small. 

In order to compare the proposed LaGMaR approach with the existing methods, we calculate the evaluation metrics via five-fold cross validation. 
In each scenario, experiments are repeated for 100 times. 
We choose tuning parameters from the candidate set $\left\{1, 5, 10, 20, 50, 100\right\}$ for the existing penalized methods.
For the matrix regression methods rSVD and MVLogistic, tuning parameters are selected by minimizing the BIC value on the training set and by maximizing the classification accuracy (CA) on another independent validation set. 
As for the tensor regression methods under various rank scenarios, the tuning parameters are set to be the same as those in \citet{zhou2014regularized}. Refer to Table \ref{tuning33} for the selected tuning parameters in various penalized methods.

\centerline{[Table \ref{tuning33} about here.]}

\begin{figure}
  \centering
  \begin{minipage}[t]{.327\linewidth}
    \includegraphics[width=5cm, height=5cm]{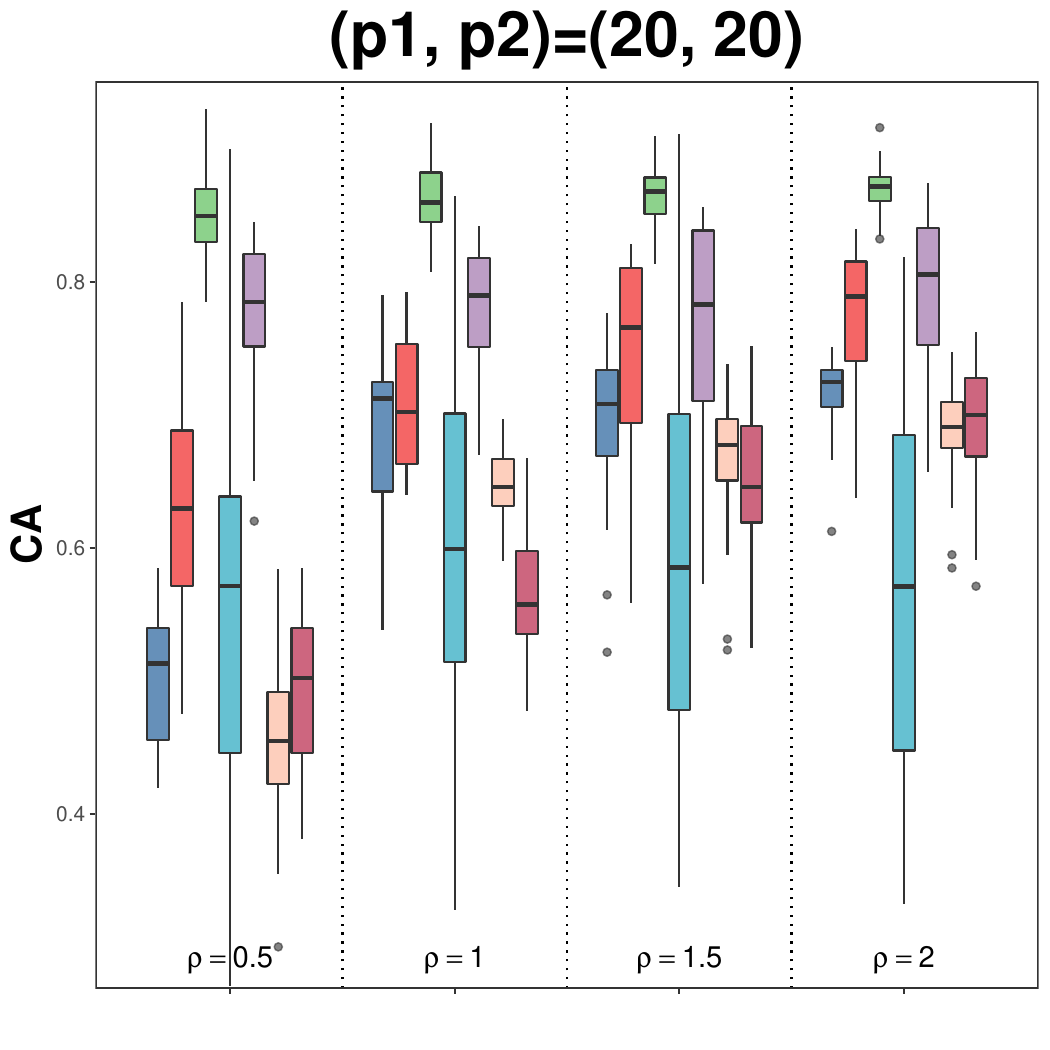}
  \end{minipage}
  \begin{minipage}[t]{.327\linewidth}
    \includegraphics[width=5cm, height=5cm]{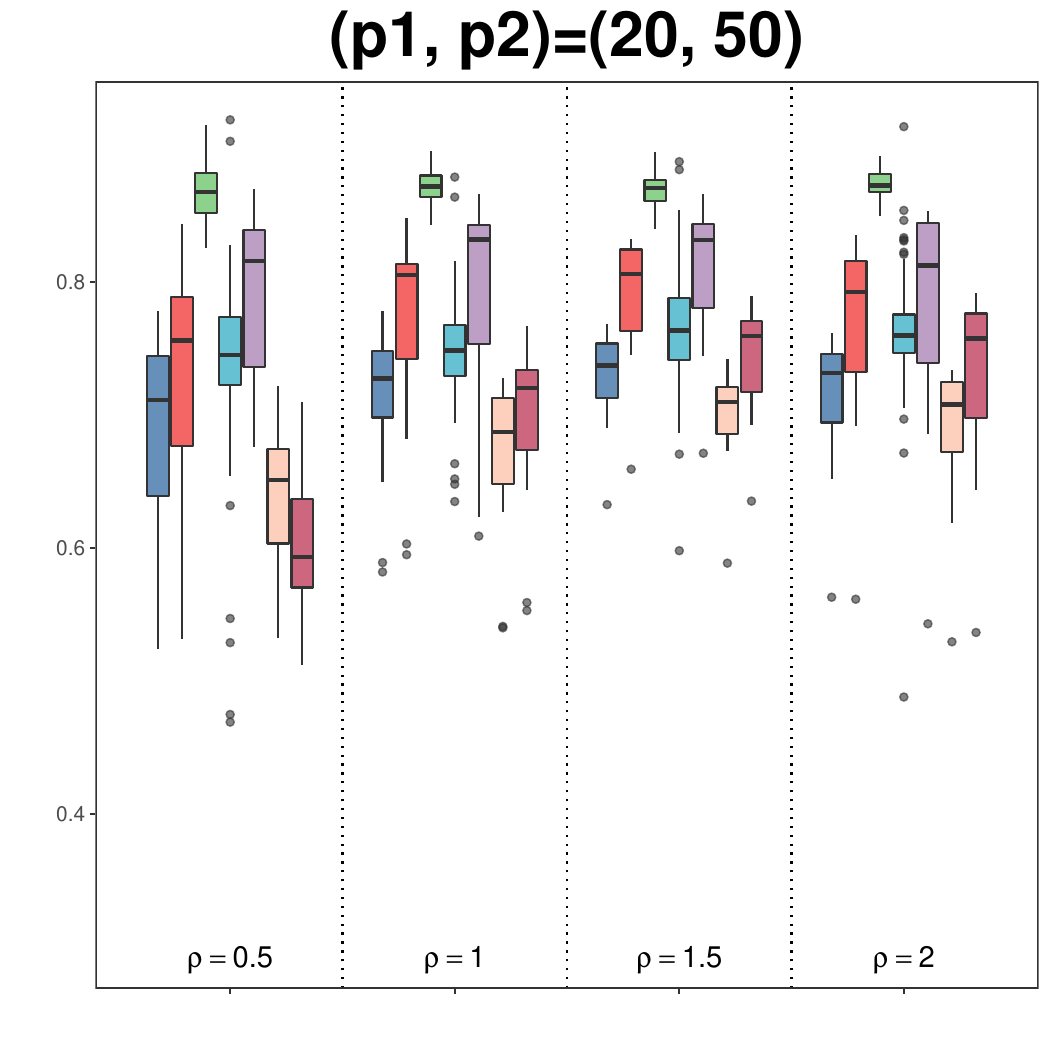}
  \end{minipage}
  \begin{minipage}[t]{.327\linewidth}
    \includegraphics[width=5cm, height=5cm]{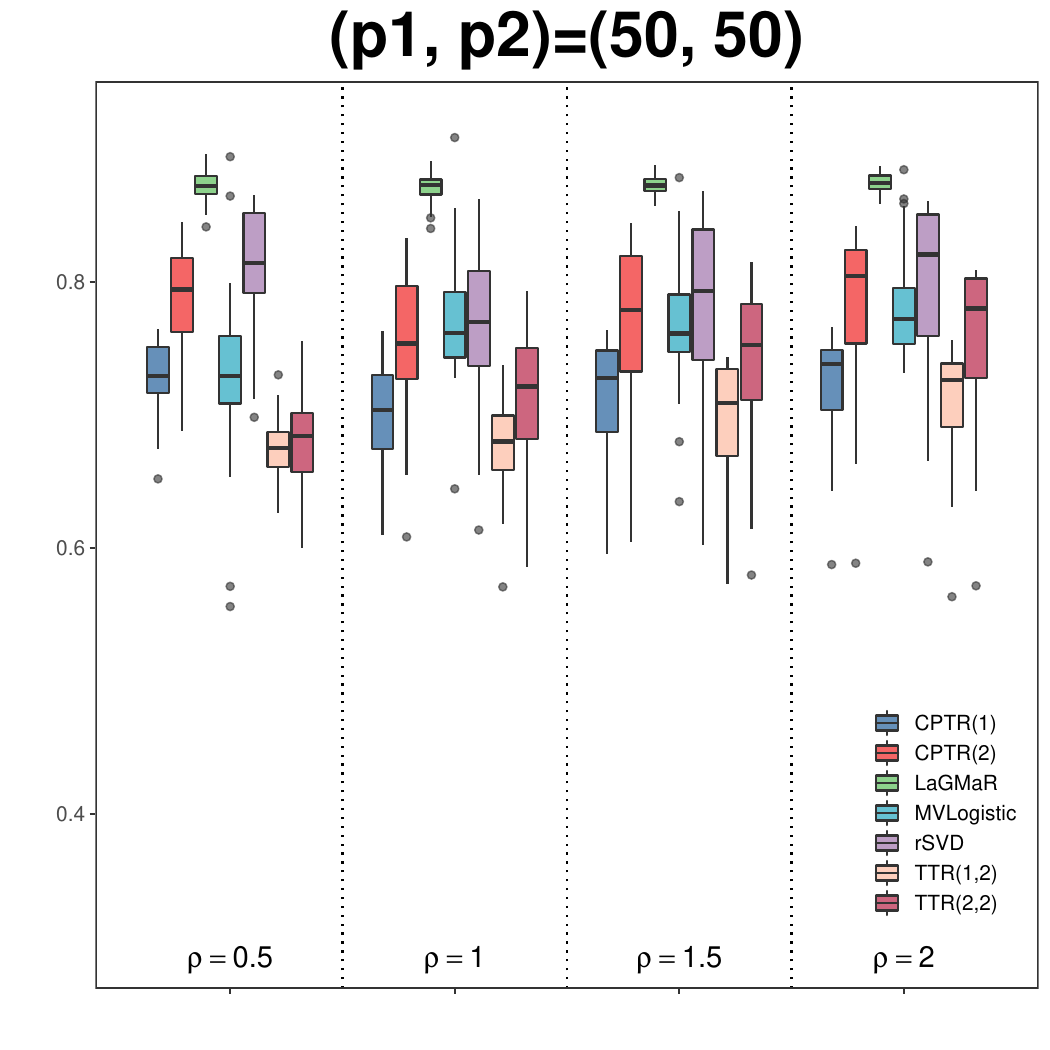}
  \end{minipage}
  
  \begin{minipage}[t]{.327\linewidth}
    \includegraphics[width=5cm, height=5cm]{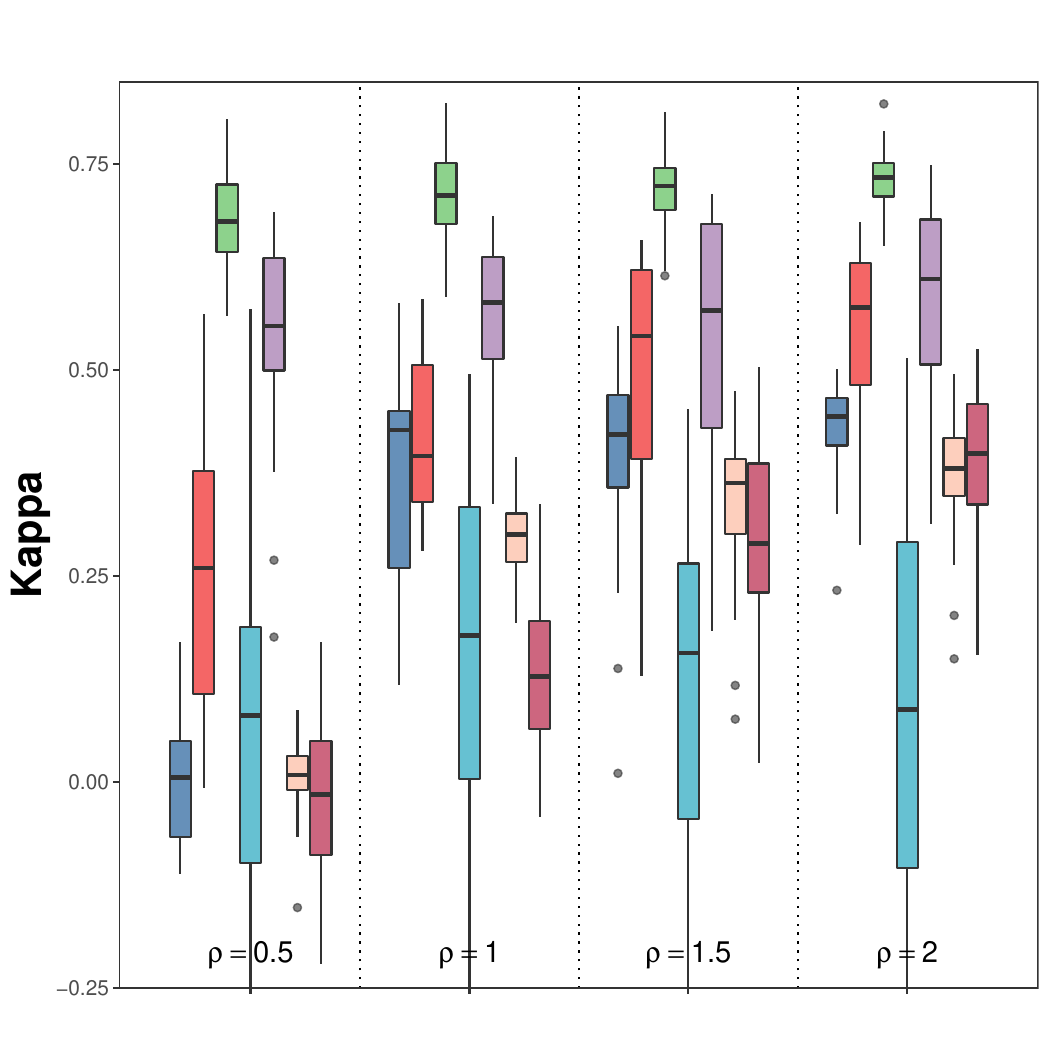}
  \end{minipage}
  \begin{minipage}[t]{.327\linewidth}
    \includegraphics[width=5cm, height=5cm]{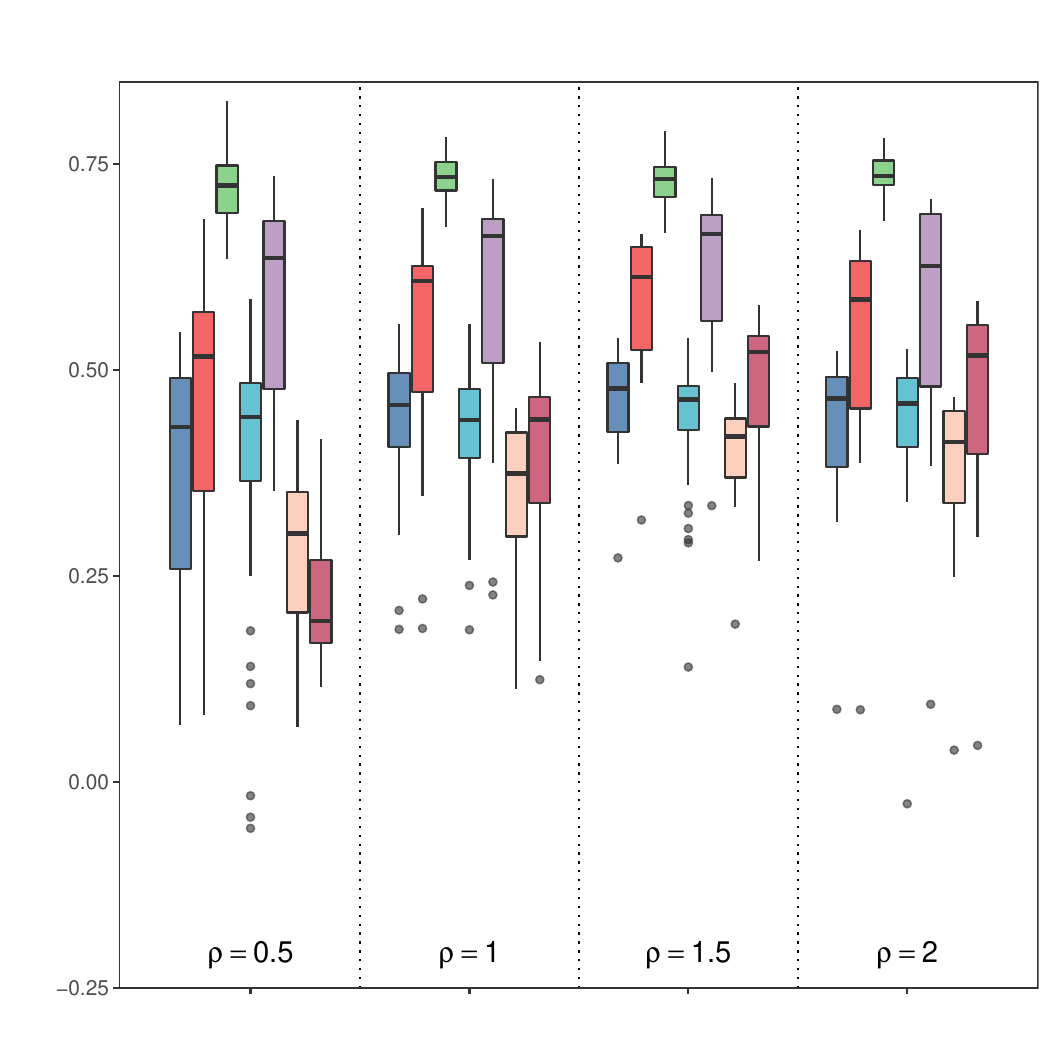}
  \end{minipage}
  \begin{minipage}[t]{.327\linewidth}
    \includegraphics[width=5cm, height=5cm]{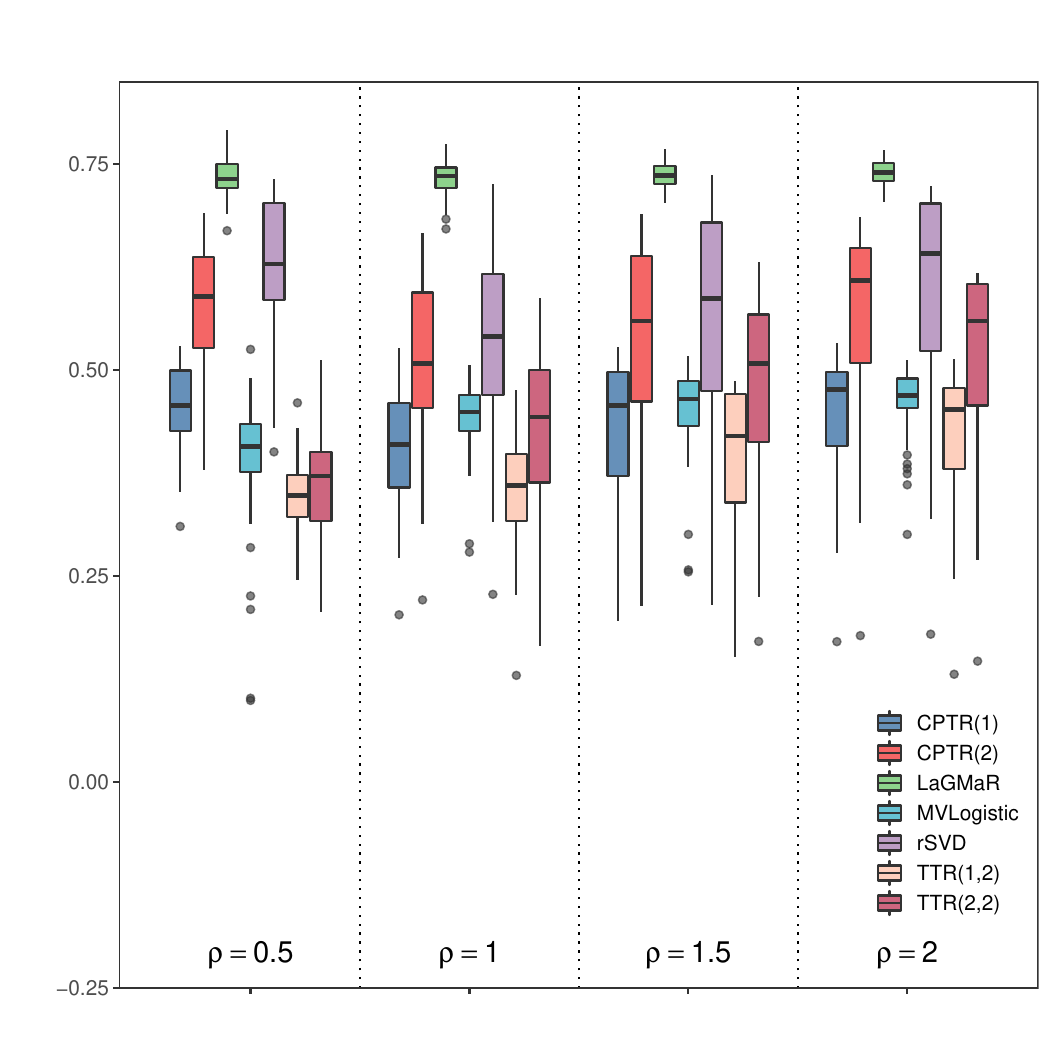}
  \end{minipage}
  
   \begin{minipage}[t]{.327\linewidth}
    \includegraphics[width=5cm, height=5cm]{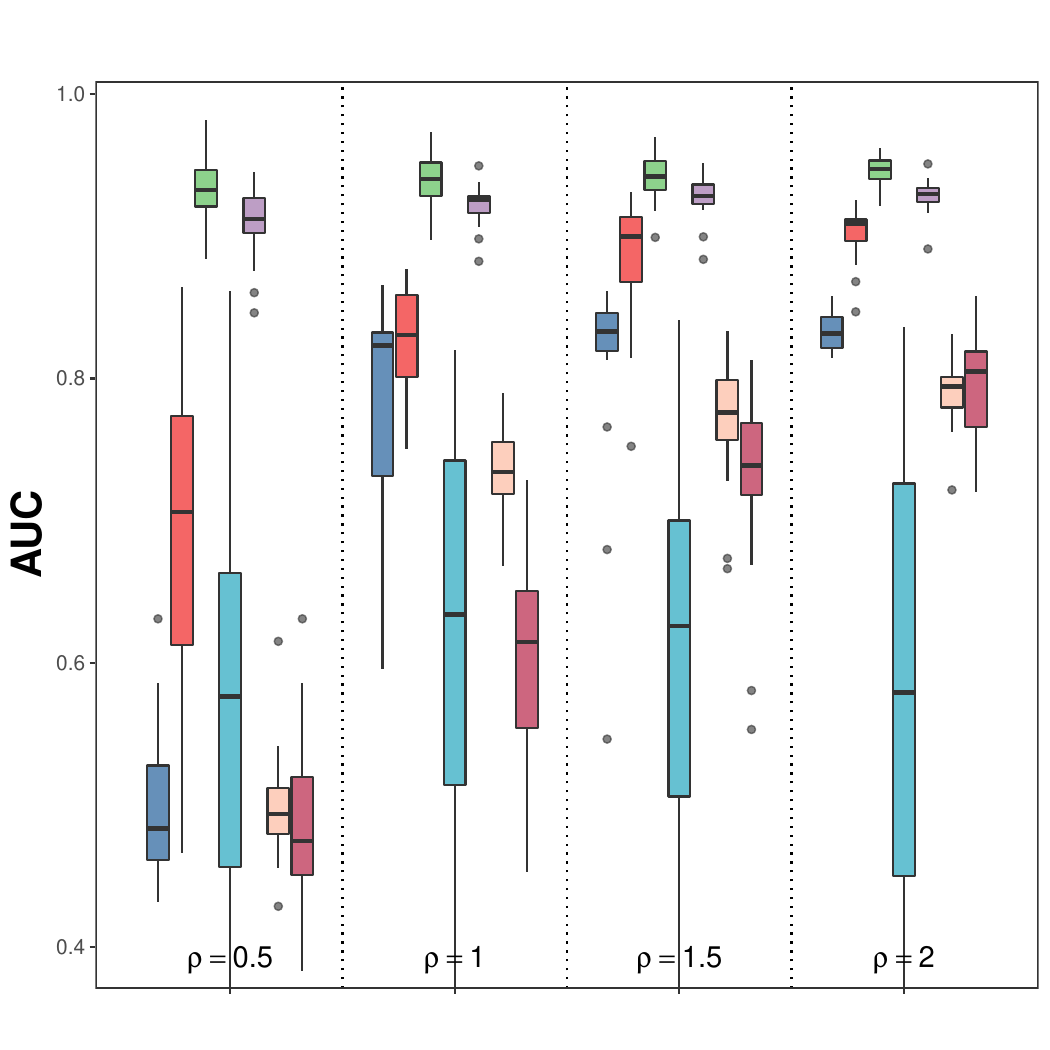}
  \end{minipage}
  \begin{minipage}[t]{.327\linewidth}
    \includegraphics[width=5cm, height=5cm]{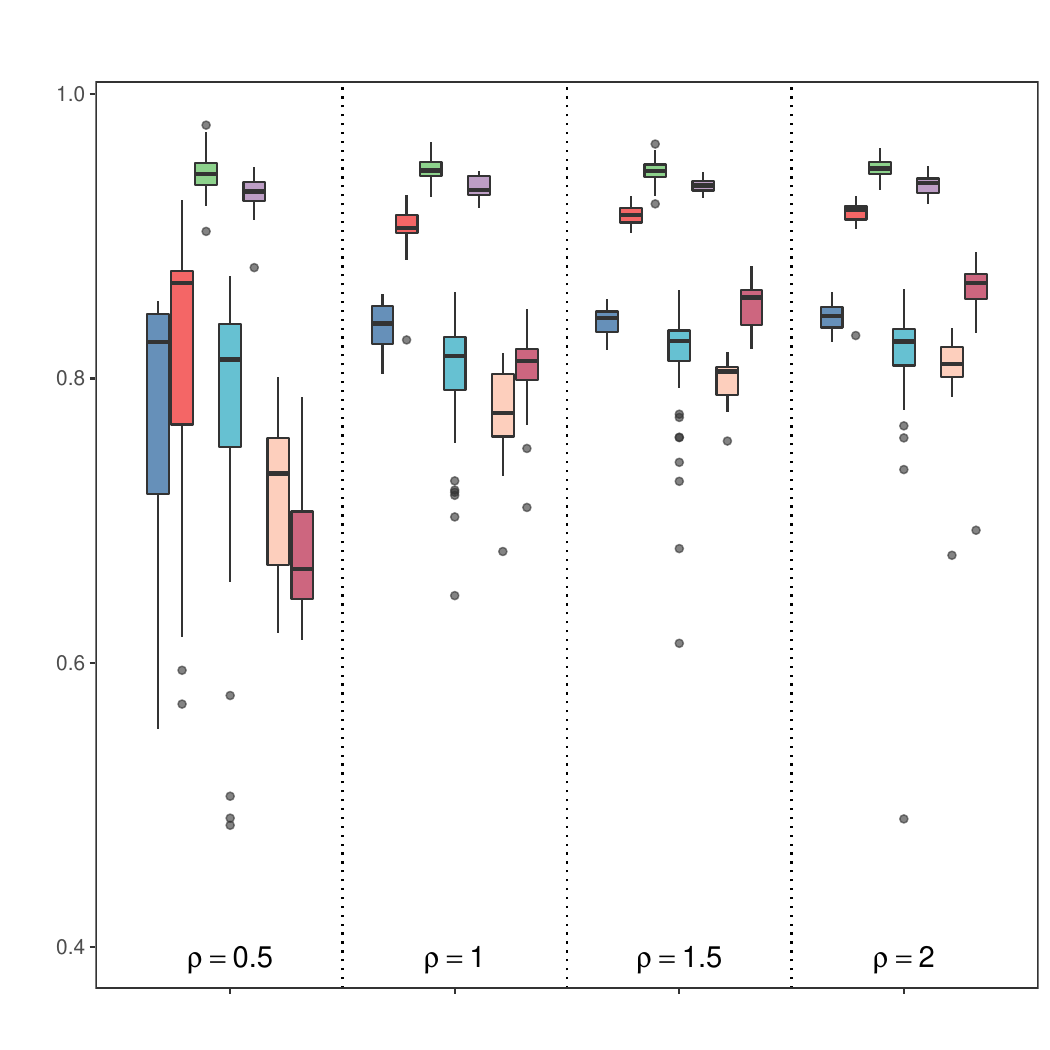}
  \end{minipage}
  \begin{minipage}[t]{.327\linewidth}
    \includegraphics[width=5cm, height=5cm]{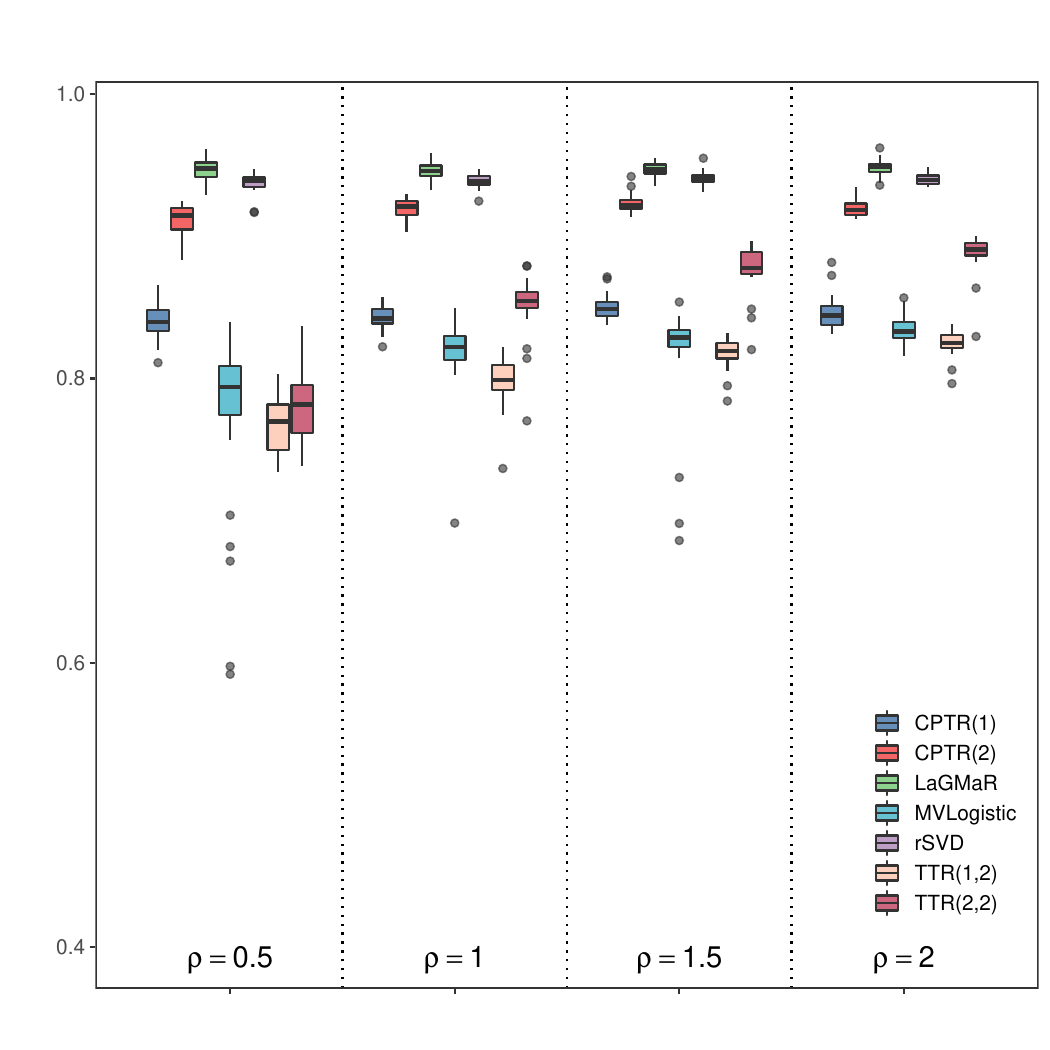}
  \end{minipage}
    \caption{ Metrics of CA, Kappa, and AUC by the seven methods under the logistic regression setting with matrix-variate covariate with different dimensionality scenarios of $(p_1, p_2)$. }
    \label{L_acc}
\end{figure}

\begin{figure}
  \centering
  \begin{minipage}[t]{.327\linewidth}
    \includegraphics[width=5cm, height=5cm]{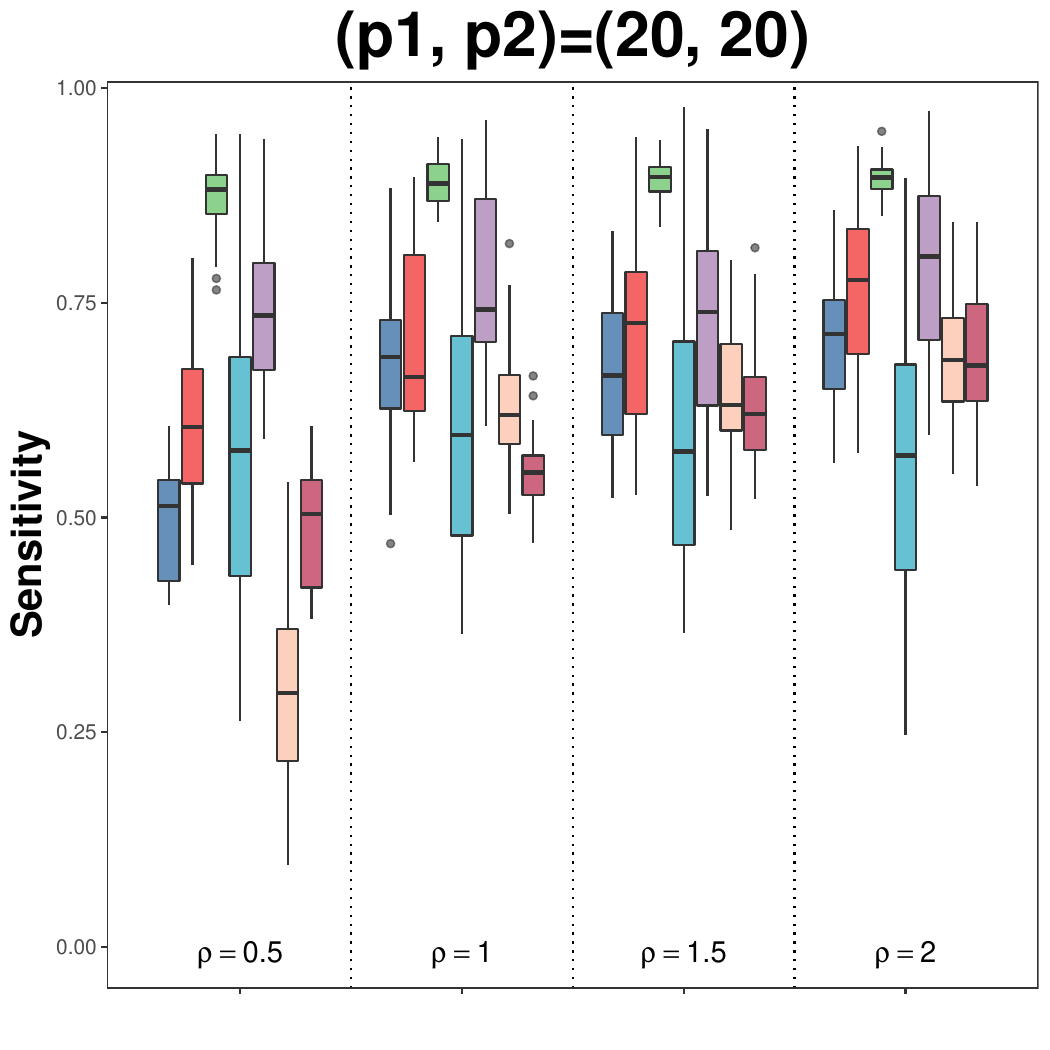}
  \end{minipage}
  \begin{minipage}[t]{.327\linewidth}
    \includegraphics[width=5cm, height=5cm]{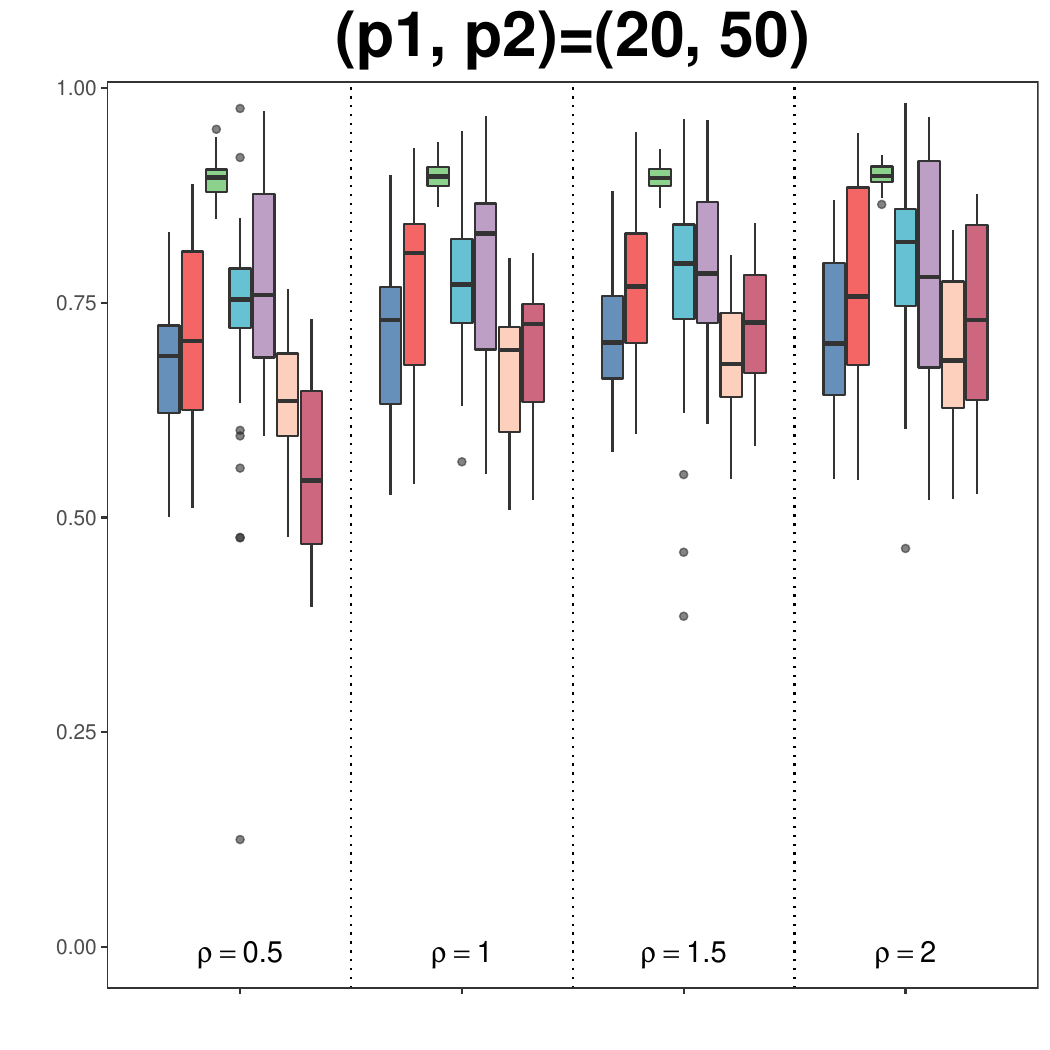}
  \end{minipage}
  \begin{minipage}[t]{.327\linewidth}
    \includegraphics[width=5cm, height=5cm]{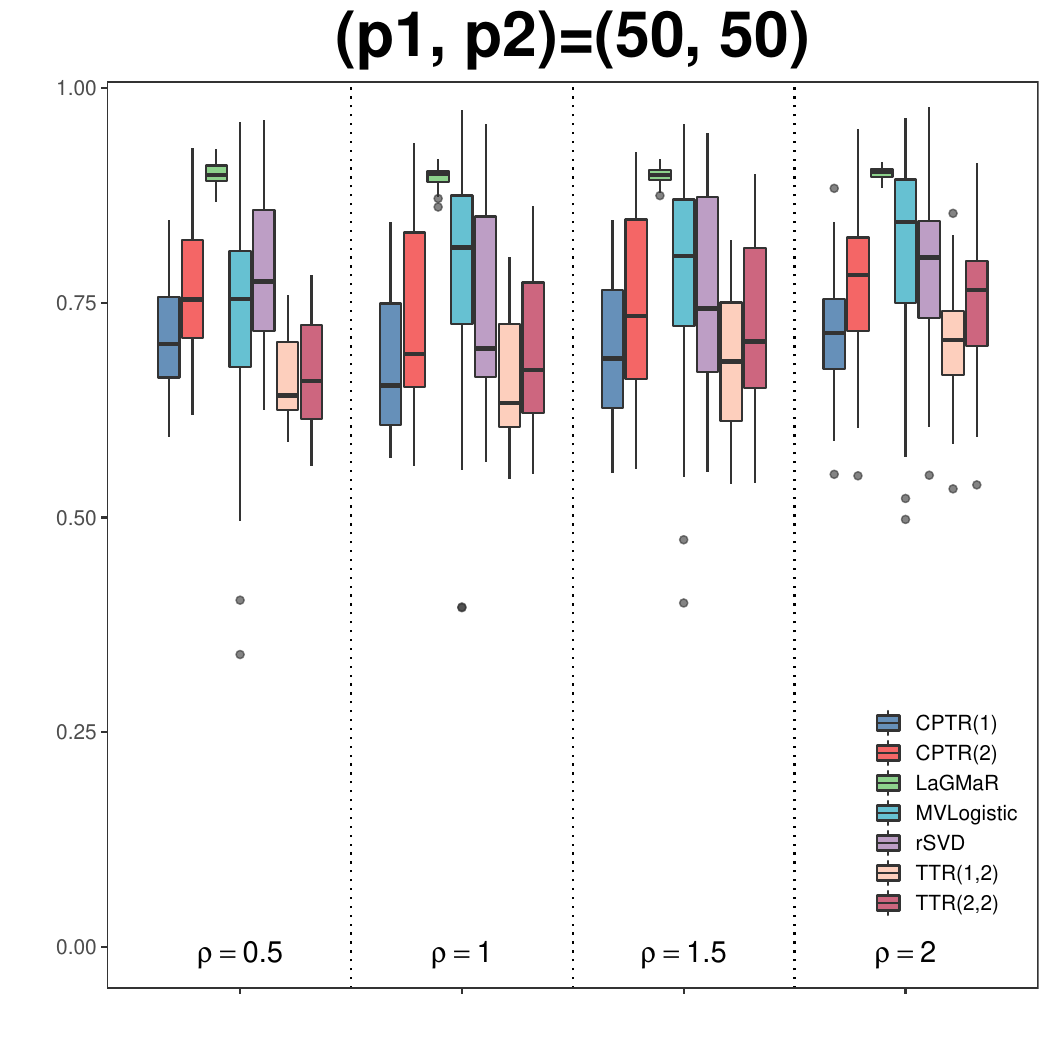}
  \end{minipage}
  
    \begin{minipage}[t]{.327\linewidth}
    \includegraphics[width=5cm, height=5cm]{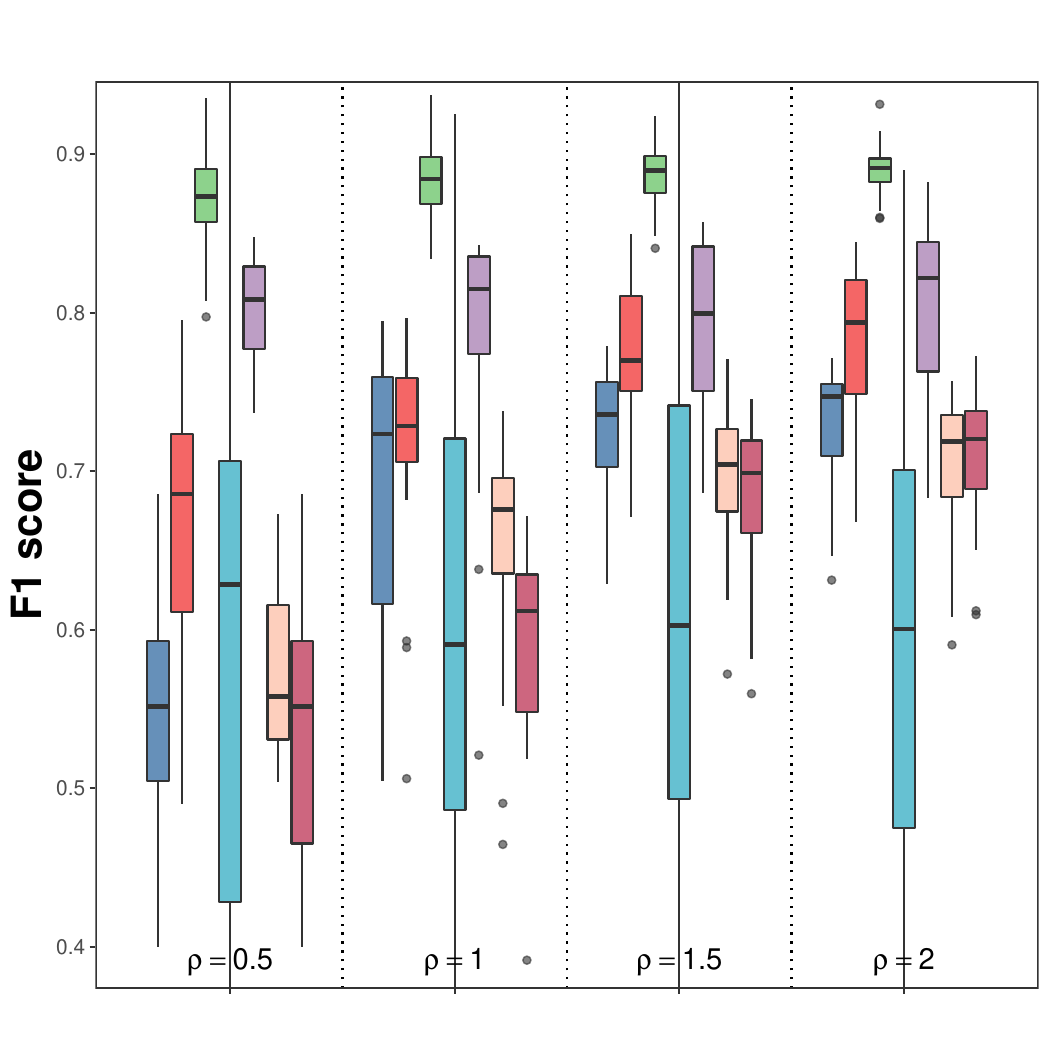}
  \end{minipage}
  \begin{minipage}[t]{.327\linewidth}
    \includegraphics[width=5cm, height=5cm]{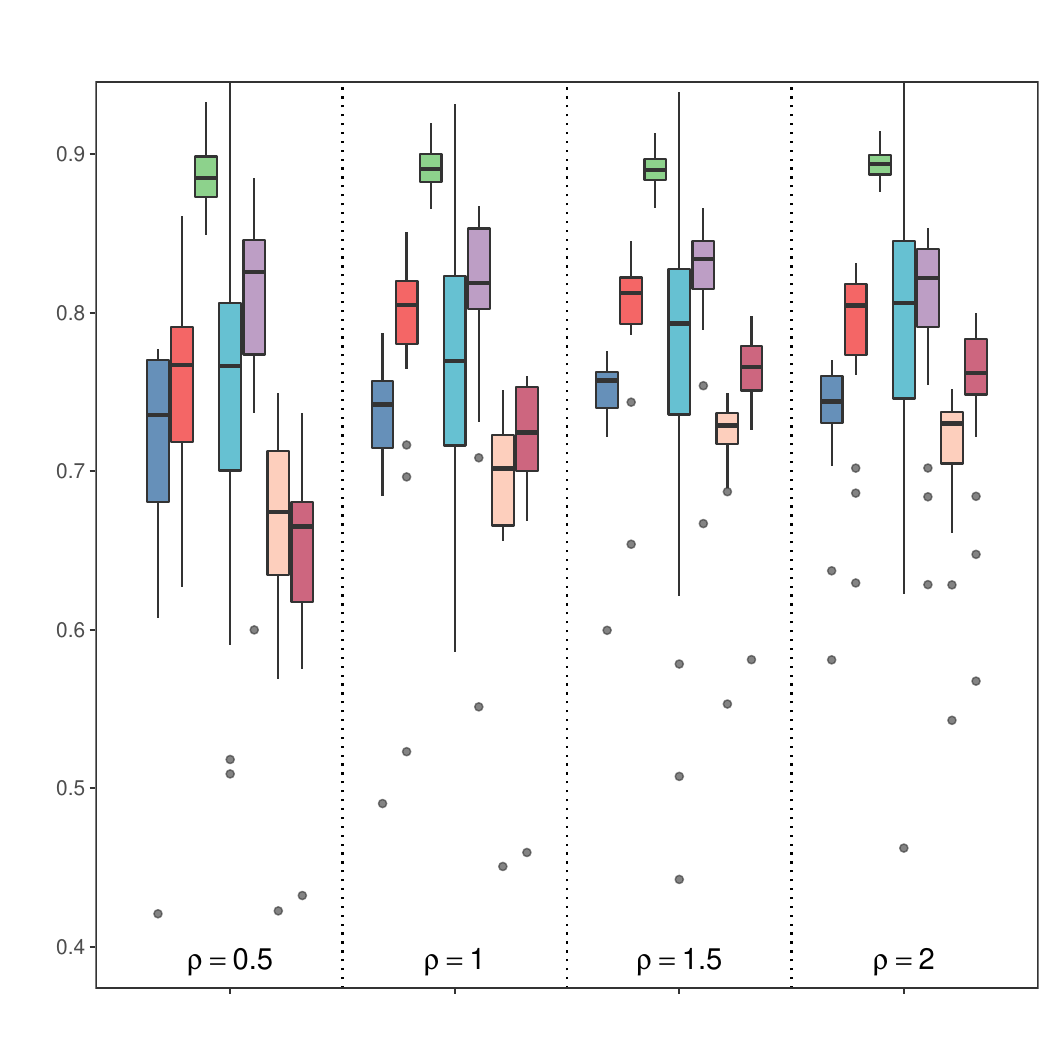}
  \end{minipage}
  \begin{minipage}[t]{.327\linewidth}
    \includegraphics[width=5cm, height=5cm]{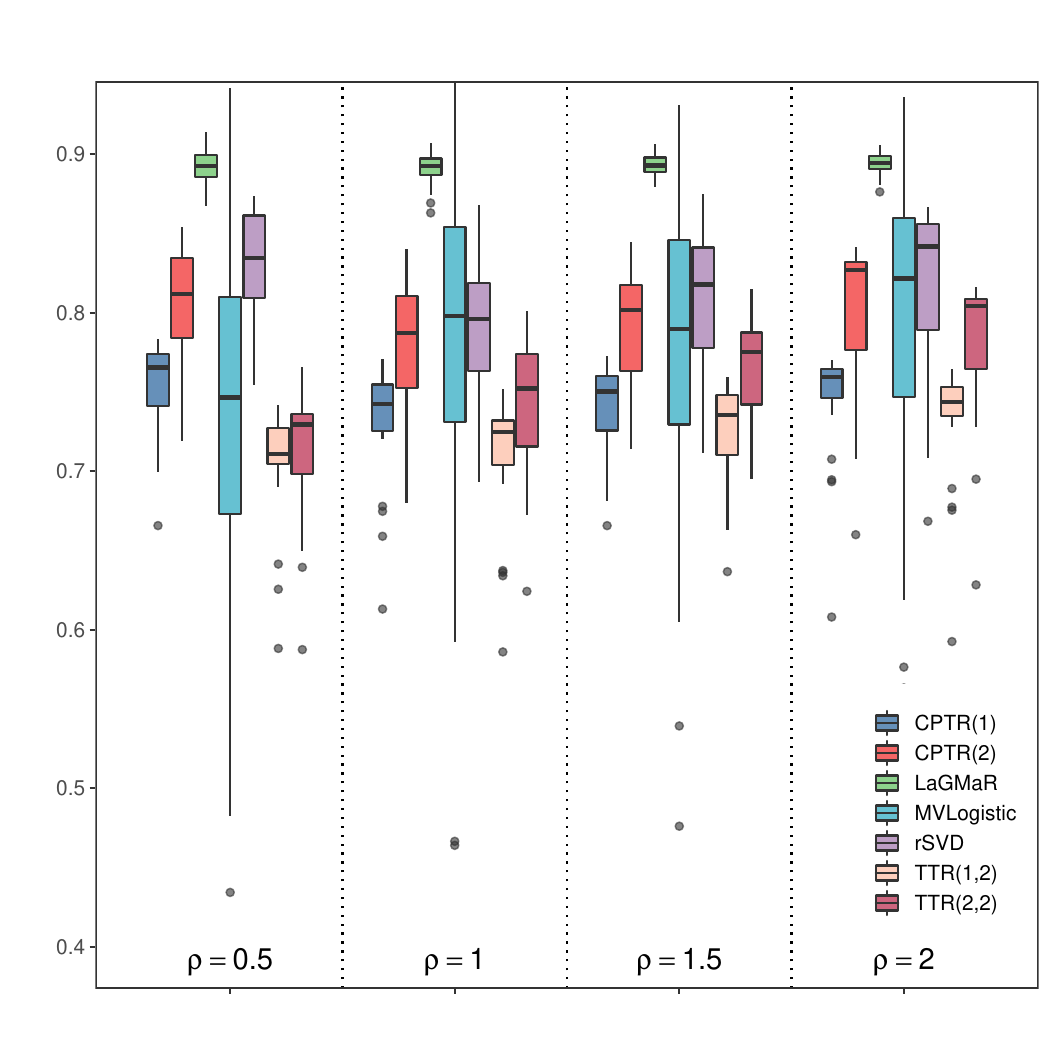}
  \end{minipage}
  
    \begin{minipage}[t]{.327\linewidth}
    \includegraphics[width=5cm, height=5cm]{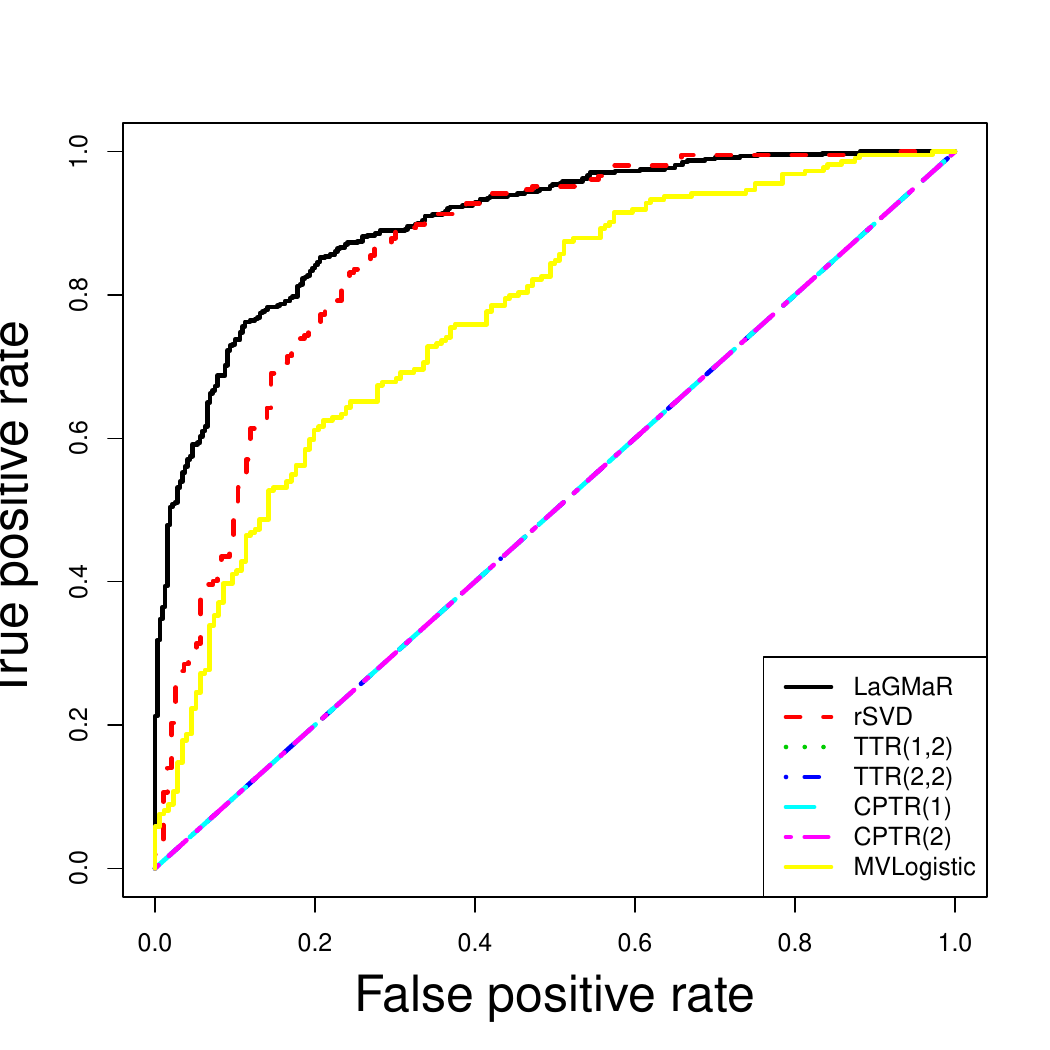}
  \end{minipage}
  \begin{minipage}[t]{.327\linewidth}
    \includegraphics[width=5cm, height=5cm]{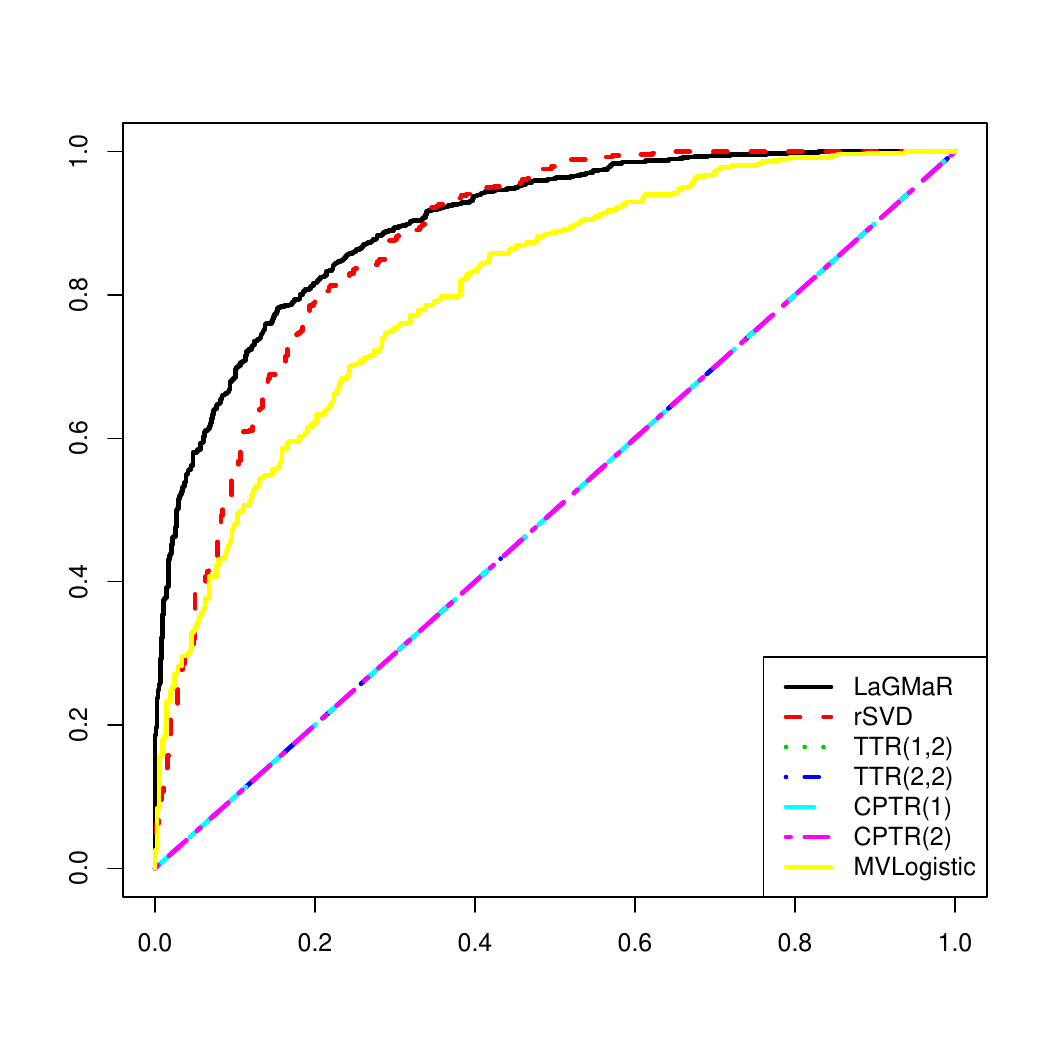}
  \end{minipage}
  \begin{minipage}[t]{.327\linewidth}
    \includegraphics[width=5cm, height=5cm]{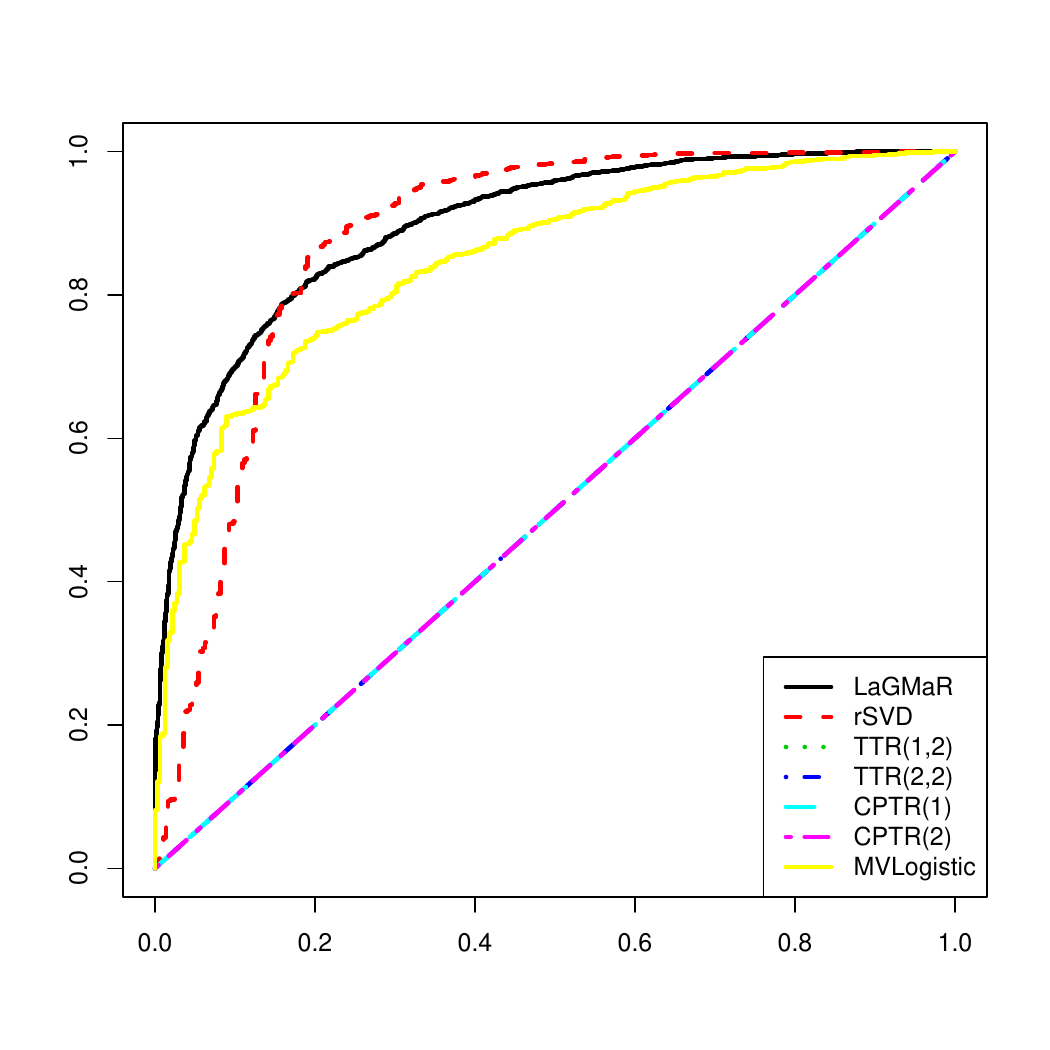}
  \end{minipage}
    \caption{ Metrics of  Sensitivity, F1 score, and the ROC curve by the seven methods under the logistic regression setting with matrix-variate covariate with different dimensionality scenarios of $(p_1, p_2)$.
    }
        \label{L_sen}
\end{figure}

\textbf{Case 1: Logistic regression}
The cutoff value is fixed at 0.5 in calculating CA, Kappa, sensitivity, and F1 score. CA is the proportion of correctly predicted observation to the total observations. Kappa measures the interrater reliability, representing the extent to which the data collected in the study are correct representations of the variables measured. Sensitivity is the proportion of correctly predicted positive observations to all observations in the actual positive class. F1 score is a tradeoff between positive predictive rate and sensitivity. The boxplots of these five metrics under different scenarios of $\left\{(p_1, p_2), \rho \right\}$ are displayed in Figures \ref{L_acc} and \ref{L_sen}. All these five metrics get closer to 1 when the sample size increases from $0.5 p_1p_2$ to $2 p_1p_2$. 
Evidently, LaGMaR outperforms rSVD on CA, Kappa and sensitivity, and the two methods perform comparably on AUC and F1 score.

\centerline{[Figure \ref{L_acc} about here.]}

\centerline{[Figure \ref{L_sen} about here.]}

Figure \ref{L_sen} shows the ROC curves, which illustrate that LaGMaR (in black solid line) and rSVD (in red dashed line) perform comparably, and they greatly outperform TTRs and CPTRs. 
With $(p_1, p_2)=(20, 20)$, the partial AUC of LaGMaR is larger than that of rSVD when the false positive rate (FPR) is smaller than 0.5, but the partial AUCs of the two methods become competitive when the FPR gets larger than 0.5. 
With $(p_1, p_2)=(20, 50)$, the partial AUC of LaGMaR is larger than that of rSVD when the FPR is smaller than 0.3, whereas the partial AUC of rSVD is larger than LaGMaR when the FPR gets larger than 0.3. 
With $(p_1, p_2)=(50, 50)$, rSVD performs slightly better than LaGMaR. Overall, LaGMaR still performs the best among six methods under the binomial GLM submodel.

\begin{figure}
  \centering
  \begin{minipage}[t]{.327\linewidth}
    \includegraphics[width=5cm, height=5cm]{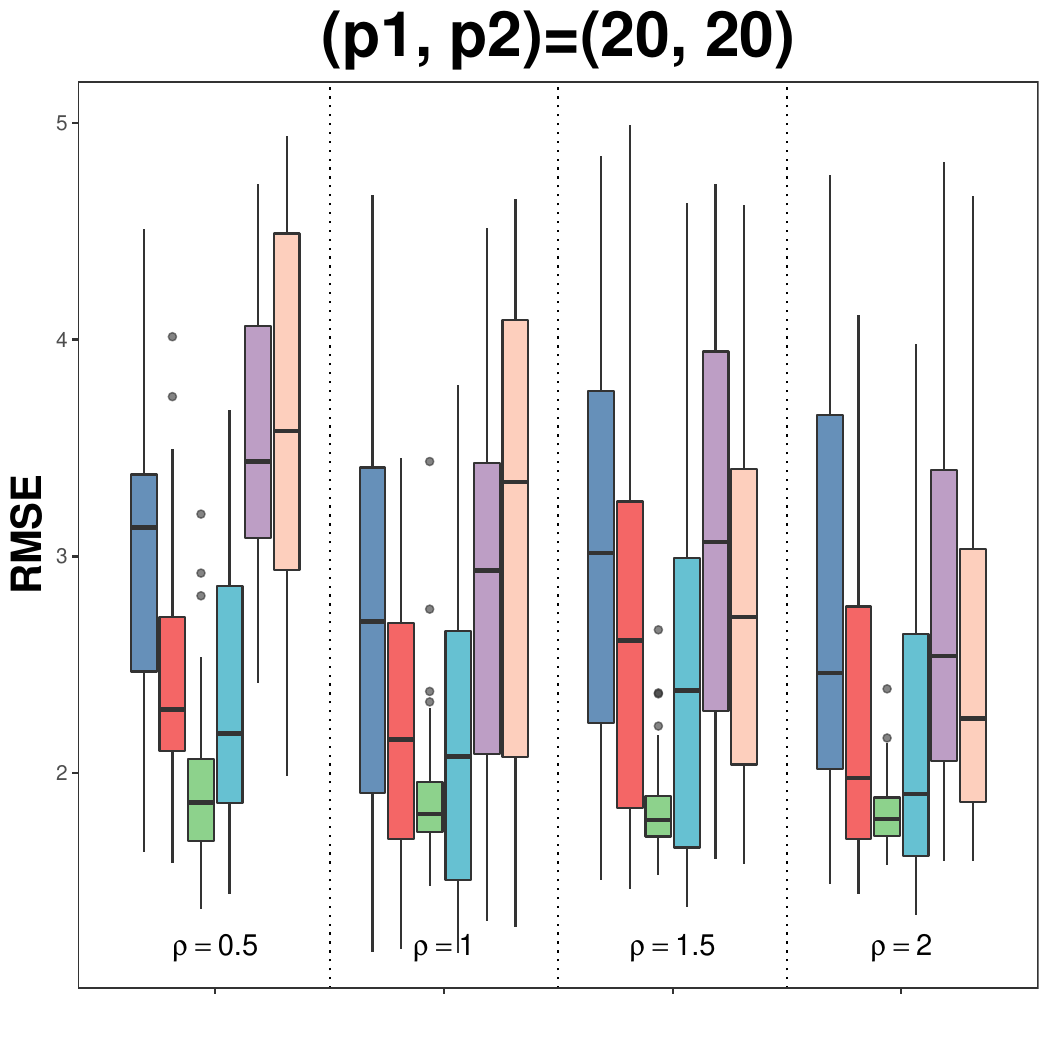}
  \end{minipage}
  \begin{minipage}[t]{.327\linewidth}
    \includegraphics[width=5cm, height=5cm]{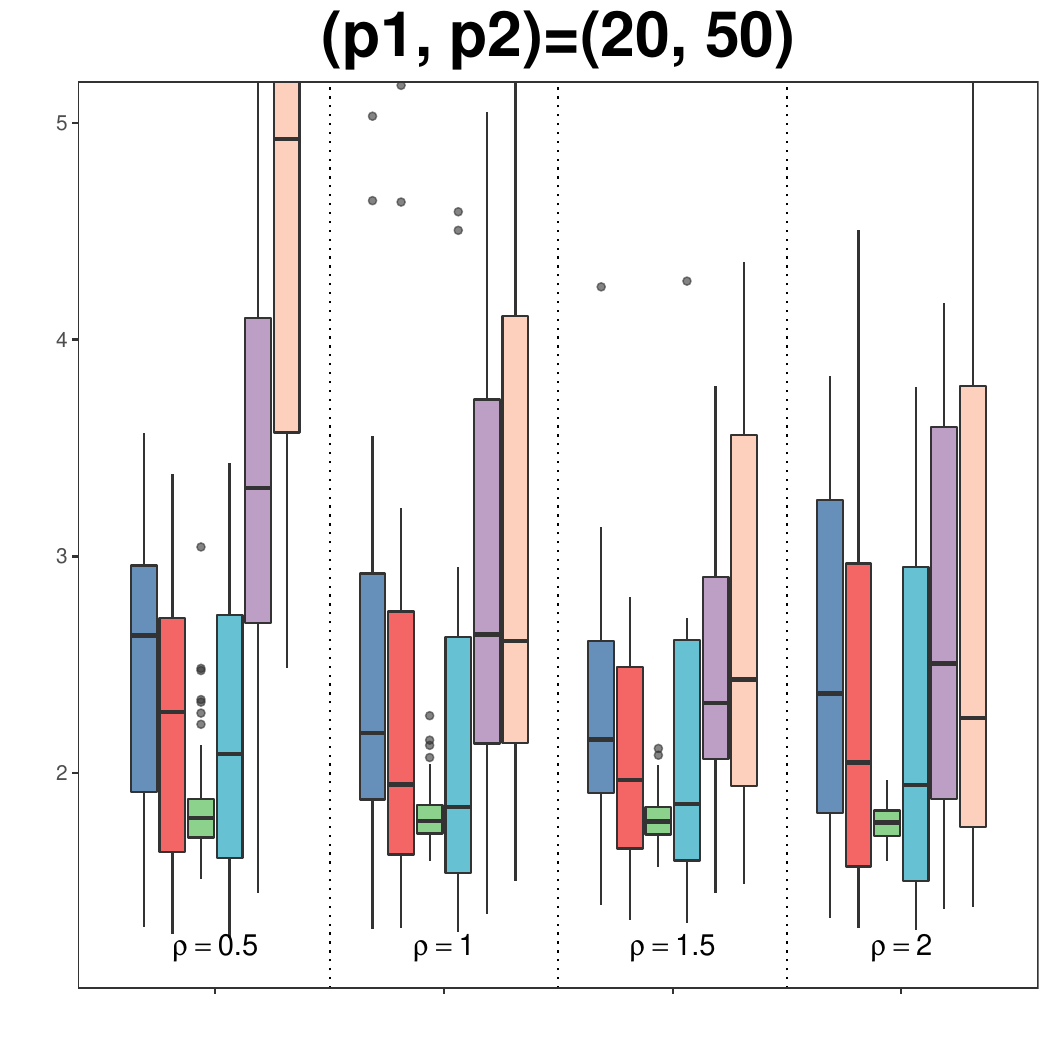}
  \end{minipage}
  \begin{minipage}[t]{.327\linewidth}
    \includegraphics[width=5cm, height=5cm]{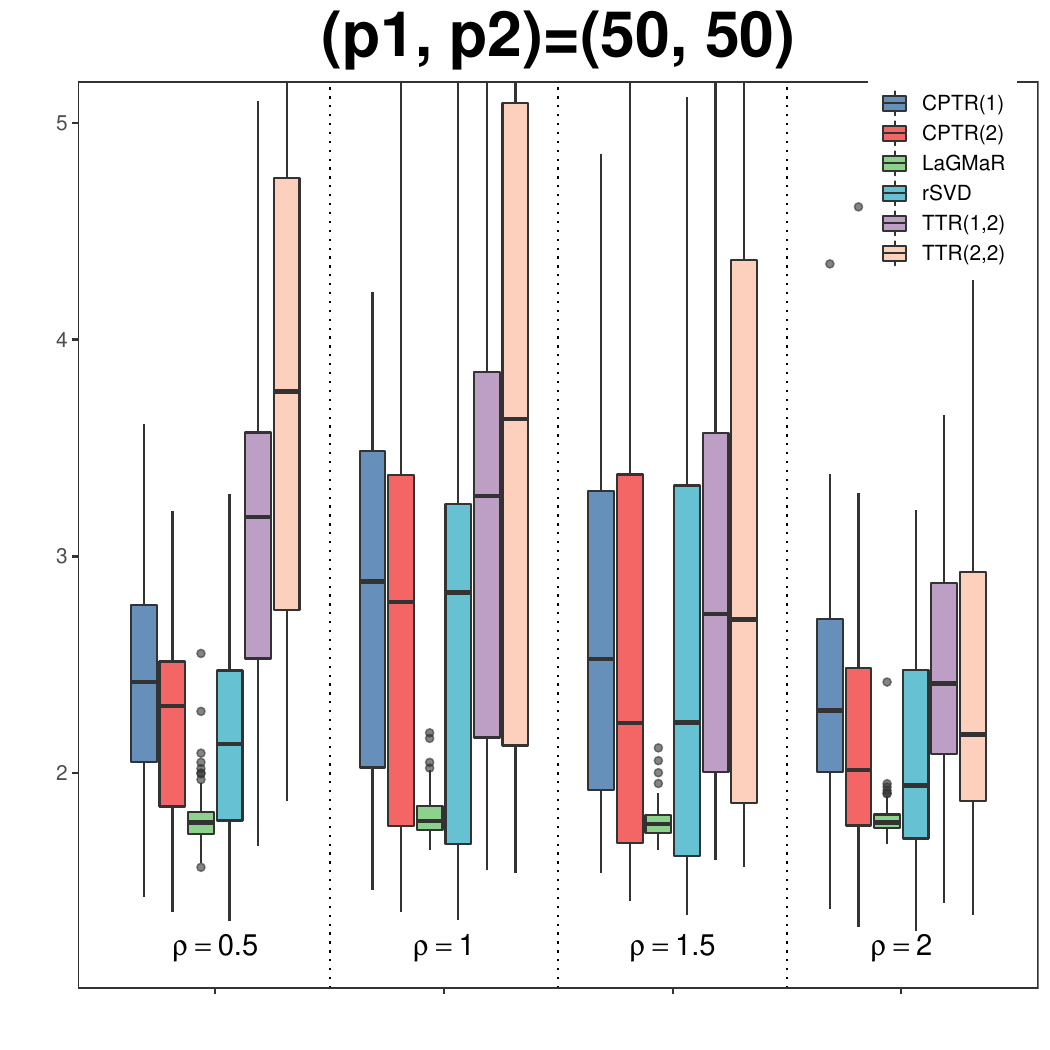}
  \end{minipage}
  
    \begin{minipage}[t]{.327\linewidth}
    \includegraphics[width=5cm, height=5cm]{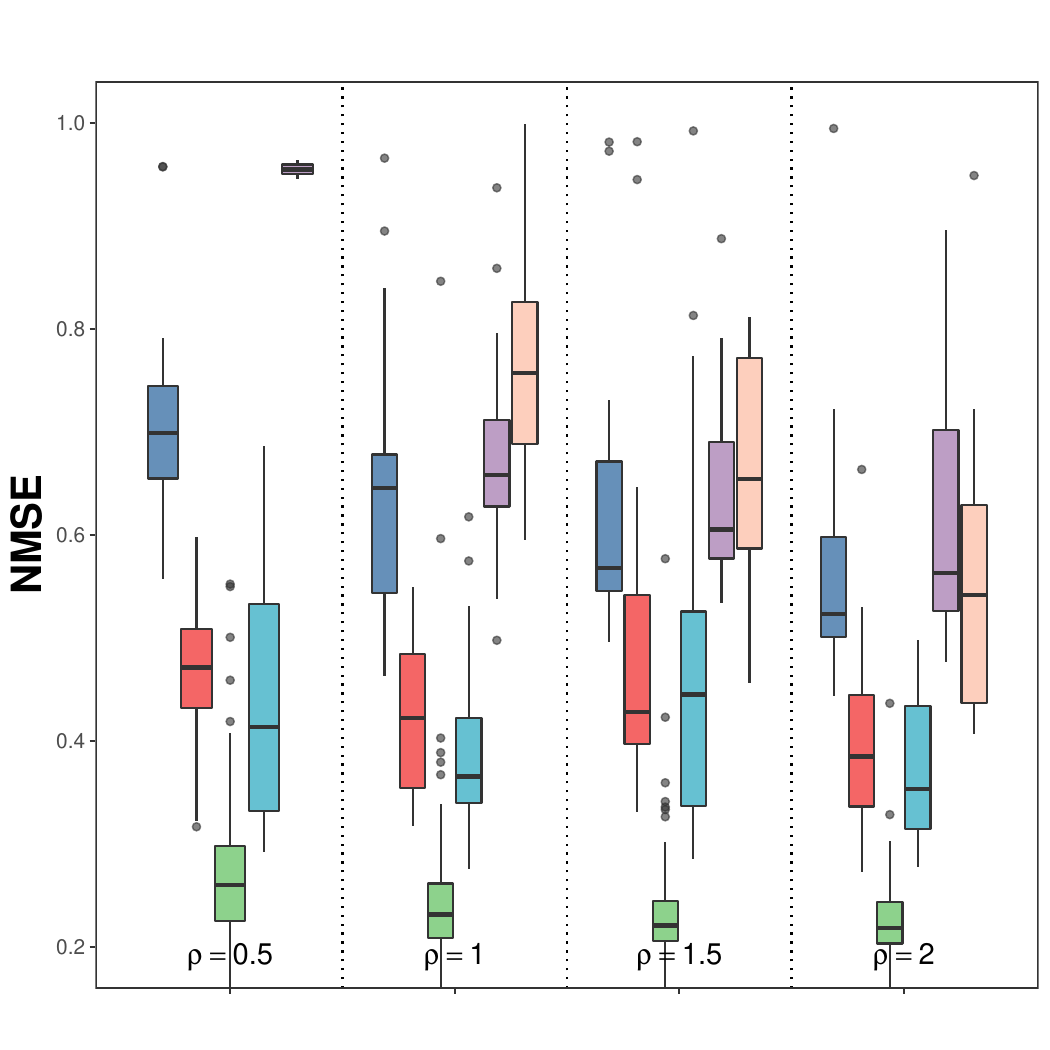}
  \end{minipage}
  \begin{minipage}[t]{.327\linewidth}
    \includegraphics[width=5cm, height=5cm]{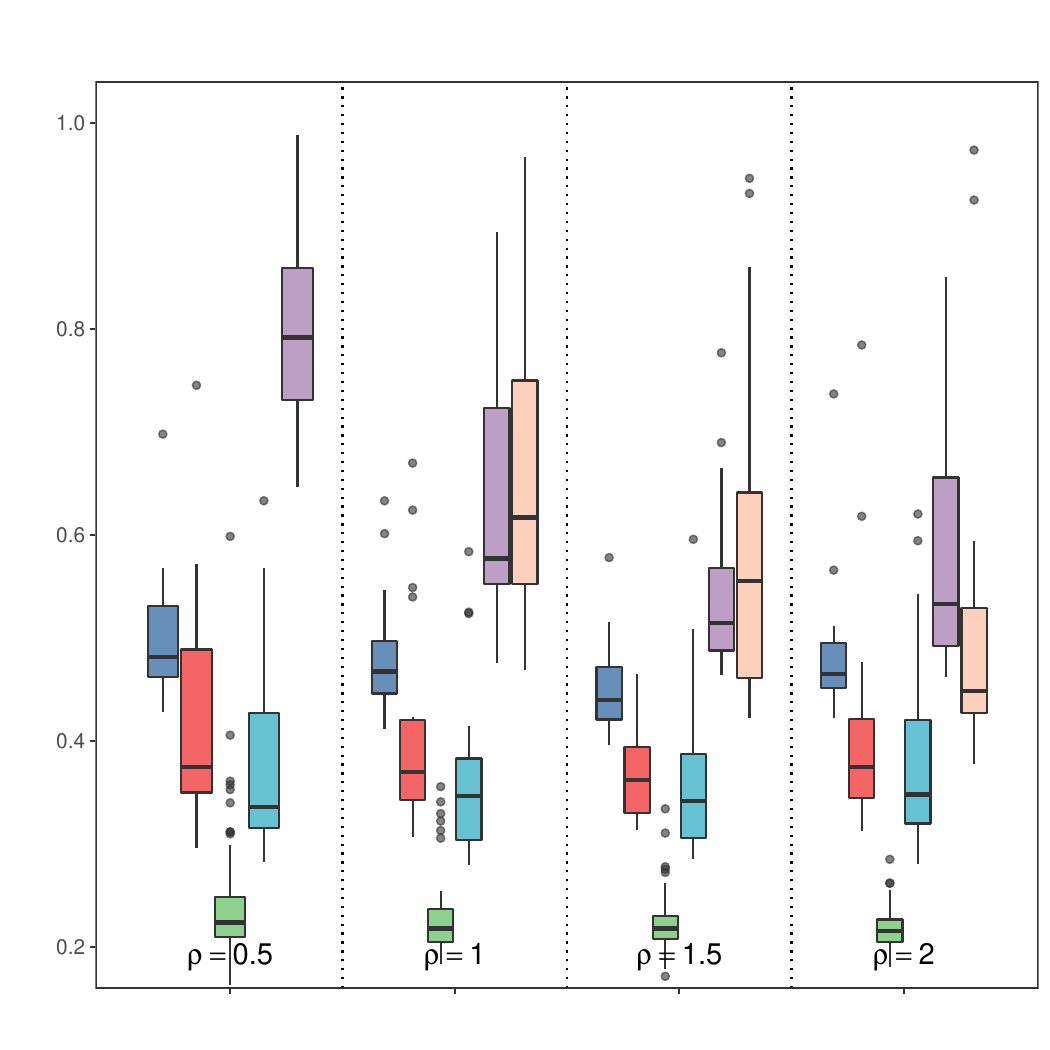}
  \end{minipage}
  \begin{minipage}[t]{.327\linewidth}
    \includegraphics[width=5cm, height=5cm]{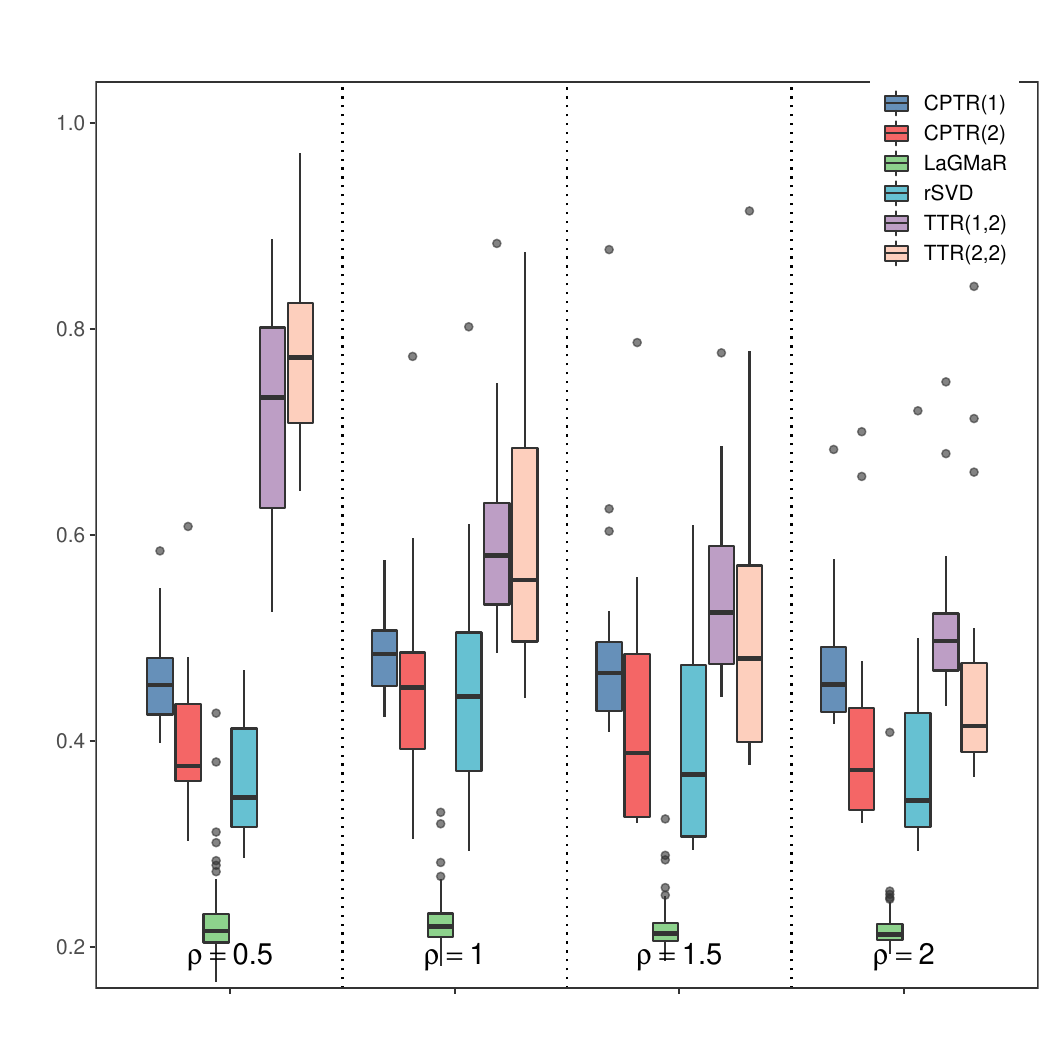}
  \end{minipage}
  
    \begin{minipage}[t]{.327\linewidth}
    \includegraphics[width=5cm, height=5cm]{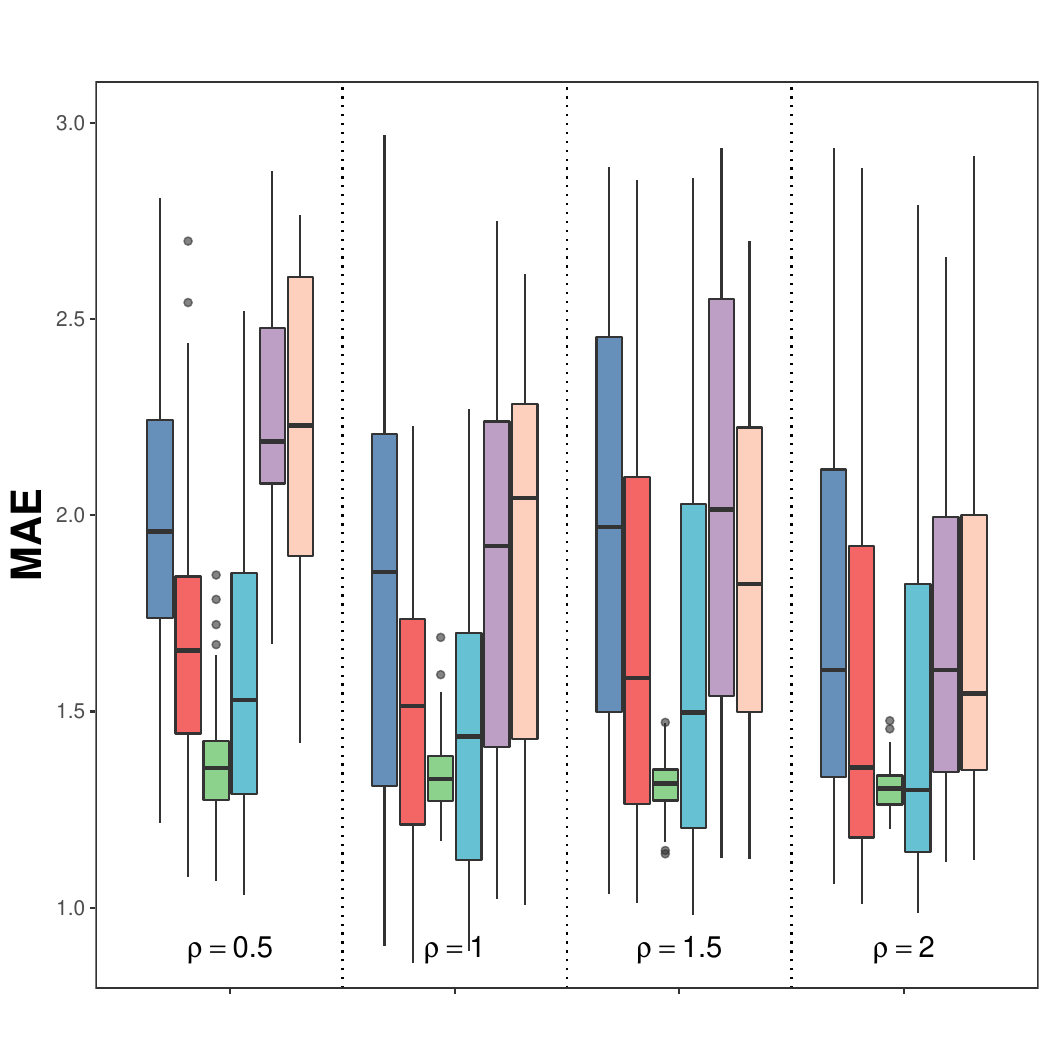}
  \end{minipage}
  \begin{minipage}[t]{.327\linewidth}
    \includegraphics[width=5cm, height=5cm]{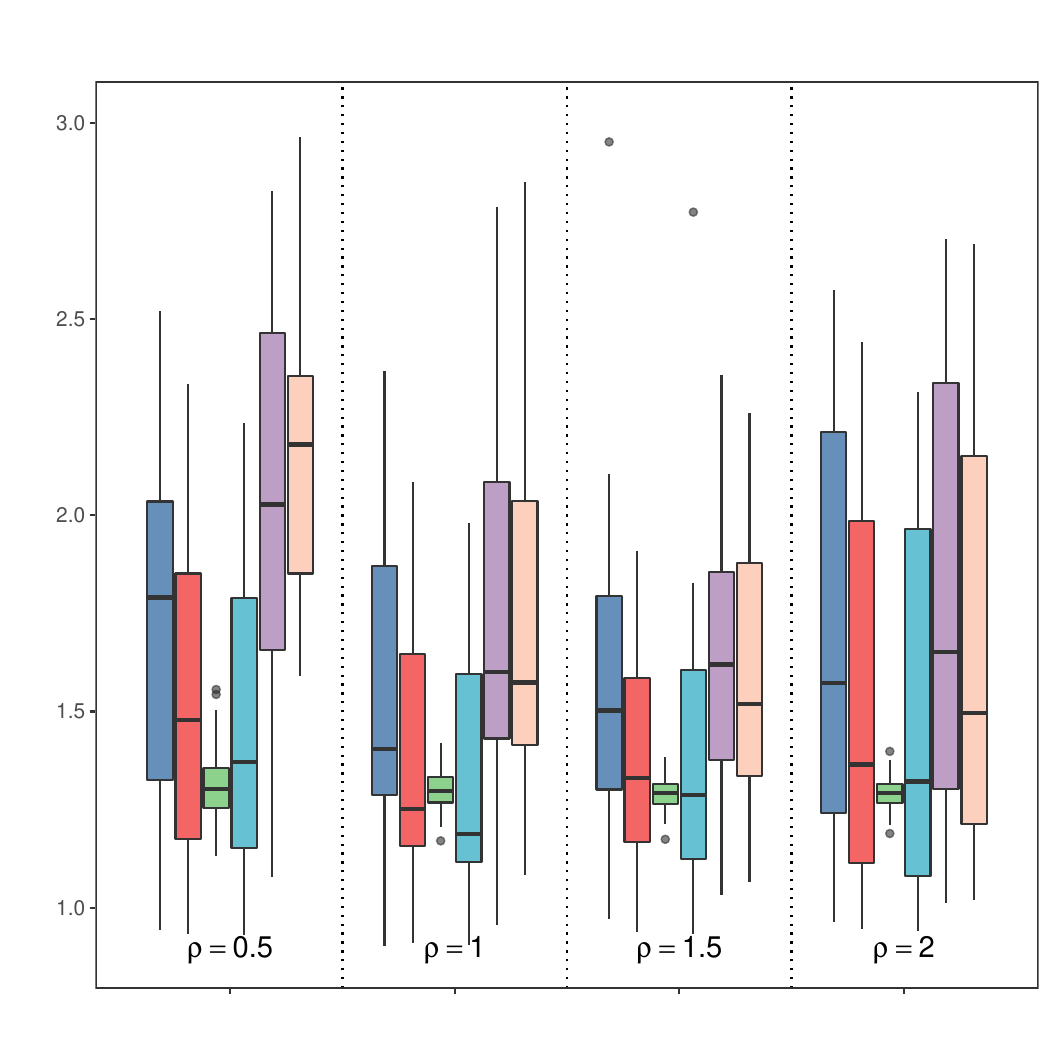}
  \end{minipage}
  \begin{minipage}[t]{.327\linewidth}
    \includegraphics[width=5cm, height=5cm]{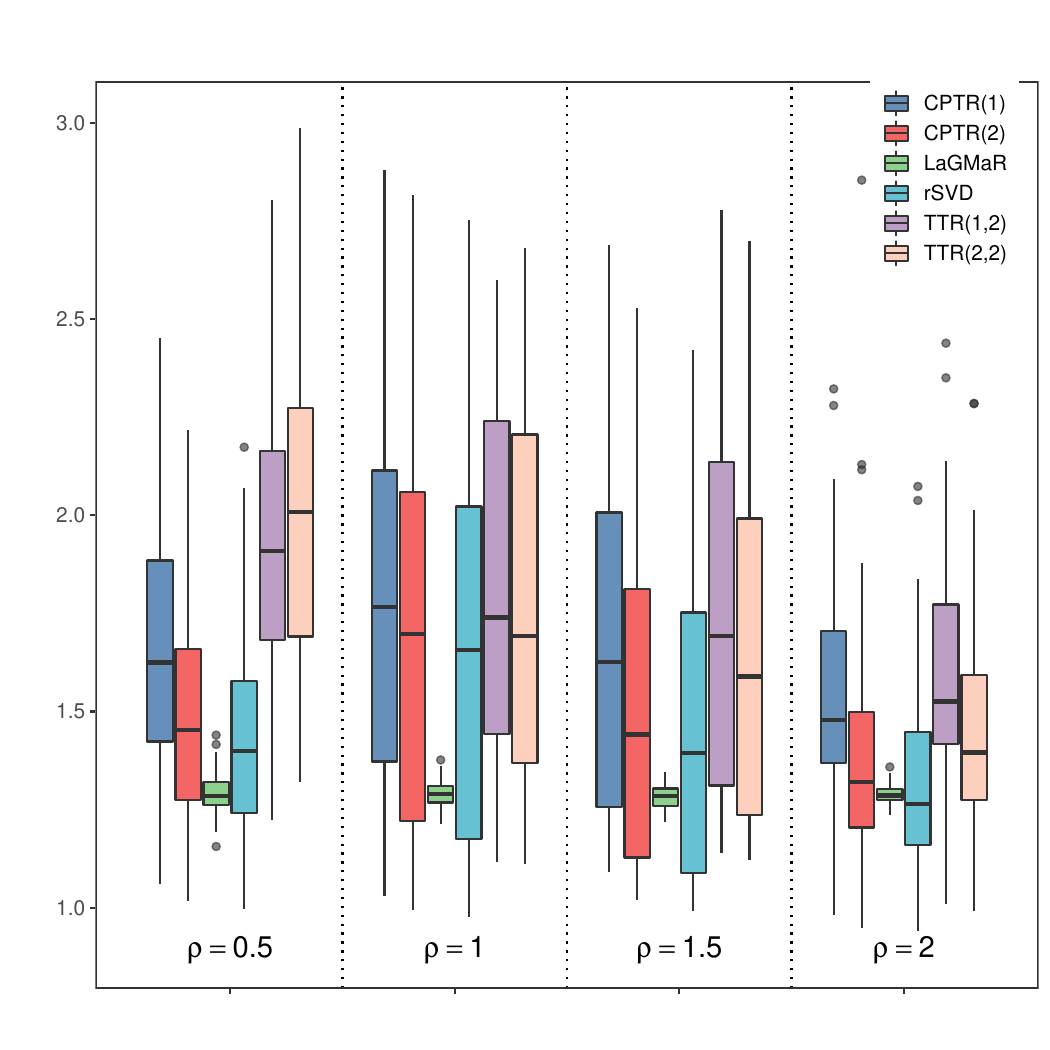}
  \end{minipage}
    \caption{(a)-(c): 
    Metrics of  RMSE, NMSE, and MAE by the six methods under the Poission regression setting with matrix-variate covariate with different dimensionality scenarios of $(p_1, p_2)$.
    }
        \label{P_rmse}
\end{figure}

\textbf{Case 2: Poisson regression} The boxplots of the three metrics under different scenarios of $\left\{(p_1, p_2), \rho \right\}$ are displayed in Figure \ref{P_rmse}. LaGMaR achieves the smallest prediction volatility among four models, that is, LaGMaR has the smallest RMSE, NMSE, and MAE under all settings of $\left\{(p_1, p_2), \rho \right\}$.

\centerline{[Figure \ref{P_rmse} about here.]}

\begin{figure}
  \centering
  \begin{minipage}[t]{.327\linewidth}
    \includegraphics[width=5cm, height=5cm]{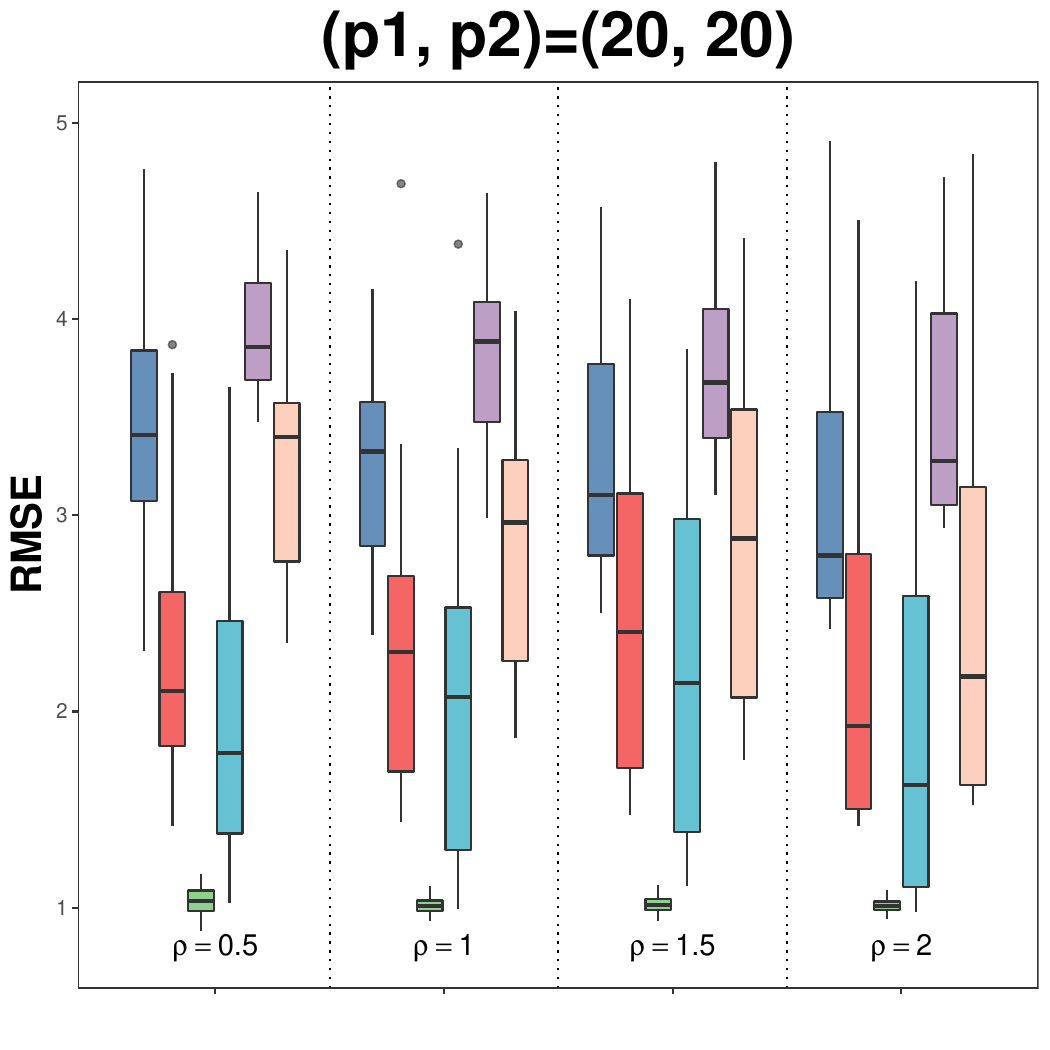}
  \end{minipage}
  \begin{minipage}[t]{.327\linewidth}
    \includegraphics[width=5cm, height=5cm]{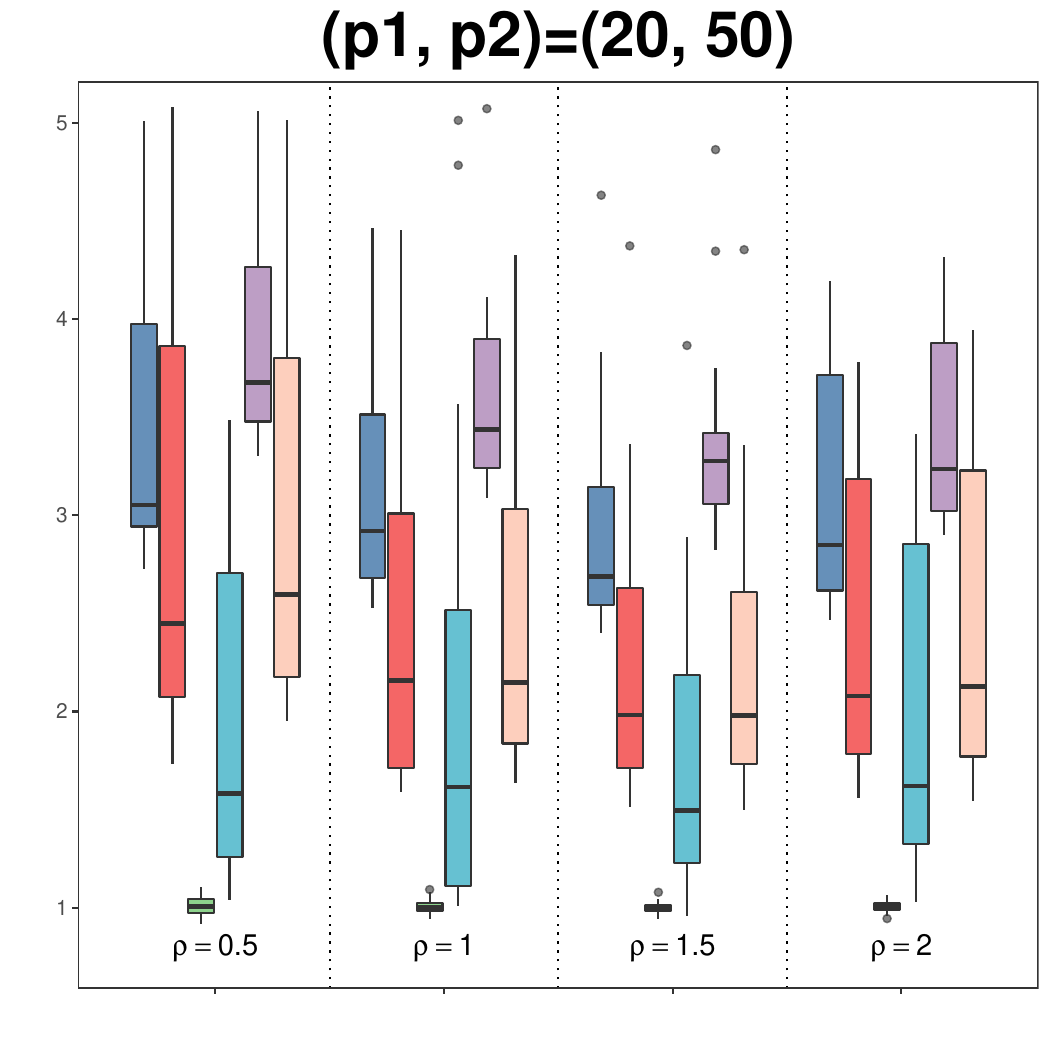}
  \end{minipage}
  \begin{minipage}[t]{.327\linewidth}
    \includegraphics[width=5cm, height=5cm]{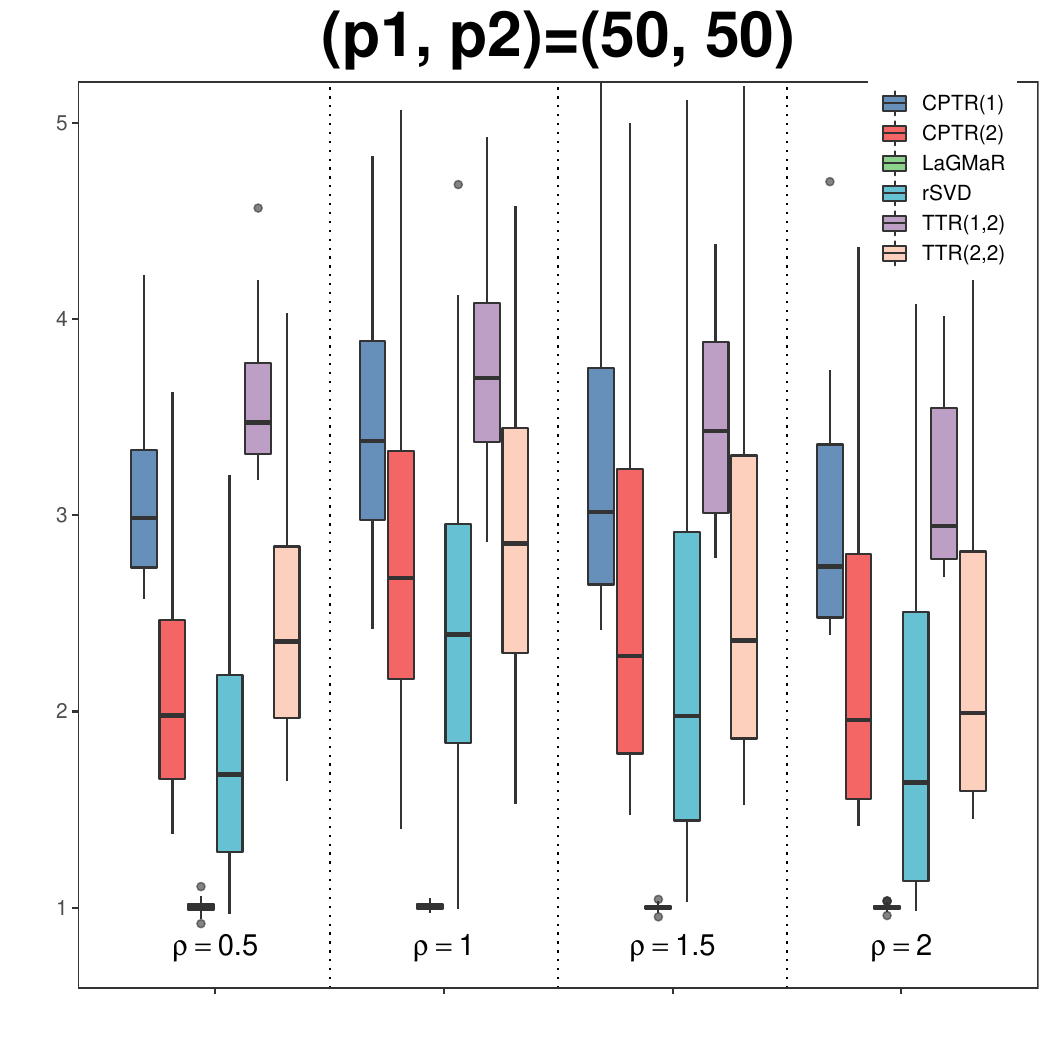}
  \end{minipage}
  
    \centering
  \begin{minipage}[t]{.327\linewidth}
    \includegraphics[width=5cm, height=5cm]{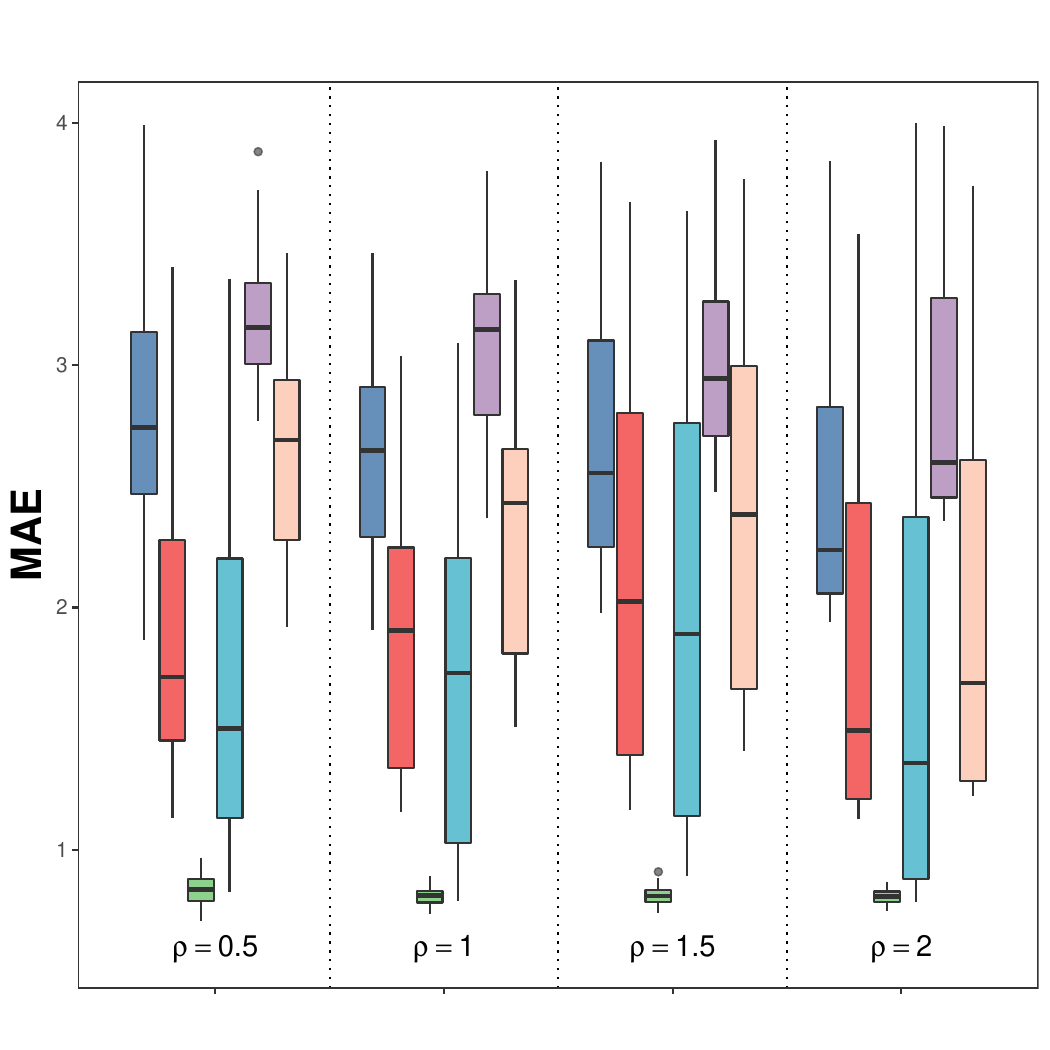}
  \end{minipage}
  \begin{minipage}[t]{.327\linewidth}
    \includegraphics[width=5cm, height=5cm]{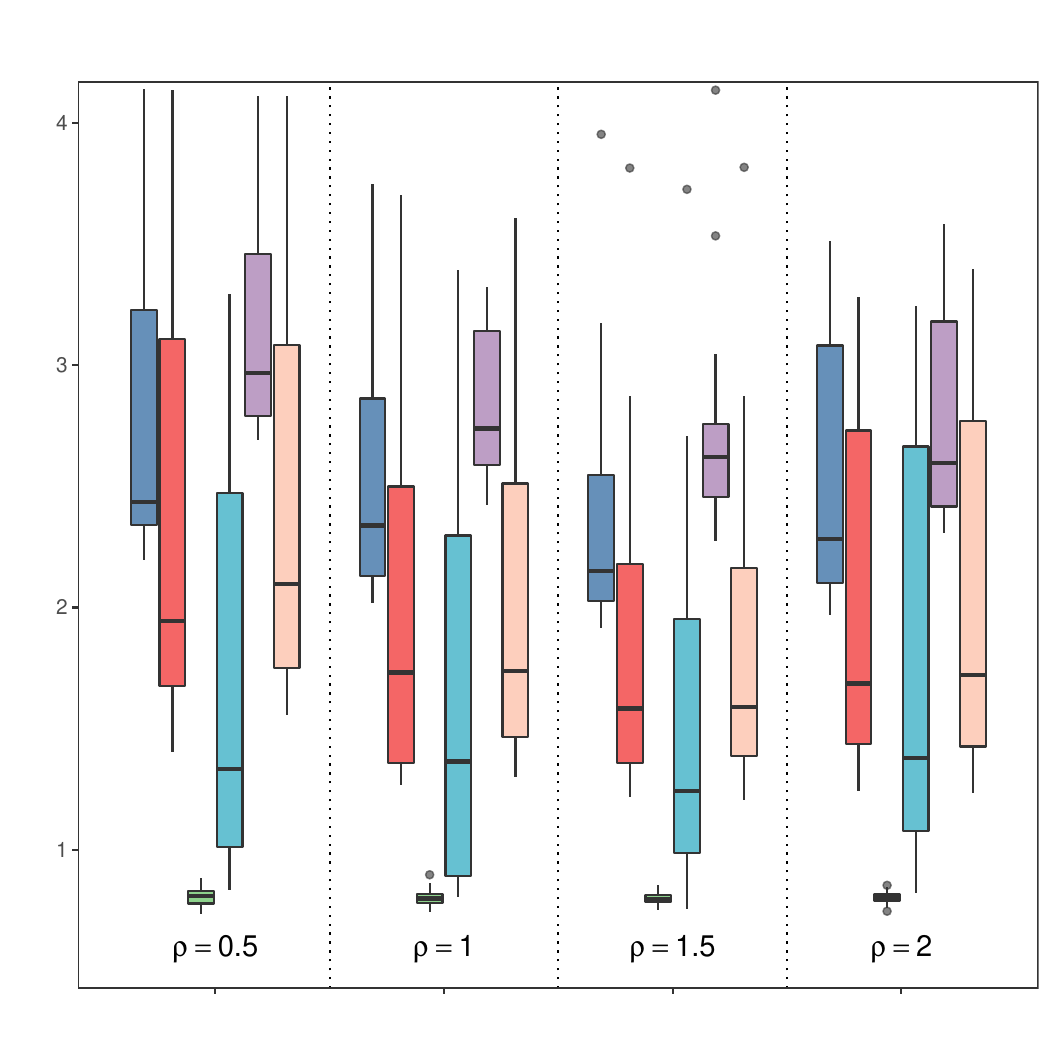}
  \end{minipage}
  \begin{minipage}[t]{.327\linewidth}
    \includegraphics[width=5cm, height=5cm]{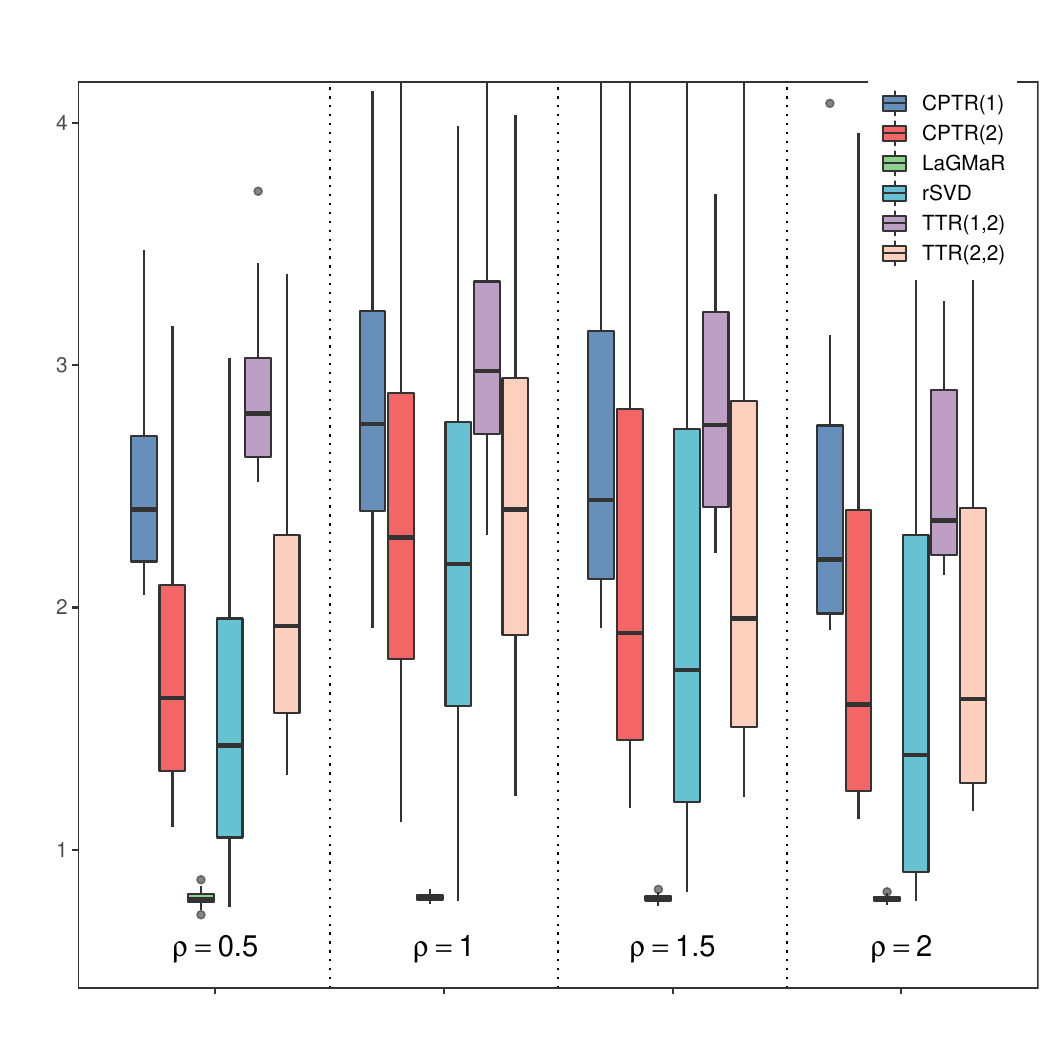}
  \end{minipage}
  \caption{
   Metrics of  RMSE and MAE by the six methods under the Poission regression setting with matrix-variate covariate with different dimensionality scenarios of $(p_1, p_2)$.
    }
      \label{O_rmse}
\end{figure}

\textbf{Case 3: Linear regression} The boxplots of RMSE and MAE under different combinations of $\left\{(p_1, p_2), \rho \right\}$ are displayed in Figure \ref{O_rmse}. It is clear that both RMSE and MAE decrease when the sample size becomes larger for fixed $(p_1, p_2)$; meanwhile, it can be seen that the RMSE of LaGMaR is the smallest among four models under all scenarios of $\left\{(p_1, p_2), \rho \right\}$, indicating the smallest prediction volatility of LaGMaR.

\centerline{[Figure \ref{O_rmse} about here.]}

\textbf{Case 4: Real COVID-CT data setting} This simulation scheme resembles the real COVID-CT data set, in which the sample size is $n=746$,  the proportion of infection is about 47\% based on the fact that there are 349 positive and 397 negative CT images, and the dimension of the observed matrix predictors is fixed with $(p_1, p_2)=(150, 150)$.  The entries of true loading matrices $\mathbf{R}$ and $\mathbf{C}$ are independently sampled from the uniform distribution $U(-\sqrt{p_1}, \sqrt{p_1})$ and $U(-\sqrt{p_2}, \sqrt{p_2})$. Each entry of the noise matrix $\mathbf{E}$ is generated from  $N(0,1)$. The latent matrix factor $\mathbf{Z}$ takes the number of factors of $(k_1, k_2)=(3,3)$ and draws from $\text{vec}(\mathbf{Z}) \sim MVN(\mathbf{\phi}, \mathbf{\Sigma})$, where $\mathbf{\phi}=(1, 0.5, 1, -0.5, 1, 0.5, 1, -0.5, 1)^{\top}$, and $\mathbf{\Sigma}_{ij}=0.5^{| i-j |},1 \leq i,j \leq k_1  k_2$.  
We generate the binary response values by $Y \sim \text{Bernoulli}(\mu)$, where $\mu=\gamma + \left<\mathbf{A}, \mathbf{Z}\right>$ with $\text{vec}(\mathbf{A})=(2, -2, \mathbf{1}_3^{\top}, -\mathbf{1}_4^{\top})^{\top}$ and $\gamma=\text{logit}(0.47)-\text{vec}(\mathbf{A})^{\top}\phi=\text{logit}(0.47)-0.5$.
We repeat simulations for 100 times, and the resulting proportion of infection is 48.9\%. Table \ref{COVID-simu} summarizes the mean and standard deviation of the five classification metrics under the aforementioned logistic regression setting. 
LaGMaR and rSVD appear to be competitive and overwelmingly superior to the other five methods, which cannot reach a satisfactory discriminant power. The possible reason is that $(p_1, p_2)$, the dimensionality of the matrices, is relatively high beyond the scope that these models can handle.

\centerline{[Table \ref{COVID-simu} about here.]}

\textbf{Estimation of factor numbers} We investigate whether LaGMaR can always select the true rank $(k_1, k_2)$ when estimating the latent factor matrix. We predefine six different combinations of true ranks $(k_1, k_2)$: $\left\{(2, 3), (2, 4), (2, 5), (3, 3), (3, 4), (3, 5)\right\}$, and compute the proportion that LaGMaR select the true rank, i.e., $(\hat{k}_1, \hat{k}_2)=(k_1, k_2)$, under different dimension $(p_1, p_2)$ and sample size. 
Simulation results based on 100 replicates are presented in Table \ref{rank_select}. It can be concluded that in latent dimension estimation, LaGMaR can always select the true rank with varying dimension $(p_1, p_2)$ and sample size. This may explain why LaGMaR behaves robust and can outplay other penalized methods which are sensitive to tuning parameters.

\begin{figure}
  \centering
  \begin{minipage}[t]{.327\linewidth}
    \includegraphics[width=4.5cm, height=5cm]{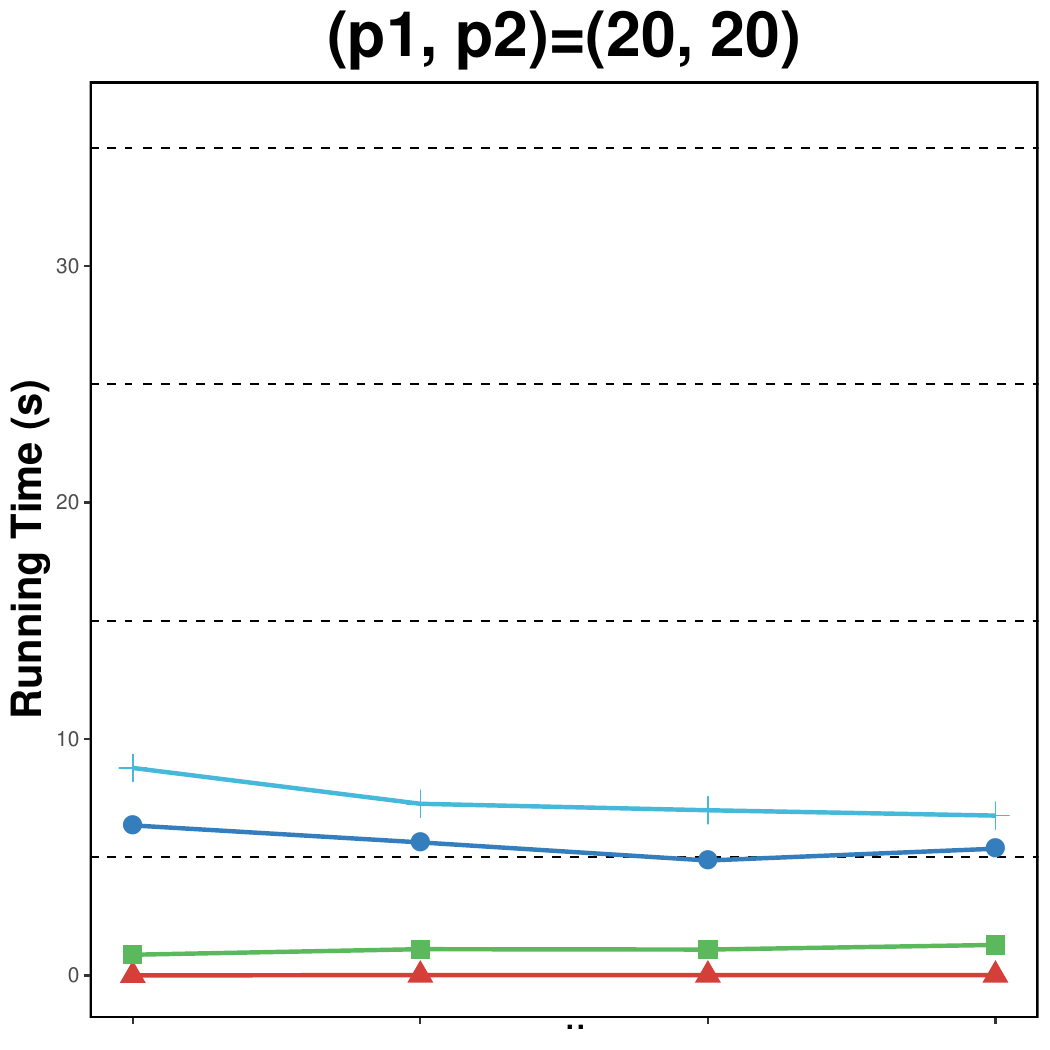}
  \end{minipage}
  \begin{minipage}[t]{.327\linewidth}
    \includegraphics[width=4.5cm, height=5cm]{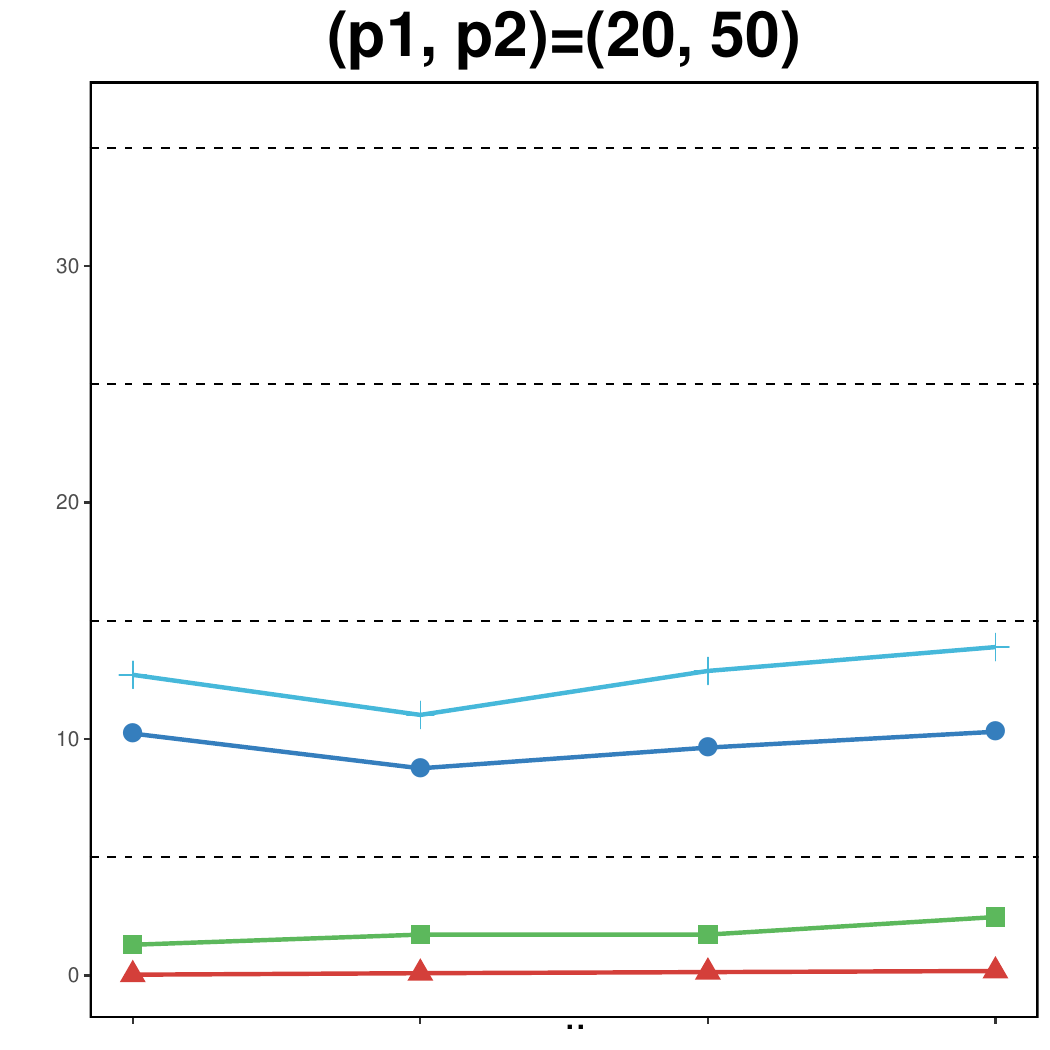}
  \end{minipage}
  \begin{minipage}[t]{.327\linewidth}
    \includegraphics[width=5.5cm, height=5cm]{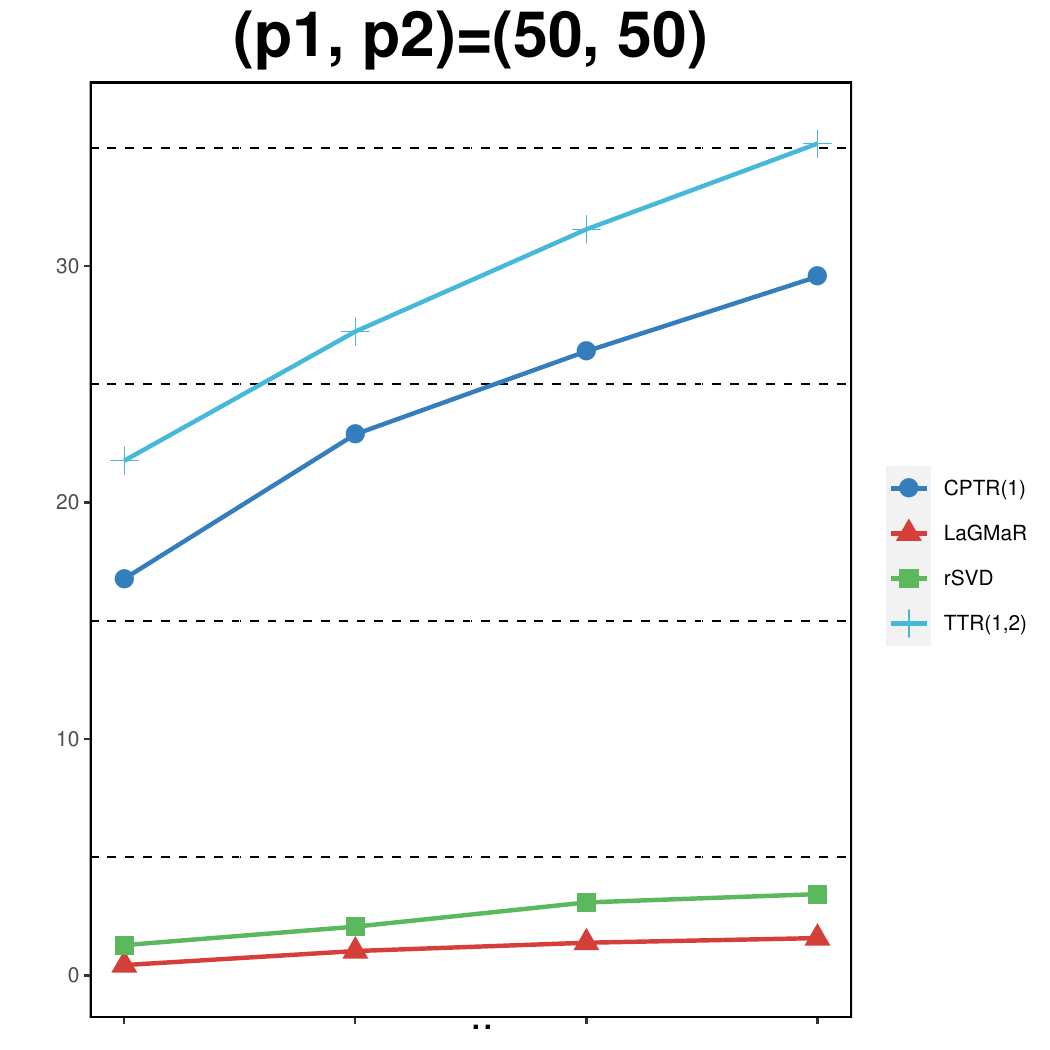}
  \end{minipage}
    \caption{Running time required to implement one replicate of fitting the logistic regression by four approaches including LaGMaR (in red triangle), CPTR(1) (in blueviolet circle), rSVD (in green square) and TTR (in cyan plus),  under different combination of $(p_1, p_2)$ when the sample size coefficient $\rho$ increases from 0.5 to 2.}
     \label{runtime}
\end{figure}

\textbf{Comparison of Computation Time} To compare computation time of LaGMaR, CPTR(1), rSVD, and TTR(1,2), we focus on the logistic regression setting. 
CPTR(2) and TTR(2,2) are not counted as it is known time consuming for fitting higher order model. 
We record the run time of fitting each model for 100 simulated datasets. 
The medians of the computation time are summarized in Figure \ref{runtime}. The four models are all implemented by Matlab2019a, and the code is tested on a Windows10 laptop with Intel(R) Core(TM) i7-1065G7 CPU @ 1.30GHz and 16GB RAM. It can be seen that TTR and CPTR are much more time consuming than LaGMaR and rSVD. Moreover, the computation time of LaGMaR does not increase significantly as $(p_1, p_2)$ gets larger.

\subsection{Additional Simulation Results with Data Generated from \citet{zhou2014regularized}}

In this subsection, we generate data under the model formulated in \citet{zhou2014regularized}, which is also the way to generate 2D-tensor (matrix) \citep{zhou2013tensor}, as follows:
$$g(\mu) = \gamma^{\top} \mathbf{v} + \left \langle \mathbf{B}, \mathbf{X} \right \rangle. $$ 
Specifically, we generate the matrix covariate $\mathbf{X}$ of size $64 \times 64$ and the five-dimensional vector covariate $\mathbf{v}$, both of which consist of independent standard normal entries. We set the sample size at $n = 500$. We set $\gamma=\mathbf{1}_5$ and generate the true array signal as $\mathbf{B}=\mathbf{B}_1 \mathbf{B}_2^{\top}$, where $\mathbf{B}_d \in \mathbb{R}^{p_d \times R}, d=1, 2$. Moreover, each entry of $\mathbf{B}$ is 0 or 1, and the percentage of non-zero entries is controlled by a sparsity level constant $s$, i.e. each entry of $\mathbf{B}_d$ is a Bernoulli distribution with the success probability being $\sqrt{1-(1-s)^{1/R}}$. We vary the rank $R = 1, 5, 10, 20$, and the level of sparsity $s = 0.01, 0.05, 0.1, 0.2, 0.5$. 

The results reported in Tables \ref{zhoulirmsepoisson}, \ref{zhouliacc}, and \ref{zhoulirmsenormal} display that LaGMaR, rSVD, and CPTR(1) outperform other methods in Poisson regression, logistic regression, and linear regression, respectively.

\section{Real Example: COVID-CT Data Set}
\label{sec4}
In this section, we apply the proposed LaGMaR approach to a real open-source chest CT data called COVID-CT, which was one of the largest publicly available COVID-19 CT scan dataset in the early pandemic. The COVID-CT dataset is available from \url{https://github.com/UCSD-AI4H/COVID-CT}. 
It contains 349 COVID-19 positive CT scans manually selected from the embedded figures out of 760 preprints on medRxiv2 and bioRxiv3, and 397 negative CT scans from four open-access databases \citep{yang2020covid}. 
The whole data set has already been split into three disjoint parts including training set, validation set, and test set. These 2D grayscale images are in the formats of JPEG and PNG. 

We compare the predictive capability of LaGMaR with some existing penalized regression methods proposed by the statistical community, including matrix regressions (MVlogistic of \citet{hung2013matrix} and rSVD of \citet{zhou2014regularized}) and penalized tensor regressions (CPTR of \citet{zhou2013tensor} and TTR of \citet{li2018tucker}). The metrics for evaluation of predictive capability include CA, Kappa, Sensitivity, and AUC, together with graphics including ROC curves and PR curves. 

\textbf{Prediction Procedure} First of all, we read the grayscale image by the \texttt{R} package \texttt{EBImage}. Each pixel value can be rescaled by dividing 255, therefore a grayscale image is converted into a matrix with all entries in the interval [0,1]. Then the images are upsampled or downsampled into squared matrices with unified size of $(\text{height}, \text{width})=(150, 150)$, resulting high-dimensional independent matrix observations $\{Y_i,\mathbf{X}_i\}_{i=1}^{746}$. Here $Y_i=1$ if a subject is positive and $Y_i=0$ otherwise, and $\mathbf{X}_i\in\mathbb{R}^{150\times 150}$ corresponds to the processed CT scan image of subject $i$. Next, we extract low-dimensional matrix-factor features based on the matrix covariates $\{\mathbf{X}_i\}_{i=1}^{746}$. We determine the number of factors $(\hat{k}_1, \hat{k}_2) = (3, 4)$ based on \eqref{rk1} and empirically choose relative larger size of $k_1 = k_2 = 5$. 
The resulting factor scores $\hat{\mathbf{Z}}_i \in \mathbb{R}^{5 \times 5}$ are based on \eqref{estZ}, in which the corresponding estimated loading matrices $\hat{\mathbf{R}}$ and $\hat{\mathbf{C}}$ are based on \eqref{ms1} and \eqref{ms2}, respectively. 
Last, we fit the LaGMaR using Algorithm \ref{lagmar_fit} and evaluate its prediction performance using Algorithm \ref{lagmar_pred}. Note that we work on the whole data set without spliting training and test sets.

\begin{figure}
  \centering
  \begin{minipage}[t]{.5\linewidth}
    \includegraphics[width=9cm, height=7cm]{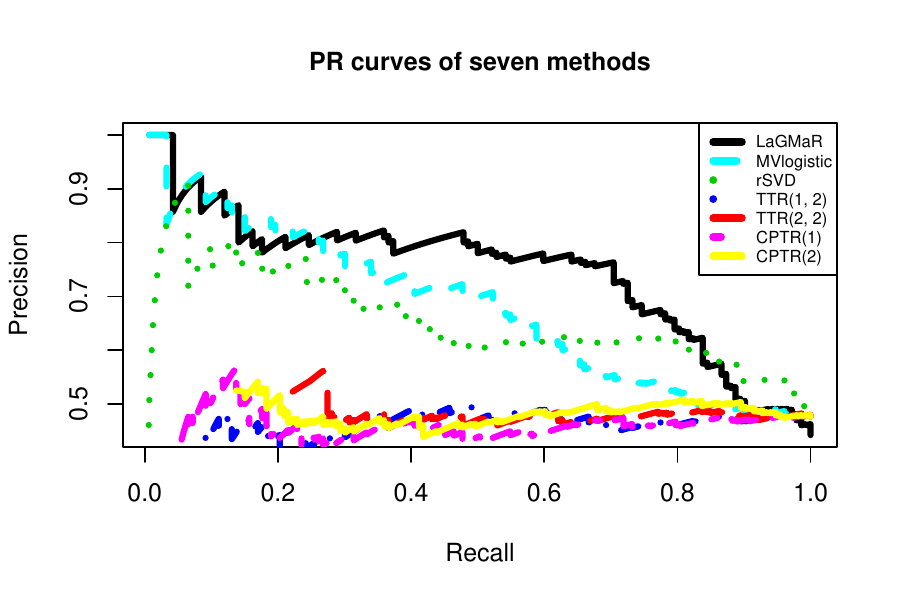}
  \end{minipage}
  \begin{minipage}[t]{.5\linewidth}
    \includegraphics[width=9cm, height=7cm]{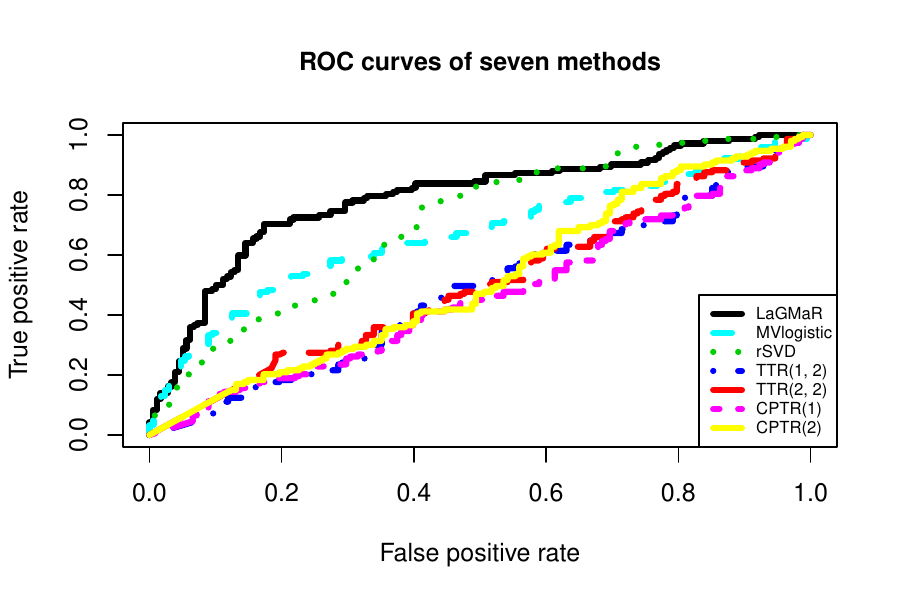}
  \end{minipage}
    \caption{ Assessment of prediction of COVID-19 under seven methods for the COVID-CT data:
    upper are the PR curves and lower are ROC curves including LaGMaR (in black solid line), MVLogistic (in cyan dashed line), rSVD (in green dotted line), TTR(1,2) (in blue dash-dotted line), TTR(2,2) (in red long dashed line), CPTR(1) (in purple two-dashed line), and CPTR(2) (in yellow solid line).}
     \label{PR}
\end{figure}

\textbf{Assessment of Prediction} We assess the prediction results of the considered methods by five-fold cross validation with 50 repetitions, and summarize all metrics in Table \ref{covid} and Figure \ref{PR}. Note that LaGMaR does not need tuning parameter for prediction.
Tuning parameters for rSVD, TTR(1, 2), TTR(2, 2), CPTR(1), and CPTR(2) are determined by the BIC criterion based on the training set, yielding magnitudes of 100, 1, 1, 1, 1, respectively. The tuning parameter for MVLogistic is 10, determined by maximizing classification accuracy on the validation set. 

\textit{Scalar Metrics} As shown in Table \ref{covid}, LaGMaR outforms the other six methods in terms of AUC, sensitivity, CA, and Kappa coefficient. For example,
in terms of AUC, LaGMaR is the most accurate in predictive modeling among all seven methods because it has an excellent capability of separating the COVID-19 patients from the healthy subjects (close to 80\%), while rSVD and MVLogistic have an acceptable degree of separability (with average AUC values 0.755 and 0.742, respectively), and tensor regression methods display no discriminative ability (with AUC values just a little bit higher than 0.5). 

Sensitivity is extremely important in prevention and control of COVID-19 disease considering its fast contagion and worldwide spread among population.
LaGMaR, MVLogistic, and rSVD have descending sensitivities with slight differences. Nevertheless, LaGMaR is the optimal choice in the situation of large scale diagnostics in unrevealing COVID-19 cases, and all four tensor regression methods perform poorly in discriminating COVID-19 cases. 

Though rSVD and MVLogistic have relatively small standard deviation, LaGMaR is the only one whose average CA score surpasses 0.7, 
indicating a better capability to correctly predict both COVID-19 cases and non COVID-19 cases compared to all other approaches. 
This is consistent with the fact that, LaGMaR is the only one that surpasses 0.4 in the Kappa coefficient, and it reaches moderate interrater reliability, while others are below 0.4 and at most can reach the fair level. In conclusion, LaGMaR has a better performance in predicting COVID-19 patients using CT data.

\centerline{[Table \ref{covid} about here.]}

\textit{Graphical Metrics} Figure \ref{PR} displays the ROC curves and PR curves of LaGMaR, together with those of the existing statistical methods.
First, let us look into the PR curves. The precisions of all seven methods are above 0.5 approximately, which means that all of them have practical utility. 
It can be seen that the PR curve of LaGMaR bulges most towards the upper right corner and achieves a notably better discriminant performance compared to other methods. 
When the recall ranges from 0.4 to 0.85, LaGMaR has the highest precision, thus resulting in a higher PR curve; meanwhile, when the recall varies from 0.4 to 0.85, the precision also varies from 0.5 to 0.85. Furthermore, rSVD also appears to be inferior to LaGMaR as it has a convex hull between recall values 0.55 to 0.8 in the PR curve.

Next, let us look into the ROC curves. LaGMaR is superior to all other methods for most FPR values, especially around FPR = 0.2. 
For the partial ROC area, when the FPR falls between 0.03 and 0.35, MVLogistic behaves better than the other methods except LaGMaR. The four tensor regression methods are unable to discriminate since their ROC curves are around inverse diagonal straightline.

\centerline{[Figure \ref{PR} about here.]}

\section{Discussion}
\label{sec5}
Since the COVID-CT dataset that inspired our research was previously analyzed by CRNet \citep{he2020sample}, we also compare the predictive capability of LaGMaR with CRNet, together with the classical ResNets \citep{2016Deep}. 
Both CRNet and ResNets are the variants of deep convolutional neural networks (CNNs). 
For the comparison between LaGMaR and ResNets, we evaluate the performance of them using repeated five-fold cross validation and repeat this process 50 times; as for LaGMaR and CRNet, we evaluate the performance of them on the test set predefined in \citet{yang2020covid} in order to make the comparison fair. The detailed experiment results are given in Table \ref{covid-resnet}. 
Though LaGMaR outplays existing statistical methods in numerical analysis, it is not a surprise that CNN variants outrun LaGMaR based on the ``All'' row cell.
Nevertheless, the gap is not that large; meanwhile, from the``Small'' row cell, we display that LaGMaR may outrun CNNs if the sample size is small by instance that a COVID-CT subset containing randomly-sampled 120 positive cases and 136 negatives in total. 

For the future work, we may consider two tracks of extension of LaGMaR considering that matrices are indeed 2nd-order tensors. One is to consider supervised learning including higher-order tensor-variate covariate. This is motivated by arising research results on unsupervised learning of various tensor factorizations \citep{ChenYangZhang2021}. It is worthwhile to investigate what kind of tensor factor scores extracted from some kind of low-rank approximation will be more representative to predict the response. 
The other is to consider the problem of predicting one imaging modality by another imaging modality. For example, in breast cancer, to avoid the health risk of the usage of contrast agent for MRI scanning, specialists would prefer to predict MRI image by the safe noninvasive fMRI imaging modality \citep{zhou2015mri}. Such problems calls for the urgent need of tensor to tensor regression.

\section{Software}
\label{sec6}

Software in the form of Rcpp, together with a sample input data set and complete documentation is available on request from the corresponding author (macliu@polyu.edu.hk).

\section*{Acknowledgments}
Zhang Yuzhe's research is partly supported by the postgraduate studentship of USTC. 
Zhang Xu's research supported in part by the National Natural Science Foundation of China (12171167).
Zhang Hong's research is partly supported by the National Natural Science Foundation of China (7209121, 12171451).
Liu Catherine's research is partly supported by General Research Funding (15301519), RGC, HKSAR.

{\it Conflict of Interest}: None declared.

\section*{Supplementary Materials}
Web Appendix is available with this paper at the Biostatistics website on Wiley Online Library.\vspace*{-8pt}

\bibliographystyle{biorefs}
\bibliography{main0930.bbl}

\newpage

\newpage

\newpage

\begin{table}[htbp]
\caption{Selected tuning parameters for rSVD, TTR(1, 2), TTR(2, 2), CPTR(1), and CPTR(2) in Subsection 3.1.}
\label{tuning33}
\begin{tabular}{cccccc}
\toprule
Response & Method & \multicolumn{4}{c}{Results for the following  dimensionality scenarios of $(p_1, p_2)$:} \cr 
\cmidrule(lr){3-6}\\
               &        & $(p_1, p_2)=(20, 20)$       & $(p_1, p_2)=(20, 50)$      & $(p_1, p_2)=(50, 50)$   \\
\toprule
binary       & rSVD  & 100  & 100 & 100   \\
               & TTR(1,2) & 50  & 20 & 100    \\
               & TTR(2,2) & 100  & 50 & 100    \\
               & CPTR(1) & 100  & 100 & 100    \\
               & CPTR(2) & 50  & 100 & 100   \\
               & MVLogistic & 10  & 50 & 20   \\
               &        &             &            &        &     \\
normal         & rSVD  & 100  & 100 & 100   \\
               & TTR(1,2) & 20  & 50 & 10    \\
               & TTR(2,2) & 100  & 100 & 20  \\
               & CPTR(1) & 20  & 1 & 10    \\
               & CPTR(2) & 100  & 10 & 100   \\
               &        &             &            &       &     \\
poisson        & rSVD  & 100  & 100 & 100   \\
               & TTR(1,2) & 5  & 10 & 5   \\
               & TTR(2,2) & 50  & 10 & 20   \\
               & CPTR(1) & 10  & 100 & 100    \\
               & CPTR(2) & 100  & 50 & 100   \\
\bottomrule
\end{tabular}
\end{table}

\newpage
\begin{table}[htbp]
    \centering
	\caption{The means (standard errors) of classification metrics based on 100 simulated COVID-CT datasets with five-fold cross validation.}
	\label{COVID-simu}
	\begin{tabular}{c|ccccc}
		\hline
		            &CA       &Kappa         &Sensitivity               &AUC    &F1 score \\
		\hline
		LaGMaR  &\textbf{0.855}(0.016) &\textbf{0.708}(0.033) &0.853(0.026) &\textbf{0.936}(0.011) &\textbf{0.851}(0.019) \\
		rSVD   &0.854(0.012) &0.707(0.024) &\textbf{0.856}(0.016) &0.935(0.008) &0.850(0.013) \\
		MVLogistic &0.489(0.043) &-0.018(0.085) &0.549(0.144) & 0.505(0.054) &0.459(0.100) \\
		TTR(1,2) &0.489(0.017) &0.000(0.000) &1.000(0.000) &0.500(0.000) &0.656(0.016) \\
		TTR(2,2) &0.489(0.017) &0.000(0.000) &1.000(0.000) &0.500(0.000) &0.656(0.016) \\
		CPTR(1) &0.489(0.017) &0.000(0.000) &1.000(0.000) &0.500(0.000) &0.656(0.016)  \\
		CPTR(2) &0.489(0.017) &0.000(0.000) &1.000(0.000) &0.500(0.000) &0.656(0.016)  \\
		\hline
	\end{tabular}
\end{table}

\begin{table}[htbp]
\caption{Frequencies of $(\hat{k}_1, \hat{k}_2)=(k_1, k_2)$ using LaGMaR under different combinations of true $(k_1, k_2)$.}
\label{rank_select}
\resizebox{\linewidth}{!}
{
\begin{tabular}{c|cccc|cccc|cccc}
\hline
                         & \multicolumn{4}{c}{$(p_1, p_2)=(20,20)$}                       & \multicolumn{4}{c}{$(p_1, p_2)=(20,50)$}                       & \multicolumn{4}{c}{$(p_1, p_2)=(50,50)$}                       \\
\hline
$(\hat{k}_1, \hat{k}_2)$ & $\rho=0.5$ & $\rho=1$ & $\rho=1.5$ & $\rho=2$ & $\rho=0.5$ & $\rho=1$ & $\rho=1.5$ & $\rho=1$ & $\rho=0.5$ & $\rho=1$ & $\rho=1.5$ & $\rho=2$ \\
\hline
(2, 3)                   & 1             & 1          & 1             & 1           & 1             & 1          & 1             & 1           & 1             & 1          & 1             & 1           \\
(2, 4)                   & 1             & 1          & 1             & 1           & 1             & 1          & 1             & 1           & 1             & 1          & 1             & 1           \\
(2, 5)                   & 1             & 1          & 1             & 1           & 1             & 1          & 1             & 1           & 1             & 1          & 1             & 1           \\
(3, 3)                    & 1             & 1          & 1             & 1           & 1             & 1          & 1             & 1           & 1             & 1          & 1             & 1           \\
(3, 4)                    & 1             & 1          & 1             & 1           & 1             & 1          & 1             & 1           & 1             & 1          & 1             & 1           \\
(3, 5)                    & 1             & 1          & 1             & 1           & 1             & 1          & 1             & 1           & 1             & 1          & 1             & 1           \\
\hline
\end{tabular}}
\end{table}

\begin{table}[htbp]
\caption{The means (standard errors) of RMSE of the Poisson response based on 100 simulated datasets with five-fold cross validation. The data generation process follows rSVD and CPTR.}
\label{zhoulirmsepoisson}
\resizebox{0.9\linewidth}{!}{
\begin{tabular}{cccccc}
\toprule
Sparsity $s$(\%) & Method & \multicolumn{4}{c}{Results for the following ranks:} \cr 
\cmidrule(lr){3-6}\\
               &        & $R=1$       & $R=5$      & $R=10$      & $R=20$      \\
\toprule
10             & LaGMaR(1,1) & 12.314(0.788)  & 13.247(0.760) & 12.512(0.856)  & 12.512(0.926)  \\
               & LaGMaR(2,2) & 10.567(0.703)  & 11.757(0.769) & 13.025(0.822)  & 11.704(0.781)  \\
               & LaGMaR(3,3) & 10.629(0.716)  & 11.824(0.783) & 13.104(0.826)  & 11.775(0.773)  \\
               & LaGMaR(4,4) & 10.715(0.724)  & 11.942(0.790) & 13.234(0.836)  & 11.889(0.782)  \\
               & LaGMaR(5,5) & 10.868(0.740)  & 12.105(0.798) & 13.409(0.842)  & 12.043(0.791)  \\
               & rSVD  & 15.403(1.200)  & 14.420(0.703) & 13.632(0.606)  & 16.228(1.092)  \\
               & TTR(1,2) & 296.559(484.901)  & 574.104(1038.117) & 846.159(1483.645)  & 990.518(940.949)   \\
               & TTR(2,2) & 34030.793(123601.330)  & Inf(Inf) & 14394.518(19986.050)  & Inf(Inf)   \\
               & CPTR(1) & 316.674(483.591)  & 368.003(255.255) & 514.172(398.211)  & 823.437(571.964)   \\
               & CPTR(2) & 84.874(61.007)  & 123.584(67.771) & 148.207(111.035)  & 260.563(217.494)   \\
               &        &             &            &             &             \\
20             & LaGMaR(1,1) & 17.388(1.194)  & 16.802(1.202) & 18.052(1.148)  & 18.611(1.179)  \\
               & LaGMaR(2,2) & 18.312(1.166)  & 19.371(1.363) & 16.078(1.111)  & 21.819(1.667)  \\
               & LaGMaR(3,3) & 18.444(1.177)  & 19.490(1.379) & 16.161(1.106)  & 21.949(1.678)  \\
               & LaGMaR(4,4) & 18.614(1.184)  & 19.676(1.379) & 16.318(1.129)  & 22.167(1.707)  \\
               & LaGMaR(5,5) & 18.853(1.205)  & 19.924(1.384) & 16.530(1.173)  & 22.457(1.704)  \\
               & rSVD  & 19.543(0.847)  & 20.741(1.329) & 21.646(0.715)  & 25.603(0.740)  \\
               & TTR(1,2) & 660.426(1071.138)  & 1761.285(1906.405) & 3435.370(6292.774)  & Inf(Inf)   \\
               & TTR(2,2) & 21305.000(58311.357)  & Inf(Inf) & Inf(Inf)  & Inf(Inf)   \\
               & CPTR(1) & 562.783(970.710)  & 4112.907(4385.487) & 2700.987(2969.006)  & Inf(Inf)   \\
               & CPTR(2) & 1325.903(4343.079)  & 3348.906(7990.895) & 3681.085(6852.666)  & 5838.914(7490.914)   \\
               &        &             &            &             &             \\
50             & LaGMaR(1,1) & 24.348(1.655)  & 40.296(2.624) & 40.850(2.697)  & 37.852(2.441)  \\
               & LaGMaR(2,2) & 28.830(1.880)  & 37.083(2.565) & 38.079(2.073)  & 40.155(2.548)  \\
               & LaGMaR(3,3) & 29.003(1.869)  & 37.304(2.582) & 38.349(2.073)  & 40.404(2.604)  \\
               & LaGMaR(4,4) & 29.319(1.847)  & 37.627(2.564) & 38.750(2.066)  & 40.772(2.638)  \\
               & LaGMaR(5,5) & 29.746(1.828)  & 38.113(2.533) & 39.289(2.083)  & 41.349(2.700)  \\
               & rSVD  & 33.619(1.788)  & 42.721(2.464) & 47.199(2.183)  & 43.833(1.175)  \\
               & TTR(1,2) & 1852.702(1096.758)  & Inf(Inf) & Inf(Inf)  & Inf(Inf)   \\
               & TTR(2,2) & Inf(Inf)  & Inf(Inf) & Inf(Inf)  & Inf(Inf)   \\
               & CPTR(1) & 2978.245(4877.704)  & Inf(Inf) & Inf(Inf)  & Inf(Inf)   \\
               & CPTR(2) & 9754.005(16252.354)  & Inf(Inf) & Inf(Inf)  & Inf(Inf)   \\
\bottomrule
\end{tabular}}
\end{table}

\begin{table}[htbp]
\caption{The means (standard errors) of classification accuracy of the binomial response based on 100 simulated datasets with five-fold cross validation. The data generation process follows rSVD  CPTR.}
\label{zhouliacc}
\resizebox{0.7\linewidth}{!}{
\begin{tabular}{cccccc}
\toprule
Sparsity $s$(\%) & Method & \multicolumn{4}{c}{Results for the following ranks:} \cr 
\cmidrule(lr){3-6}\\
               &        & $R=1$       & $R=5$      & $R=10$      & $R=20$      \\
\toprule
10             & LaGMaR(1,1) & 0.507(0.025)  & 0.499(0.025) & 0.502(0.025)  & 0.497(0.025)  \\
               & LaGMaR(2,2) & 0.503(0.025)  & 0.502(0.025) & 0.501(0.028)  & 0.506(0.025)  \\
               & LaGMaR(3,3) & 0.506(0.022)  & 0.504(0.025) & 0.500(0.025)  & 0.506(0.024)  \\
               & LaGMaR(4,4) & 0.508(0.022)  & 0.503(0.024) & 0.507(0.026)  & 0.507(0.025)  \\
               & LaGMaR(5,5) & 0.506(0.025)  & 0.503(0.026) & 0.505(0.025)  & 0.508(0.025)  \\
               & rSVD  & 0.770(0.028)  & 0.621(0.024) & 0.575(0.038)  & 0.571(0.030)  \\
               & TTR(1,2) & 0.726(0.072)  & 0.508(0.020) & 0.503(0.019)  & 0.501(0.021)   \\
               & TTR(2,2) & 0.524(0.031)  & 0.523(0.025) & 0.518(0.031)  & 0.523(0.028)   \\
               & CPTR(1) & 0.521(0.033)  & 0.523(0.025) & 0.518(0.031)  & 0.523(0.028)   \\
               & CPTR(2) & 0.521(0.033)  & 0.523(0.025) & 0.519(0.033)  & 0.523(0.028)   \\
               & MVLogistic & 0.646(0.108)  & 0.542(0.048) & 0.542(0.028)  & 0.515(0.032)   \\
               &        &             &            &             &             \\
20             & LaGMaR(1,1) & 0.499(0.023)  & 0.505(0.024) & 0.498(0.025)  & 0.498(0.030)  \\
               & LaGMaR(2,2) & 0.502(0.027)  & 0.500(0.023) & 0.496(0.028)  & 0.507(0.026)  \\
               & LaGMaR(3,3) & 0.503(0.024)  & 0.501(0.022) & 0.504(0.024)  & 0.505(0.023)  \\
               & LaGMaR(4,4) & 0.503(0.026)  & 0.502(0.025) & 0.502(0.024)  & 0.507(0.025)  \\
               & LaGMaR(5,5) & 0.504(0.024)  & 0.504(0.023) & 0.506(0.024)  & 0.508(0.023)  \\
               & rSVD  & 0.757(0.021)  & 0.609(0.023) & 0.593(0.030)  & 0.579(0.031)  \\
               & TTR(1,2) & 0.736(0.085)  & 0.494(0.035) & 0.500(0.025)  & 0.496(0.016)   \\
               & TTR(2,2) & 0.520(0.025)  & 0.505(0.036) & 0.505(0.027)  & 0.514(0.019)   \\
               & CPTR(1) & 0.515(0.024)  & 0.505(0.036) & 0.505(0.026)  & 0.513(0.019)   \\
               & CPTR(2) & 0.515(0.024)  & 0.505(0.036) & 0.505(0.027)  & 0.514(0.019)   \\
               & MVLogistic & 0.818(0.072)  & 0.533(0.045) & 0.511(0.028)  & 0.521(0.030)   \\
               &        &             &            &             &             \\
50             & LaGMaR(1,1) & 0.504(0.025)  & 0.504(0.022) & 0.500(0.028)  & 0.499(0.023)  \\
               & LaGMaR(2,2) & 0.501(0.026)  & 0.503(0.028) & 0.504(0.024)  & 0.501(0.025)  \\
               & LaGMaR(3,3) & 0.502(0.027)  & 0.504(0.026) & 0.504(0.023)  & 0.500(0.024)  \\
               & LaGMaR(4,4) & 0.502(0.027)  & 0.505(0.024) & 0.503(0.021)  & 0.502(0.024)  \\
               & LaGMaR(5,5) & 0.504(0.026)  & 0.506(0.026) & 0.505(0.022)  & 0.504(0.022)  \\
               & rSVD  & 0.757(0.023)  & 0.681(0.028) & 0.657(0.044)  & 0.630(0.022)  \\
               & TTR(1,2) & 0.670(0.067)  & 0.544(0.054) & 0.522(0.024)  & 0.500(0.030)   \\
               & TTR(2,2) & 0.499(0.031)  & 0.497(0.028) & 0.495(0.026)  & 0.502(0.029)   \\
               & CPTR(1) & 0.495(0.027)  & 0.496(0.028) & 0.496(0.027)  & 0.502(0.029)   \\
               & CPTR(2) & 0.495(0.027)  & 0.496(0.028) & 0.496(0.027)  & 0.502(0.029)   \\
               & MVLogistic & 0.789(0.070)  & 0.603(0.058) & 0.534(0.036)  & 0.571(0.051)   \\
\bottomrule
\end{tabular}}
\end{table}

\begin{table}[htbp]
\caption{The means (standard errors) of RMSE of the normal response based on 100 simulated datasets with five-fold cross validation. The data generation process follows rSVD and CPTR.}
\label{zhoulirmsenormal}
\resizebox{0.8\linewidth}{!}{
\begin{tabular}{cccccc}
\toprule
Sparsity $s$(\%) & Method & \multicolumn{4}{c}{Results for the following ranks:} \cr 
\cmidrule(lr){3-6}\\
               &        & $R=1$       & $R=5$      & $R=10$      & $R=20$      \\
\toprule
10             & LaGMaR(1,1) & 21.020(0.926)  & 22.914(0.863) & 21.561(0.944)  & 21.519(0.999)  \\
               & LaGMaR(2,2) & 18.152(0.772)  & 20.370(0.895) & 22.309(1.002)  & 19.981(0.856)  \\
               & LaGMaR(3,3) & 18.253(0.783)  & 20.482(0.899) & 22.436(1.021)  & 20.098(0.862)  \\
               & LaGMaR(4,4) & 18.405(0.769)  & 20.641(0.887) & 22.601(1.053)  & 20.273(0.888)  \\
               & LaGMaR(5,5) & 18.606(0.769)  & 20.882(0.904) & 22.840(1.060)  & 20.497(0.885)  \\
               & rSVD  & 21.725(0.673)  & 20.164(0.628) & 19.239(0.523)  & 22.661(0.809)  \\
               & TTR(1,2) & 19.590(2.365)  & 24.046(1.198) & 23.573(0.842)  & 28.208(1.352)   \\
               & TTR(2,2) & 5.295(5.795)  & 35.103(2.205) & 35.649(2.035)  & 42.866(2.433)   \\
               & CPTR(1) & 1.166(0.047)  & 18.738(0.978) & 23.463(2.222)  & 29.349(1.952)   \\
               & CPTR(2) & 1.602(0.099)  & 31.910(1.707) & 35.902(1.894)  & 43.707(2.394)   \\
               &        &             &            &             &             \\
20             & LaGMaR(1,1) & 29.698(1.314)  & 28.869(1.170) & 31.045(1.456)  & 32.066(1.410)  \\
               & LaGMaR(2,2) & 31.480(1.370)  & 33.270(1.399) & 27.671(1.226)  & 37.478(1.621)  \\
               & LaGMaR(3,3) & 31.697(1.367)  & 33.452(1.414) & 27.803(1.215)  & 37.699(1.651)  \\
               & LaGMaR(4,4) & 31.965(1.412)  & 33.700(1.382) & 28.031(1.226)  & 38.007(1.659)  \\
               & LaGMaR(5,5) & 32.319(1.486)  & 34.027(1.395) & 28.325(1.270)  & 38.376(1.699)  \\
               & rSVD  & 27.586(0.785)  & 28.524(0.915) & 29.646(1.177)  & 35.067(0.985)  \\
               & TTR(1,2) & 23.194(1.989)  & 33.629(1.749) & 36.299(1.378)  & 42.995(1.988)   \\
               & TTR(2,2) & 5.830(5.871)  & 48.025(2.930) & 53.437(3.565)  & 63.285(3.336)   \\
               & CPTR(1) & 1.183(0.041)  & 25.920(1.927) & 31.551(2.024)  & 39.469(2.710)   \\
               & CPTR(2) & 1.659(0.102)  & 45.151(4.135) & 52.381(3.458)  & 63.941(3.098)   \\
               &        &             &            &             &             \\
50             & LaGMaR(1,1) & 41.756(1.736)  & 69.717(3.016) & 69.739(2.911)  & 64.858(2.561)  \\
               & LaGMaR(2,2) & 49.560(2.026)  & 64.005(2.659) & 65.265(2.770)  & 69.474(3.217)  \\
               & LaGMaR(3,3) & 49.842(2.029)  & 64.377(2.692) & 65.628(2.753)  & 69.867(3.240)  \\
               & LaGMaR(4,4) & 50.324(2.059)  & 64.871(2.656) & 66.262(2.704)  & 70.426(3.251)  \\
               & LaGMaR(5,5) & 50.857(2.026)  & 65.554(2.642) & 66.960(2.738)  & 71.164(3.331)  \\
               & rSVD  & 46.854(1.592)  & 60.084(1.987) & 65.190(1.576)  & 60.524(2.041)  \\
               & TTR(1,2) & 39.884(3.906)  & 63.090(4.053) & 72.670(4.370)  & 70.567(4.748)   \\
               & TTR(2,2) & 3.985(5.647)  & 70.462(12.828) & 93.087(10.573)  & 99.016(8.980)   \\
               & CPTR(1) & 1.188(0.063)  & 35.407(1.748) & 47.145(2.044)  & 52.010(2.117)   \\
               & CPTR(2) & 1.708(0.112)  & 55.669(3.526) & 79.877(5.880)  & 90.378(5.461)   \\
\bottomrule
\end{tabular}}
\end{table}

\begin{table}[htbp]
    \centering
	\caption{The means (standard errors) of classification metrics on COVID-CT data with 50 replicates of five-fold cross validations.}
	\label{covid}
	\begin{threeparttable}
	\begin{tabular}{c|cccc}
		\hline
		            &AUC     &Sensitivity  &CA  &Kappa \\
		\hline
		LaGMaR    & \textbf{0.785}(0.017)  & \textbf{0.669}(0.027) & \textbf{0.729}(0.016) & \textbf{0.449}(0.032)\\
		MVLogistic  & 0.742(0.006)  & 0.656(0.013) & 0.690(0.008) & 0.377(0.017) \\
		rSVD         & 0.755(0.005)  & 0.641(0.011) & 0.679(0.007) & 0.356(0.014) \\
		TTR(1,2)    &0.563(0.027)  &0.541(0.029) &0.547(0.022) & 0.092(0.044) \\
		TTR(2,2)    &0.550(0.025)  &0.539(0.030) &0.537(0.023) & 0.073(0.047) \\
		CPTR(1)    & 0.557(0.020)  & 0.542(0.028)  & 0.543(0.019) & 0.086(0.039) \\
		CPTR(2)    & 0.550(0.019)  & 0.542(0.024) & 0.539(0.016) & 0.079(0.031) \\
		\hline
	\end{tabular}
    \end{threeparttable}
\end{table}

\newpage
\begin{table}[htbp]
    \centering
	\caption{The means (standard errors) of classification metrics on complete and small COVID-CT data set with 50 replicates of five-fold cross validations.}
	\label{covid-resnet}
	\begin{threeparttable}
	\begin{tabular}{|cc|c|cccc|}
		\hline
		            &Sample   &Model   &CA       &Kappa          &Sensitivity     &AUC    \\
		\hline
&All &LaGMaR & 0.729(0.016) & 0.449(0.032)  & 0.669(0.027)   & 0.785(0.017) \\
		          & &ResNet18+FC(512, 2)  & 0.759(0.018) & 0.510(0.039)   & 0.682(0.063)        & 0.850(0.012)  \\
		          & &ResNet34+FC(512, 2)  & \textbf{0.782}(0.011) & \textbf{0.558}(0.020) & \textbf{0.721}(0.019)       & \textbf{0.869}(0.012)  \\
		          & &ResNet50+FC(2048, 2)   &0.777(0.010) & 0.547(0.021)  &0.684(0.036)       &\textbf{0.869}(0.005)  \\
		\hline
&Small	&LaGMaR & \textbf{0.684}(0.031) & \textbf{0.358}(0.063)  & \textbf{0.640}(0.049)   & 0.731(0.030) \\
		                & &ResNet18+FC(512, 2)  & 0.634(0.030) & 0.256(0.054)   & 0.436(0.095)        & 0.737(0.028)  \\
		                & &ResNet34+FC(512, 2)  & 0.620(0.027) & 0.216(0.062) & 0.346(0.104)       & 0.734(0.027)  \\
		                & &ResNet50+FC(2048, 2)   &0.606(0.032) & 0.178(0.055)  &0.327(0.141)       &\textbf{0.752}(0.015)  \\
		\hline
	\end{tabular}
	\begin{tablenotes}
	\footnotesize
	\item[1] ResNet18, ResNet34, and ResNet50 represent different ResNet backbones.
	
	\item[2] A fully-connected layer $\text{FC}(m, n)$ can be considered as a learnable linear mapping, which maps a vector $\mathbf{v} \in \mathbb{R}^{m}$ onto a vector $\mathbf{u} \in \mathbb{R}^{n}$.
	\end{tablenotes}
    \end{threeparttable}
\end{table}

\begin{table}[htbp]
\centering
\caption{The classification metrics on the test set of COVID-CT data of LaGMaR and CRNets. CRNet-Rand means that the weights of CRNet are randomly initialized, and CRNet-Trans means that the weights of CRNet are pretrained on a large-scale dataset called ImageNet.}
\begin{tabular}{cccc}
\hline
            & CA   & F1   & AUC  \\
\hline
LaGMaR      & 0.68 & 0.65 & 0.74 \\
CRNet-Rand  & 0.72 & 0.76 & 0.77 \\
CRNet-Trans & 0.73 & 0.76 & 0.79 \\
\hline
\end{tabular}
\end{table}

\label{lastpage}
\end{document}